\documentclass[
preprint2
]{aastex}

\usepackage[fleqn]{amsmath}
\usepackage{graphicx}
\usepackage[FIGBOTCAP]{subfigure}
\usepackage{txfonts}
\usepackage{natbib}
\bibpunct{(}{)}{;}{a}{}{,}
\usepackage[pdftex=true,colorlinks=true]{hyperref}
\usepackage[all]{hypcap}

\newcommand{\vONE}[1]{#1}

\title{Protostellar Outflows and Radiative Feedback from Massive
  Stars} \author{Rolf Kuiper} \affil{ Max Planck Institute for
  Astronomy, K\"onigstuhl 17, D-69117 Heidelberg, Germany\\ University
  of T\"ubingen, Institute for Astronomy and Astrophysics,
  Computational Physics, Auf der Morgenstelle 10, D-72076 T\"ubingen,
  Germany} \email{kuiper@mpia.de} \and \author{Harold W.~Yorke}
\affil{Jet Propulsion Laboratory, California Institute of Technology,
  4800 Oak Grove Drive, Pasadena, CA 91109, USA}
\email{Harold.W.Yorke@jpl.nasa.gov} \and \author{Neal J. Turner}
\affil{Jet Propulsion Laboratory, California Institute of Technology,
  4800 Oak Grove Drive, Pasadena, CA 91109, USA}
\email{Neal.J.Turner@jpl.nasa.gov}

\begin{abstract}
We carry out radiation hydrodynamical simulations of the formation of
massive stars 
\vONE{in the super-Eddington regime}
including both 
\vONE{their}
radiative feedback and protostellar
outflows.  
The calculations start from a prestellar core of dusty gas
and continue until the star stops growing.  
The accretion ends when the remnants of the core are ejected, 
mostly by the force of the direct stellar radiation in the polar direction and elsewhere by the reradiated thermal infrared radiation.  
How long the
accretion persists depends on whether the protostellar outflows are
present.  We set the mass outflow rate to 1\% of the stellar sink
particle's accretion rate.  The outflows open a bipolar cavity
extending to the core's outer edge, through which the thermal
radiation readily escapes.  The radiative flux is funneled into the
polar directions while the core's collapse proceeds near the equator.
The outflow thus extends the ``flashlight effect'', or anisotropic
radiation field, 
\vONE{found in previous studies} 
from the few hundred AU scale of the circumstellar
disk up to the 0.1-parsec scale of the core.  The core's flashlight
effect allows core gas to accrete on the disk for longer, in the same
way that the disk's flashlight effect allows disk gas to accrete on
the star for longer.  Thus although the protostellar outflows remove
material near the core's poles, causing slower stellar growth over the
first few free-fall times, they also enable accretion to go on longer
in our calculations.  The outflows ultimately lead to stars of
somewhat higher mass.
\end{abstract}

\keywords{
accretion, accretion disks ---
methods: numerical ---
stars: formation ---
stars: jets ---
stars: massive ---
stars: winds, outflows
\\
\copyright\ 2014. All rights reserved
}

\begin{document}
\maketitle

\clearpage
\section{Introduction}
\label{sect:introduction}
Protostellar outflows are ubiquitous in high-mass star forming regions
and are believed to be a direct consequence of the stars' growth.  The
outflows from massive protostars range broadly in appearance, from
highly collimated and jet-like \citep[e.g.][]{Forbrich:2009p20734,
  Linz:2010p4356, Beuther:2010p10561, Sepulveda:2011p12644,
  Wang:2011p3204, Beltran:2011p5712, 
  Moscadelli:2013p17691, Cesaroni:2013p17688, Palau:2013p16699} to
somewhat collimated \citep{Hennemann:2009p635} or confined outflow
structures \citep{Wang:2011p3204}.  \citet{Beltran:2011p5712}, for
instance, report an outflow with a poorly collimated blue-shifted lobe
and a better collimated red-shifted lobe in their source~A.  Poorly
collimated outflows sometimes surround collimated jets and hence could
result from the jet entraining ambient gas \citep[e.g.][]{
  Fallscheer:2009p245, 
  Beuther:2010p359, 
  Sepulveda:2011p12644}.  Furthermore, recent observations of a couple
of massive young stars show collimated fast jets in combination with
disk winds roughly perpendicular to Keplerian-like accretion disks
\citep{Murakawa:2013p23340}, a picture quite similar to that for
Solar-mass stars.  The observed range of outflow properties might be
an evolutionary sequence \citep{Beuther:2005p142}; a recent
observational study of the evolution of molecular outflows in
high-mass star forming regions and the associated SiO excitation
conditions is in \citet{SanchezMonge:2013p21495}.  Broader reviews of
massive stars' outflow properties are in \citet{Shepherd:2005p11163},
\citet{Arce:2007p10514}, and \citet{Bally:2008p20721}.

Theoretical studies of outflows can be divided between those focusing
on the launching, acceleration and collimation, and those dealing with
the feedback effects of outflows on the circumstellar environment.  On
the launching side of this division, tight collimation is generally
attributed to Lorentz forces.  However, the complex outflow
morphologies in high-mass star forming regions could mean that more
than one acceleration and collimation mechanism is at work.  Magnetic
fields \citep[e.g.][]{Seifried:2012p5262, Sheikhnezami:2012p21519},
molecular line radiation pressure \citep{Vaidya:2011p14600},
ionization \citep{Peters:2012p16740}, and photoevaporation of disks
\citep{Yorke:2002p22373} have all been suggested as contributing to
outflows.

Only a few studies include both the launching physics and the feedback
at large scales.  These follow the magneto-hydrodynamical (MHD)
evolution of a collapsing molecular core either with a cooling
prescription \citep{Banerjee:2007p691} or a barotropic equation of
state \citep{Hennebelle:2011p5748}.  The time span covered in such
calculations is limited by the need to simultaneously resolve the MHD
launching physics on small scales and span the large-scale collapse.
So far it has only been possible to study the early stages of collapse
--- up to half the free-fall time of the pre-stellar core.  To
circumvent the small timesteps and make it possible to follow the
outflow feedback longer, several studies have used a sub-grid scheme
to inject protostellar outflows into the computational domain.  
These
include MHD simulations \citep{Wang:2010p2487}, radiation-hydrodynamics (RHD) studies \citep{Cunningham:2011p953}, and multiple outflow interactions  \citep{Peters:2014p27736}.
Whereas \citet{Wang:2010p2487} followed the MHD collapse for roughly
2-3 free-fall times, the RHD simulations of
\citet{Cunningham:2011p953} were still limited to the initial
0.8~free-fall times, i.e.\ not long enough to determine the net effect
of outflows on the evolution.
\citet{Peters:2014p27736} studied the effect of multiple outflows from low-mass companion stars in massive star forming regions, explaining the morphology and power of observed outflows, but neglecting the effect of radiation pressure from the massive star.

Here we focus on the long-term evolution of the collapsing pre-stellar environment.
A protostar forms, grows through accretion, initiates an
outflow, and continues to grow until it becomes so luminous that its
radiative forces shape the dynamics of the surrounding gas.
The high optical depth of a massive accretion disk surrounding a protostar significantly reduces the radiation pressure onto the stellar accretion flow by causing an anisotropy of the reemitted thermal radiation field, a mechanism also known as the ``flashlight effect''  \citep{Yorke:1999p156, Yorke:2002p735, Krumholz:2005p406, Kuiper:2010p541}.

\vONE{
In general, protostellar outflows impact the stellar environment via their momentum feedback \citep[see][for reviews]{Ray:2007p10756, Arce:2007p10514, Frank:2014p28089}.
Hence, outflows are assumed to have a negative SFE feedback via decreasing the accretion rate and, hence, limiting the accreted mass.
On the other hand, for massive stars in the super-Eddington regime (i.e.~their environment is severely affected by radiative forces), early protostellar outflows may alter the later radiative feedback epoch of these stars.
Indeed, a low-density / optically thin cavity produced by protostellar outflows might decrease the radiative feedback on the accretion flow, i.e.~outflows also have a positive SFE feedback.
Conceptually, this effect was already shown in purely radiation transport simulations of a static configuration with a star, a disk, an envelope, and an outflow cavity in \citet{Krumholz:2005p406}.
To finally conclude, which of these effects is stronger - the negative or the positive one - dynamical studies of the star formation process are essential.
But previous radiation-hydrodynamical studies including the effect of radiative forces \citep[such as][]{Krumholz:2007p416, Cunningham:2011p953, Commercon:2011p1547} have been performed only for the sub-Eddington / marginal-Eddington regime.
Here, we go beyond this barrier, to quantitatively determine the efficiency of both effects and their interplay for substantially massive stars.

Such super-Eddington massive stars are assumed to form naturally via global cloud collapse as presented in recent observations by \citet{Peretto:2013p26092}.
}

When the high optical depth cannot further be sustained by ongoing accretion from the envelope onto the disk,
ultimately, the radiation forces overcome the gravitational attraction of the protostar and
reverse the infall.  
We follow the evolution of the core through the
entire stellar accretion phase, stopping the calculations only after
the initial mass within our computational domain is either accreted
onto the central star or expelled through the outer boundary at a
radius of 0.1~pc.  The simulations run 6.7 and 5.7 free-fall times
starting with pre-stellar cores having initial density profiles $\rho
\propto r^{-1.5}$ and $\rho \propto r^{-2.0}$, respectively.  To
separate the consequences of protostellar outflows from radiation
feedback, we also perform comparison calculations without the
outflows.  
The difference between these two is expected to be the cavities opened by the early protostellar outflows and these will affect the radiation field.

We treat the radiation pressure feedback using a hybrid radiative
transport scheme, calculating the stellar irradiation by a ray-tracing
method and the thermal dust emission by flux-limited diffusion.
For details, we refer the reader to the methods section below.

In Sect.~\ref{sect:methods} we briefly describe our star formation
modeling framework and the sub-grid model for the protostellar
outflows, and in Sect.~\ref{sect:setup} the initial state of the
pre-stellar cores.  The results are presented in four sections
focusing on the dominant forces (Sect.~\ref{sect:results-forces}), the
core evolution and mass loss (Sect.~\ref{sect:results-core})), the
outflows (Sect.~\ref{sect:results-outflow}) and the disk and stellar
accretion (Sect.~\ref{sect:results-disk}).  In
Sect.~\ref{sect:observations} we discuss the observational
implications and in Sect.~\ref{sect:limitations} reiterate the
limitations of this study and outline the prospects for follow-up
work.  A summary is in Sect.~\ref{sect:summary}.

\section{Methods}
\label{sect:methods}
\subsection{General Star Formation Framework}
To follow the evolution of the gas and dust in a collapsing, slowly
rotating pre-stellar core, we solve the equations of compressible
radiation-hydrodynamics, including self-gravity and shear viscosity.
The basic numerical code is the same as in our previous studies
\citep{Kuiper:2010p541, Kuiper:2011p21204, Kuiper:2012p1151}.  The
hydrodynamical solver we use is the open source magneto-hydrodynamics
code Pluto \citep{Mignone:2007p544}.  The coordinate system is
spherical with radial and colatitude coordinates $(r,\theta)$.

A special feature of the code is our hybrid radiation transport
scheme, which combines a highly accurate frequency-dependent
ray-tracing step for the stellar irradiation with a gray flux-limited
diffusion (FLD) approximation for the dust thermal (re-)emission.
Detailed description of the derivation, numerical implementation, and
benchmarking of the radiation transport solver are in
\citet{Kuiper:2010p586}.  Our hybrid approach is a major improvement
over the pure gray FLD approximation
\citep[e.g.][]{Bodenheimer:1990p615, Yorke:1995p550}, in which not
only the thermal dust emission but also the stellar irradiation is
computed in the FLD approximation.  
Especially when radiation dominates the energy budget of an optically thin outflow
cavity, the stellar irradiation must be treated accurately to
correctly model the dynamical evolution \citep{Kuiper:2012p1151}.
Most importantly, the much higher opacity of the stellar irradiation in contrast to the rather low opacity of the thermal radiation field has to be taken into account properly.
Furthermore, the technique takes into
account the irradiation's anisotropy in the optically thin limit
(relevant for studies of molecular outflows) and renders the formation
of shadows behind highly optically thick regions (relevant for studies
of circumstellar disks).  For a setup including a star, a slightly
flared disk, and an envelope, the hybrid method is nearly as accurate
as modern Monte-Carlo radiative transfer codes (excluding the effect
of scattering) in all optical depth regimes \citep{Kuiper:2013p19458},
but much faster.

The spectrum of the stellar irradiation is treated using 79~frequency
bins.  Across these bins the dust opacity's frequency dependence is
drawn from \citet{Laor:1993p358}.  Frequency-averaged (gray) Planck
and Rosseland mean opacities for the thermal dust emission are
computed from the frequency-dependent dust opacities.  In dust-free
regions the gas opacity is used instead of dust and is fixed to a
constant value $\tau = 0.01 \mbox{ cm}^2 \mbox{ g}^{-1}$.

The central star evolves following a sub-grid model computed by
fitting the stellar evolutionary tracks of \citet{Hosokawa:2009p23}.
These tracks include the effects of accretion onto forming high-mass
stars.
Including the effects of protostellar outflows is a straightforward
extension of the self-gravity radiation-hydrodynamical star formation
framework of \citet{Kuiper:2010p541}.

\subsection{Protostellar Outflows Sub-Grid Model}
A consistent study of the launching mechanism(s) and the developing
outflow structure within the collapsing pre-stellar core would require
computing the MHD evolution of the collapse down to a small fraction
of an AU resolution to guarantee meaningful results.  Due to limited
computing resources, we can currently perform such a self-consistent
study in 2D for about one free-fall time of the pre-stellar core.  To
enable this investigation of protostellar outflow feedback over the
course of the protostar's entire accretion phase, we adopt the
following approach.  Rather than modeling the launching mechanism(s)
and structures of protostellar outflows from basic principles, we use
a sub-grid model, whereby the momentum injection and mass loss rate of
the protostellar outflows follow an analytical prescription \citep{Cunningham:2011p953}, and depend
on the protostellar parameters (mass, radius, and
accretion rate).

In these simulations a protostar forms and evolves at the center of
our grid, i.e.\ within a sink cell of radius $R_\mathrm{min}$, the
inner radial boundary of the computational domain.  The protostellar
outflow associated with the protostar can therefore be implemented in
the numerical scheme as an inner boundary condition at
$R_\mathrm{min}$, which we set equal to 10~AU.  Because the
protostellar outflow's launching is generally attributed to
magneto-centrifugal acceleration and/or twisting of magnetic field
lines near the accretion disk's inner rim, we turn on the outflow when
a rotation-supported disk first forms in the calculation.  The exact
time to initiate the outflow is chosen based on the corresponding
reference simulation without an outflow.  The switching-on occurs when
the central protostar has grown to $8 \mbox{ M}_\odot$ and $25 \mbox{
  M}_\odot$ in the cases with initial core density profiles $\rho
\propto r^{-\beta}$ with power law indexes $\beta = 1.5$ and $\beta =
2$, respectively.

Once the outflow turns on, the outward radial velocity at the inner
radial boundary is set proportional to the Keplerian velocity at the
stellar radius
\begin{equation}
  v_\mathrm{BC} = f_\mathrm{v} \times v_\mathrm{Kepler}(R_*)
\end{equation}
with $v_\mathrm{Kepler}(R_*) = \sqrt{G M_* / R_*}$, where $G$ is the
gravitational constant and $M_*$ and $R_*$ are the protostellar mass
and radius.  Here we use $f_\mathrm{v} = 1/3$.  The parametrization
above can also be given in terms of the escape speed
$v_\mathrm{escape}$ at the injection radius $R_\mathrm{min}$
\begin{equation}
  v_\mathrm{BC} = f_\mathrm{v} \times \sqrt{\frac{R_\mathrm{min}}{2
    R_*}} \times v_\mathrm{escape}(R_\mathrm{min}).
\end{equation}
At all epochs of the simulations, the injection velocity is higher
than the corresponding escape velocity.  The mean ratio $v_\mathrm{BC}
\approx 3.5~v_\mathrm{escape}(R_\mathrm{min})$.

The density at the inner radial boundary is set proportional to the
actual accretion rate $\dot{M}_*$ onto the protostar
\begin{equation}
  \rho_\mathrm{BC} = f_{\dot{M}} \times f(\theta) \times
  \frac{\dot{M}_*}{4 \pi~R_\mathrm{min}^2~v_\mathrm{BC} }
\end{equation}
The mass loss rate $\dot{M}_\mathrm{outflow} = 4 \pi r^2 \rho(r) v(r)$
in the outflow is
\begin{equation}
  \dot{M}_\mathrm{outflow} = f_{\dot{M}} \times f(\theta) \times \dot{M}_*.
\end{equation}

\begin{figure}[htbp]
  \includegraphics[width=0.47\textwidth]{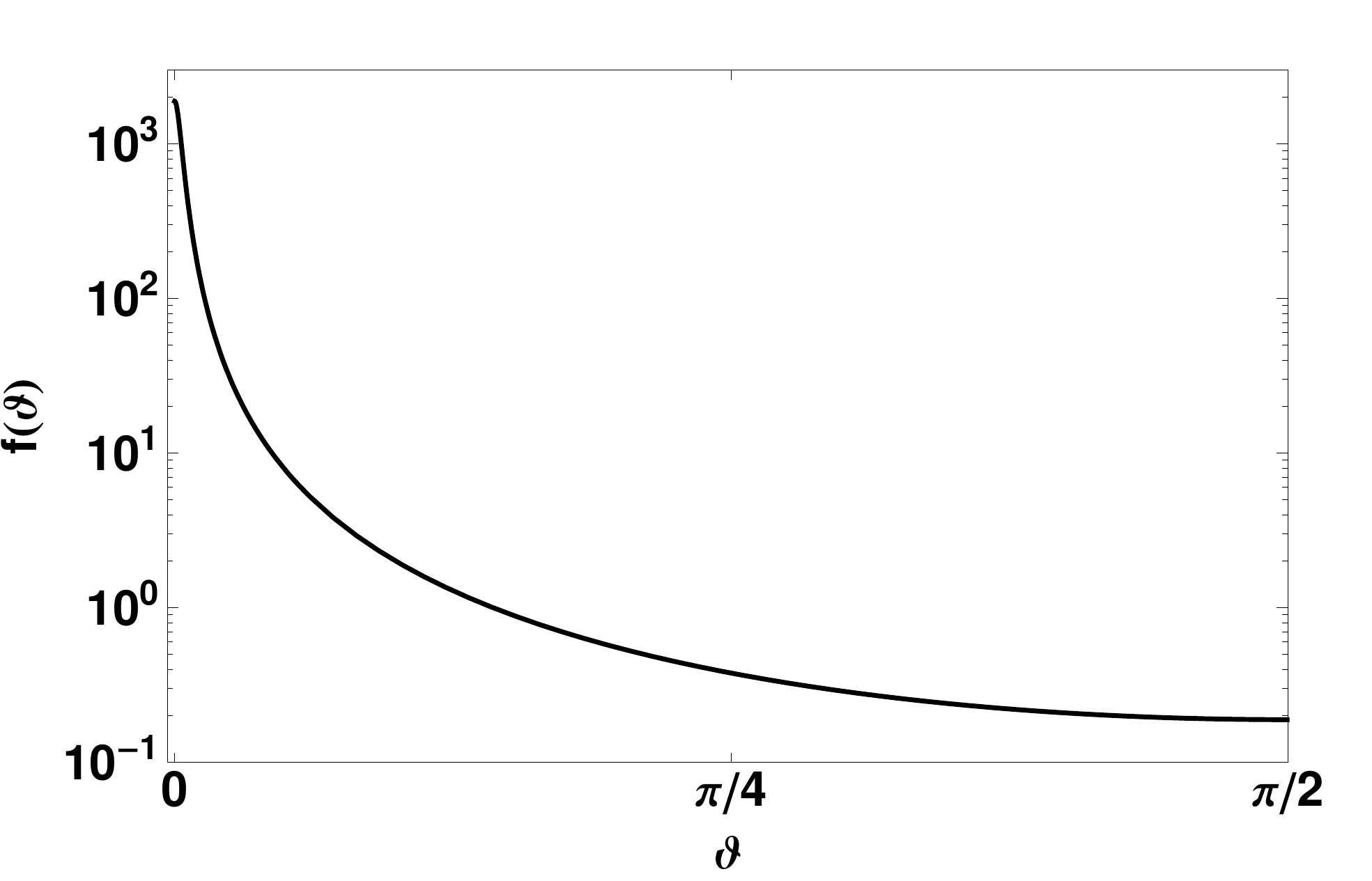}
  \caption{ Angular weighting of the protostellar outflow injection as
    function of the angle with respect to the rotation axis.  }
  \label{fig:AngularDependence}
\end{figure}

The outflow is concentrated near the polar axis.  The angular
distribution $f(\theta)$ is conveniently parametrized using a
formulation from \citet{Matzner:1999p13489}
\begin{equation}
\label{eq:angularweighting}
  f(\theta) = \left(\ln \left(2/\theta_0 \right) * \left( \sin^2
  \theta + \theta_0^2 \right)\right)^{-1}.
\end{equation}
This yields collimated flows up and down the rotation axis and broad
wings toward the disk's mid-plane $\theta = \pi/2$, resembling the
jets observed within broader molecular outflows.  The so-called
flattening of the distribution is given by the constant $\theta_0$,
which we set to 0.01 in accordance with \citet{Cunningham:2011p953},
giving the angular weighting in Fig.~\ref{fig:AngularDependence}.
The low value of $\theta_0 \ll 1$ yields an outflow force feedback
focussed near the rotation axis, most strongly impacting the poles.
The momentum flux injected at ($\theta=0$) is four orders of magnitude
greater than in the disk mid-plane ($\theta=\pi/2$).

Protostellar outflows have several feedback effects.  First, the
accretion onto the star is reduced by $f_\mathrm{\dot{M}}$ (``mass
loss feedback'').  Second, the outflow changes the dynamics of the
stellar environment due to additional outward momentum
$\rho_\mathrm{outflow}~v_\mathrm{outflow}$ (``kinematic feedback'').
Third, the outflows clear material away from the poles, making the
ambient density more anisotropic.  This is especially important for
high-mass stars, where the starlight radiative forces eventually
dominate (``radiative feedback'').  For simplicity and to allow
straightforward comparison with simulations lacking the protostellar
outflows, here we focus on the kinematic and radiative feedback,
setting $f_{\dot{M}}=0.01$ so that mass loss feedback is minor.

\subsection{Comparison to Cunningham et al.~(2011)}
While the analytical formulae we use to parametrize the outflows are
identical to \citet{Cunningham:2011p953}, the numerical implementation
differs because of the particulars of the computational grids.
\citet{Cunningham:2011p953} use a 3D Cartesian adaptive mesh with
maximum resolution $\Delta x \times \Delta y \times \Delta z = (10
\mbox{ AU})^3$.  We use a 2D static grid in spherical coordinates with
maximum resolution $\Delta r \times \Delta \theta= 1.27 \times 1.04
\mbox{ AU}^2$.  In \citet{Cunningham:2011p953}, forming protostars are
represented by sink particles.  A protostellar outflow associated with
a sink particle is injected into the ambient gas via a kernel
function, which spreads the outflow injection over four to eight grid
cells radially around the sink particle, i.e.~within $r = 40 - 80$~AU.
In our treatment, the protostar forms in a sink cell at the center of
the spherical grid.  Hence, its interaction with the ambient medium
can be described via ordinary boundary conditions at the innermost
grid point, at $R_\mathrm{min} = 10$~AU.

We aim to run the simulations long enough to follow the protostar's
whole accretion phase, and to investigate the protostellar outflows'
effects at late epochs when the star grows so luminous that radiative
forces dominate the dynamics.  This is made possible by choosing a
simpler setup than \citet{Cunningham:2011p953}: our models are
symmetric around the rotation axis and about the midplane.  High
spatial resolution is important especially in the innermost regions,
where the protostellar outflow is injected and first interacts with
the envelope medium, and where radiative forces act on the inner rim
of the accretion disk.  High resolution is achieved by using spherical
coordinates with a logarithmically-spaced radial grid.  The pressure
scale height of the accretion disk is resolved by at least one grid
cell in our simulations.  In \citet{Cunningham:2011p953} the disk
scale height was unresolved and therefore -- to avoid an artificial
disruption of the forming disk by the protostellar outflow -- the
outflow injection was switched off in an equatorial belt at least one
cell thick ($\Delta x \ge 10$~AU).  No such switch is needed in our
calculations, allowing us to investigate the accretion flow's
interaction with the outflow and starlight, both as the disk is
forming and in late evolutionary stages.

\section{Initial Conditions}
\label{sect:setup}
We consider four cases.  All start with a pre-stellar core containing
$100 \mbox{ M}_\odot$, 0.1~pc in radius, with a uniform temperature
20~K and in slow solid body rotation at $\Omega = 2 \times
10^{-13}$~s$^{-1}$.  For each of two different initial density
distributions, $\rho \propto r^{-1.5}$ and $\rho \propto r^{-2}$, we
carry out runs with and without protostellar outflows.  The four cases
are listed in Table~\ref{tab:run-table}.
\begin{table*}[tb]
\begin{center}
\begin{tabular}{l | c c | c c c}
Run label & $\rho \propto r^{-\beta}$ & Protostellar Outflow & $\overline{\dot{M}_*}~[10^{-4}\mbox{ M}_\odot \mbox{ yr}^{-1}]$ & $t_\mathrm{acc}$~[kyr] & $M_*^\mathrm{final}~[\mbox{M}_\odot]$ \\
\hline
rho1.5 		& 1.5 & No	& 1.2		& 350		& 42 \\
rho1.5-PO 	& 1.5 & Yes	& 1.0		& 460		& 47 \\
rho2.0 		& 2.0 & No	& 5.8 	& 90			& 52 \\
rho2.0-PO 	& 2.0 & Yes	&1.5 		& 360		& 55
\end{tabular}
\end{center}
\caption{ Overview of simulations performed.  The first three columns
  specify the run label, the initial slope $\beta$ of the density, and
  whether the protostellar outflow feedback is included in the
  simulation.  The last three columns give the resulting mean
  accretion rate, the duration of the accretion epoch, and the final
  mass of the protostar.  }
\label{tab:run-table}
\end{table*}

\section{Dominant Forces}
\label{sect:results-forces}
Protostars' masses are ultimately determined by what happens to the
reservoir of potentially accretable material.  In our calculations the
growth ends when the remaining envelope is expelled either by
radiation forces or by entrainment in outflows.  To gain insight into
these feedback processes, in this section we examine the main forces
along the radial direction.

Initially the flow is governed by collapse under self-gravity.  As
collapse proceeds, centrifugal forces increase near the center of the
core due to angular momentum conservation, until a rotation-supported
disk forms in orbit around the protostar.  Once the disk is in place,
a protostellar outflow is launched and injects outward momentum into
the ambient medium.  The protostar's luminosity grows until eventually
radiative forces overcome the star's gravitational attraction.
Although thermal pressure builds up as the core contracts and the star
brightens, its contribution to the radial forces remains limited.
Note, however, that we neglect photoionization of the
ambient medium and heating by photoelectrons, both of which increase thermal pressure forces.  
UV feedback is most likely limited to later epochs of stellar evolution
(than we study here), after the accreting protostar reaches the main
sequence \citep{Hosokawa:2009p23, Hosokawa:2010p690, Kuiper:2013p19987}.  

\begin{figure*}[p]
\begin{center}
\hspace{10mm}\includegraphics[width=0.89\textwidth]{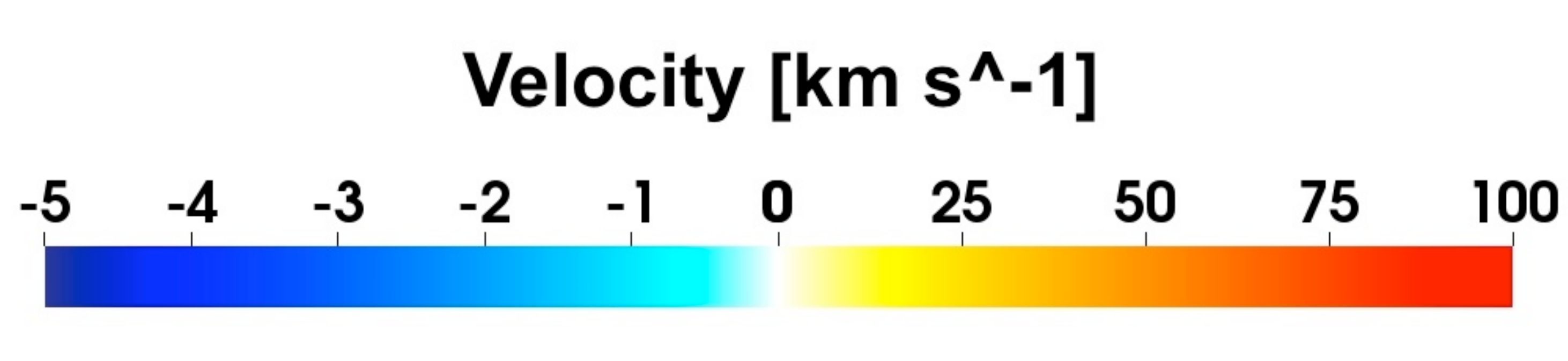}\\
\includegraphics[width=0.95\textwidth]{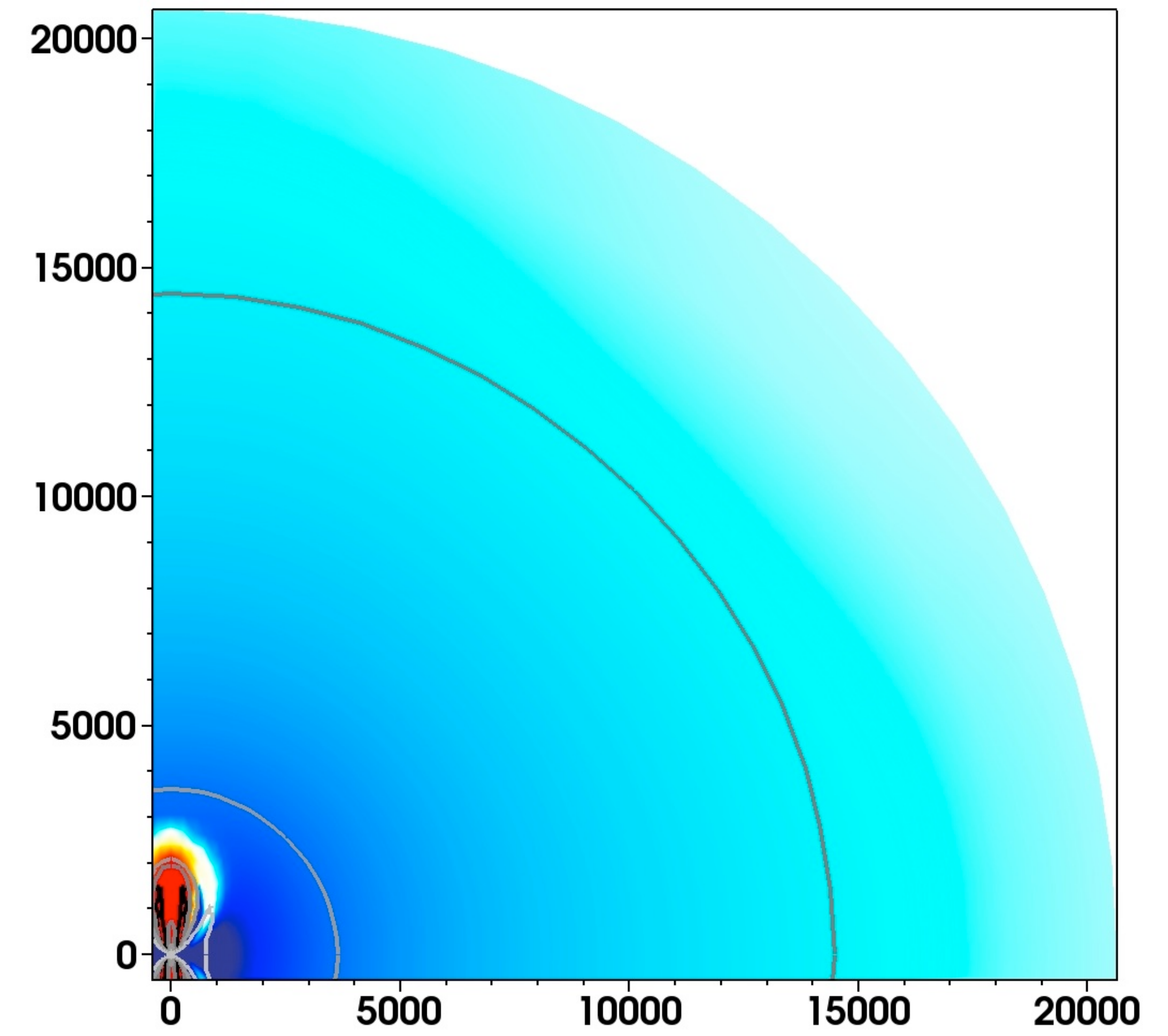}
\caption{ 
Infall and outflow velocities across the computational domain extending to 0.1~pc from the protostar.  
Light- to dark-blue colors denote radial infall and yellow to red colors radial outflow.
Four isodensity contours are overplotted in the intermediate- to low-density gas 
($10^{-19} \mbox{ g cm}^{-3} \le \rho \le 10^{-16} \mbox{ g cm}^{-3}$). 
The axes are labeled in AU. 
The snapshot is from run rho2.0-PO at $t = 15$~kyr 
($M_* \approx 24 \mbox{ M}_\odot$). 
\vONE{
According to the classification of Sect.~\ref{sect:classification}, this point in time belongs to the ``launching phase'' (stage I).
}
}
\label{fig:VisIt-WholeDomain1}
\end{center}
\end{figure*}

\begin{figure*}[p]
\begin{center}
\includegraphics[width=0.95\textwidth]{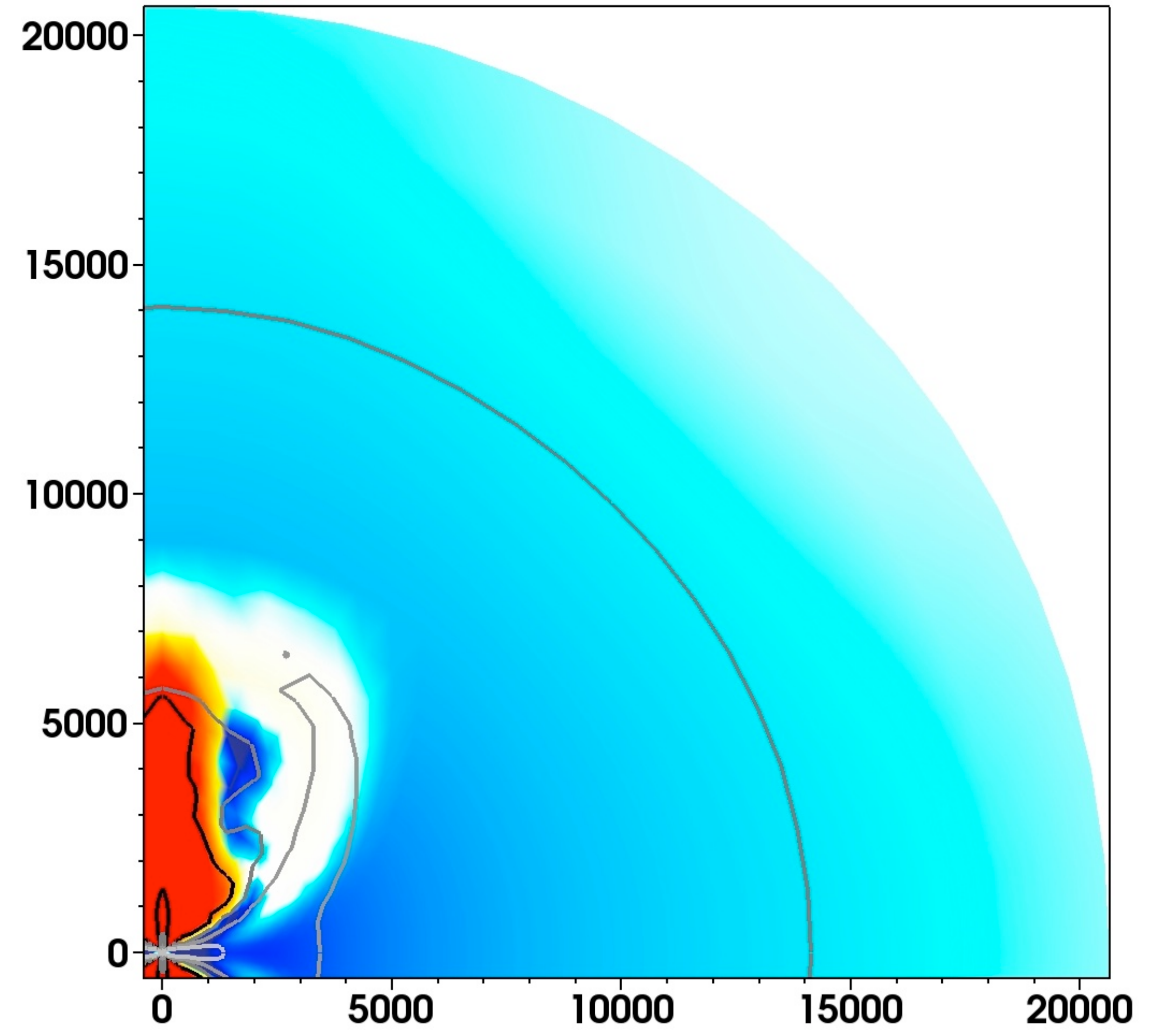}
\caption{ Same as Fig.~\ref{fig:VisIt-WholeDomain1} but at $t=18$~kyr ($M_* \approx 28 \mbox{ M}_\odot$).
\vONE{
According to the classification of Sect.~\ref{sect:classification}, this point in time marks the end of the ``launching phase'' (stage I).
}
}
\label{fig:VisIt-WholeDomain2}
\end{center}
\end{figure*}

\begin{figure*}[p]
\begin{center}
\includegraphics[width=0.95\textwidth]{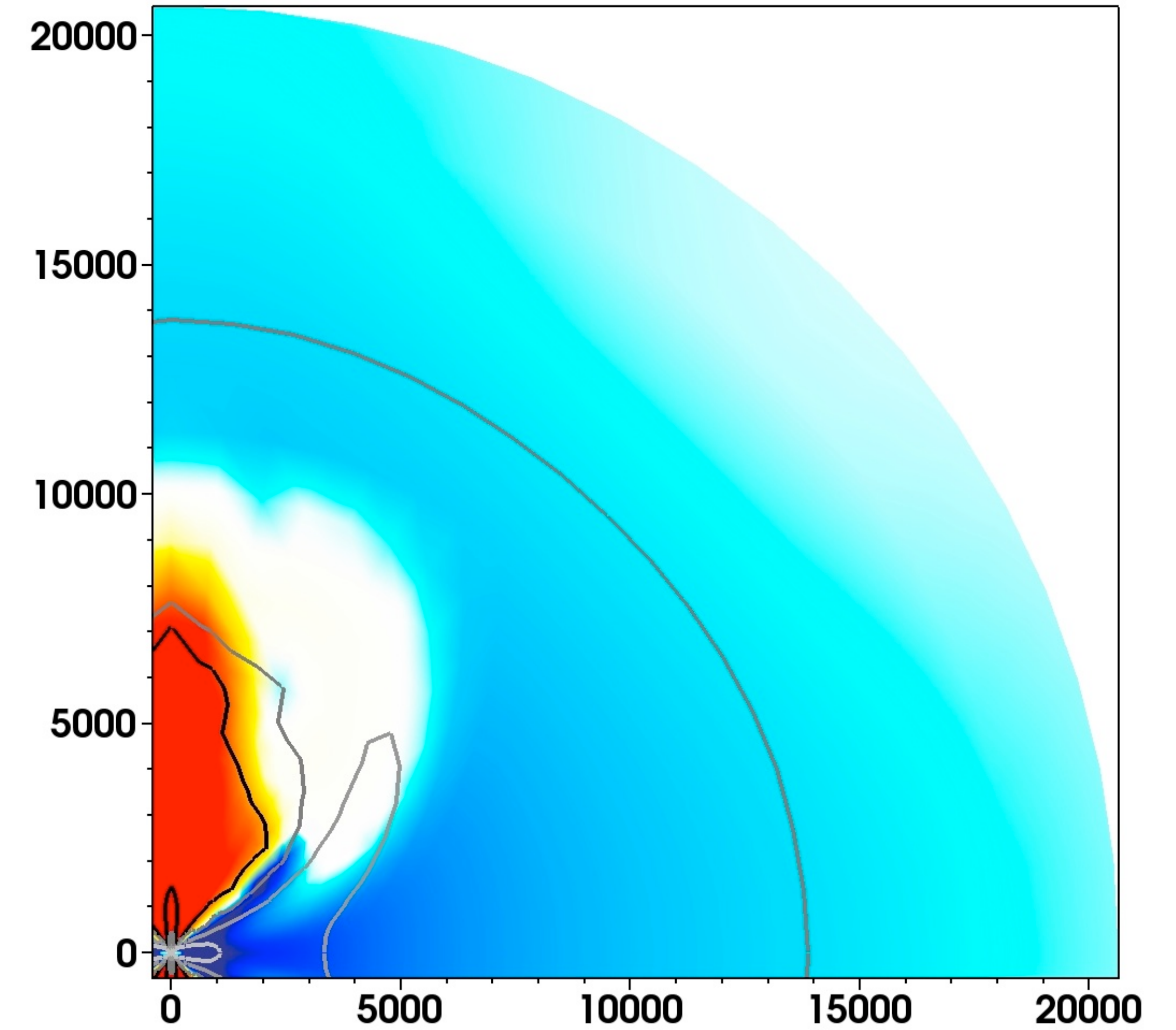}
\caption{ Same as Fig.~\ref{fig:VisIt-WholeDomain1} but at $t=20$~kyr ($M_* \approx 30 \mbox{ M}_\odot$).
\vONE{
According to the classification of Sect.~\ref{sect:classification}, this point in time belongs to the ``quasi-stationary expansion phase'' (stage II).
}
}
\label{fig:VisIt-WholeDomain3}
\end{center}
\end{figure*}

\begin{figure*}[p]
\begin{center}
\includegraphics[width=0.95\textwidth]{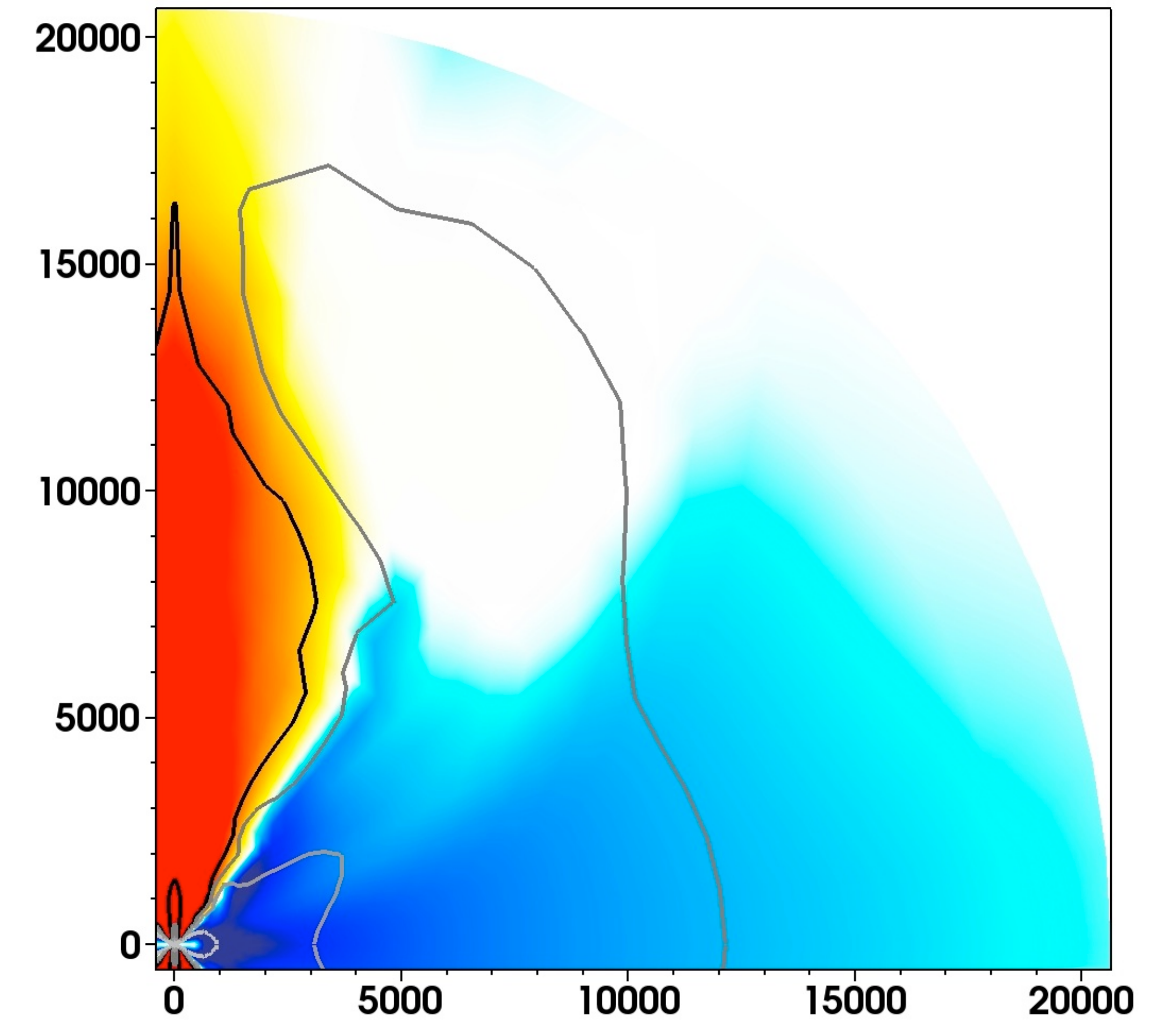}
\caption{ Same as Fig.~\ref{fig:VisIt-WholeDomain1} but at $t=30$~kyr ($M_* \approx 35 \mbox{ M}_\odot$).
\vONE{
According to the classification of Sect.~\ref{sect:classification}, this point in time marks the end of the ``quasi-stationary expansion phase'' (stage II).
}
}
\label{fig:VisIt-WholeDomain5}
\end{center}
\end{figure*}

\begin{figure*}[p]
\begin{center}
\includegraphics[width=0.95\textwidth]{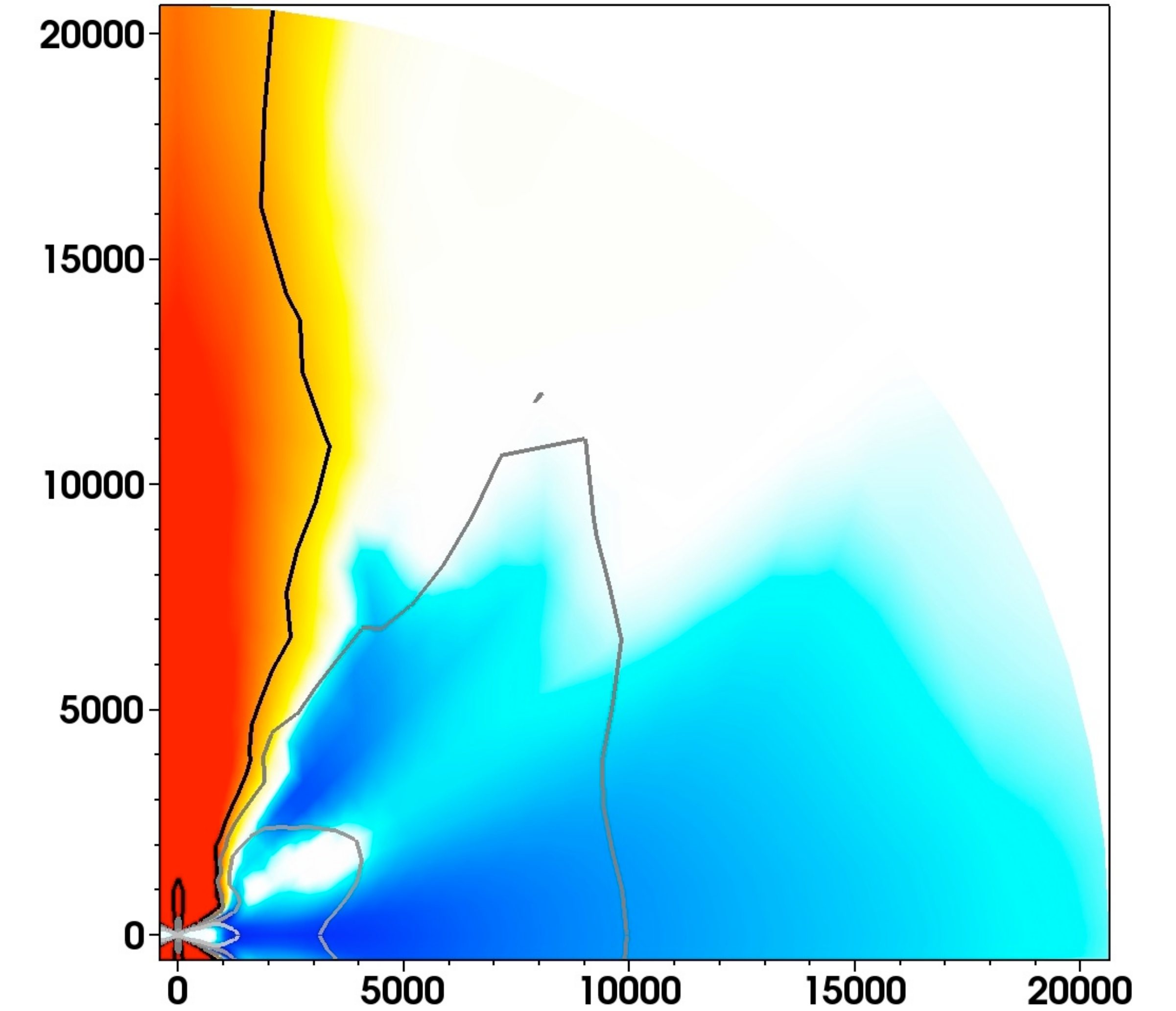}
\caption{ Same as Fig.~\ref{fig:VisIt-WholeDomain1} but at $t=40$~kyr ($M_* \approx 40 \mbox{ M}_\odot$).
\vONE{
According to the classification of Sect.~\ref{sect:classification}, the system is still at the beginning of the ``radiation-pressure-dominated phase'' (stage III).
}
}
\label{fig:VisIt-WholeDomain7}
\end{center}
\end{figure*}

\begin{figure*}[p]
\begin{center}
\includegraphics[width=0.95\textwidth]{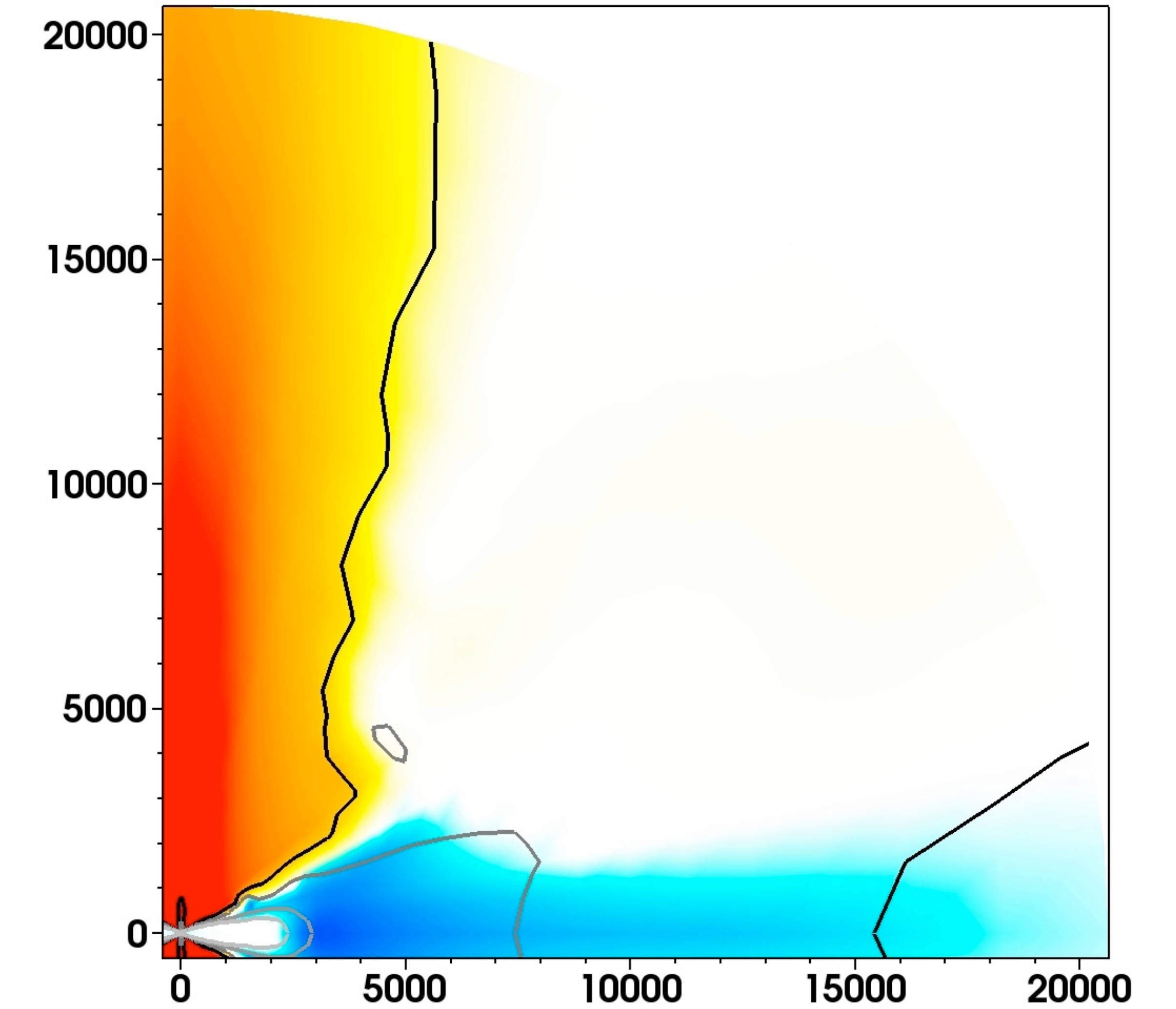}
\caption{ Same as Fig.~\ref{fig:VisIt-WholeDomain1} but at $t=60$~kyr ($M_* \approx 46 \mbox{ M}_\odot$).
\vONE{
According to the classification of Sect.~\ref{sect:classification}, the point in time belongs to the ``radiation-pressure-dominated phase'' (stage III).
}
}
\label{fig:VisIt-WholeDomain9}
\end{center}
\end{figure*}

Thus the radial dynamics of the polar regions during the pre-main
sequence phases are dominated by gravity, outflows and stellar radiative
forces.  The resulting velocity patterns in the pre-stellar core
collapse are shown in Figs.~\ref{fig:VisIt-WholeDomain1} to
\ref{fig:VisIt-WholeDomain9}.  The figures distinguish the inward
accretion flow from the outward motion due to protostellar outflow and both stellar and thermal
radiation pressure feedback.  The time sequence is chosen to display
the clearing of the bipolar region.  In the accretion disk the radial
flows are controlled by gravity approximately balanced by centrifugal
forces, with radiative forces contributing at certain times.

\section{Core Evolution and Mass Loss}
\label{sect:results-core}
In all four runs, with and without the feedback from protostellar
outflows, the radiative forces remove a substantial fraction of the
envelope.  The epoch of radiative feedback starts once the protostar
becomes super-Eddington with respect to the opacities of the accreting
material.  In the two runs without protostellar outflows, radiative
and centrifugal forces reverse the flow closely above (and below) the
accretion disk, launching a wide-angle wind.  
Closer to the rotation
axis, the stellar radiative forces reverse the flow, launching a bipolar
outflow.  
At the same time the high midplane optical depth of the
accretion disk means the thermal dust reemission is strongly
anisotropic due to the flashlight effect, enhancing the radiative forces in the bipolar outflow while reducing
them at lower latitudes.

The two runs including the protostellar outflows differ once the disk
forms and the outflows are injected.  Accretion onto the protostar is
then slower than in the versions without the outflows, a direct result
of the outflows entraining and carrying away envelope material.  We
designate this the ``kinematic'' feedback to contrast it with the
``radiative'' feedback.

\begin{figure*}[htbp]
\begin{center}
\subfigure{\includegraphics[width=0.44\textwidth]{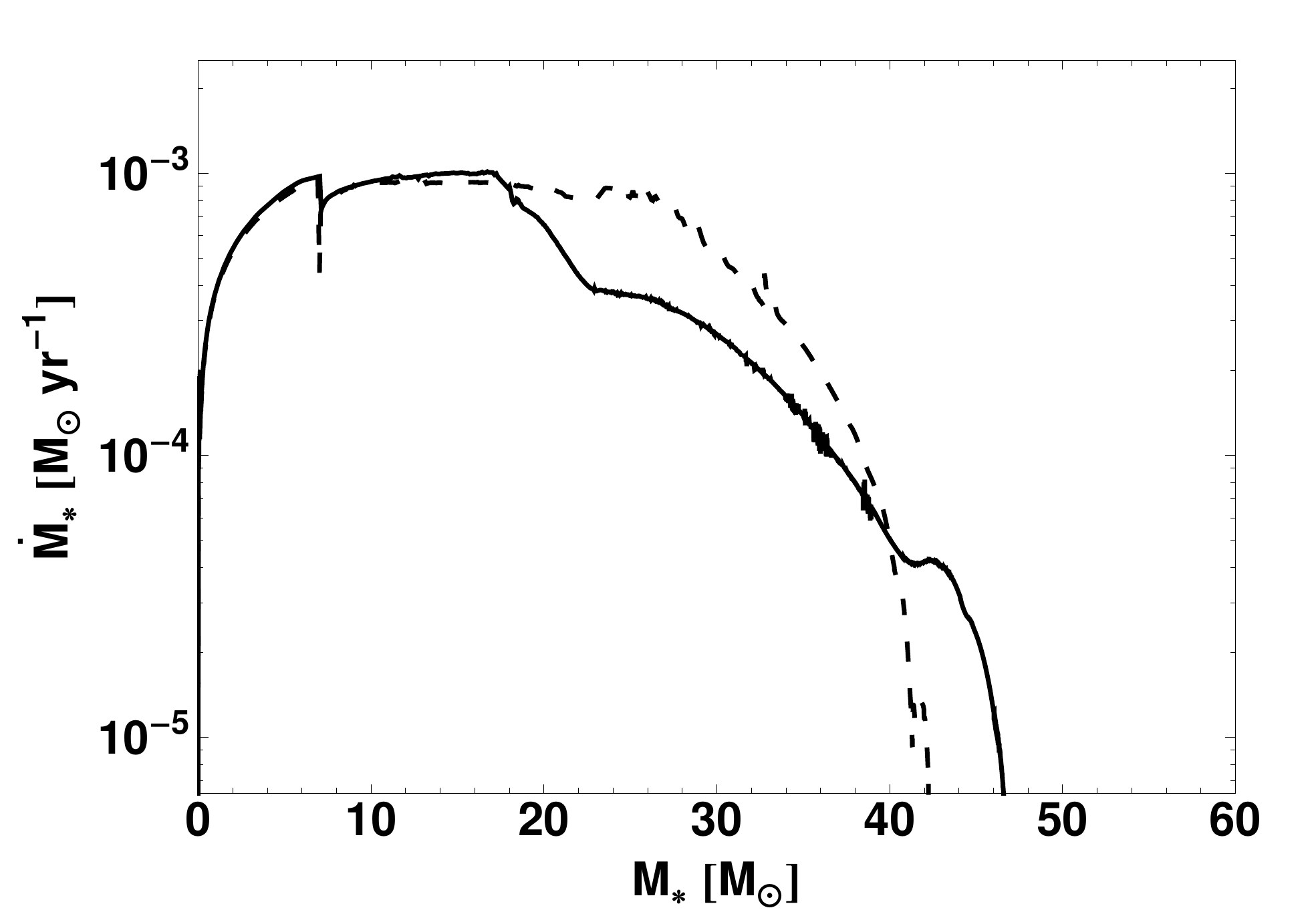}}
\subfigure{\includegraphics[width=0.44\textwidth]{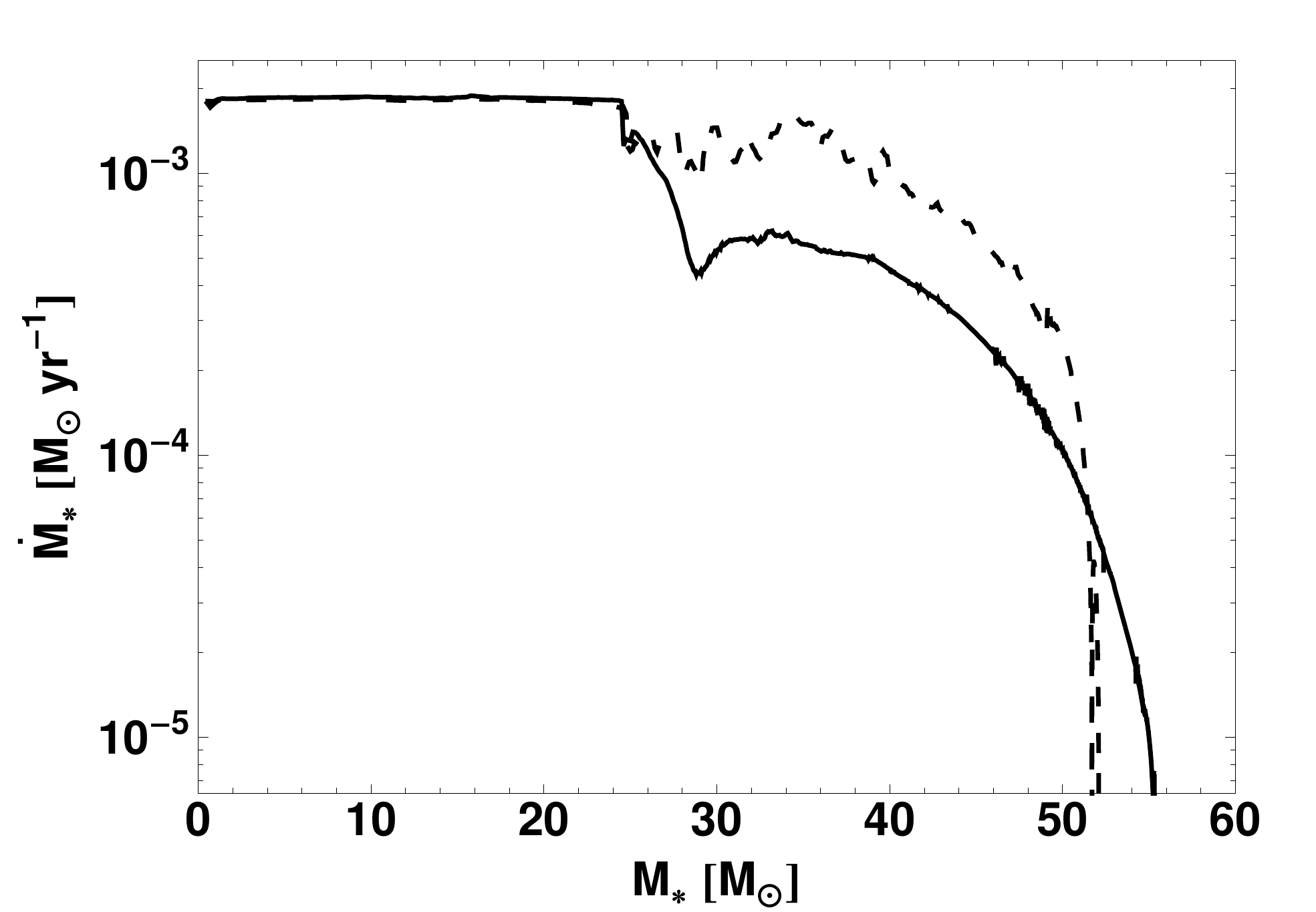}}\\
\hspace{3.2mm}\subfigure{\includegraphics[width=0.40\textwidth]{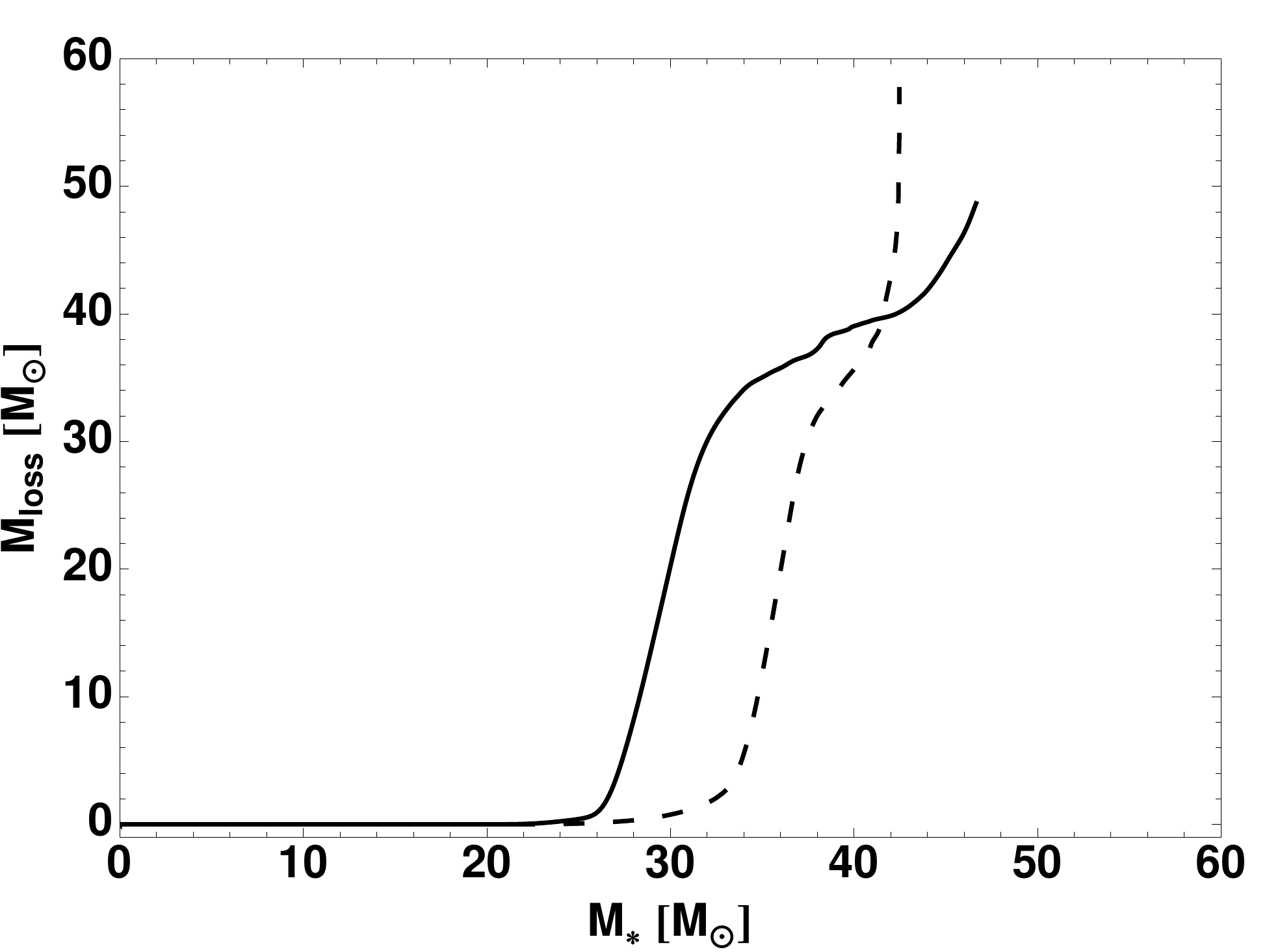}}
\hspace{6.5mm}\subfigure{\includegraphics[width=0.40\textwidth]{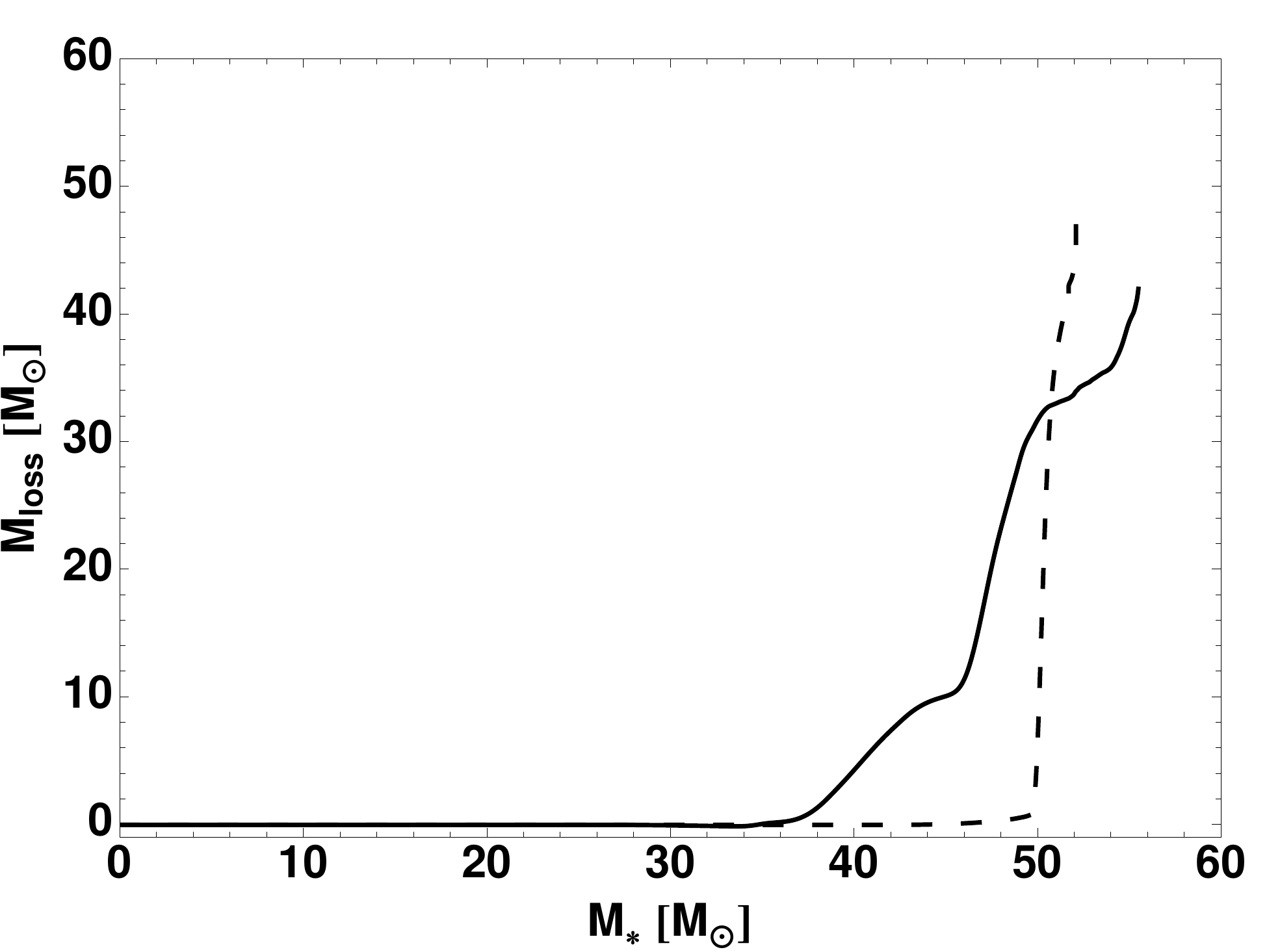}}
\caption{ Stellar accretion rate $\dot{M}_*$ (upper panels) and mass
  loss from the protostellar cores $M_\mathrm{loss}$ (lower panels) as
  functions of the accreted stellar mass $M_*$ for the two cases $\rho
  \propto r^{-1.5}$ (left panels) and $\rho \propto r^{-2}$ (right
  panels).  Solid (dashed) lines depict the cases with (without)
  protostellar outflows.  $M_\mathrm{loss}$ is the integrated outward
  mass flux through a sphere of radius 0.1~pc.  }
\label{fig:scalars}
\end{center}
\end{figure*}

By ejecting envelope material from the system, the outflows alter the
evolution of the disk and protostar.  Fig.~\ref{fig:scalars} shows the
rates of stellar accretion $\dot{M}_*$ (upper panels) and mass loss
from the prestellar core $M_\mathrm{loss}$ (lower panels) as functions
of the accreted stellar mass $M_*$.
The kinematic outflow feedback coincides with a decline in the star's
accretion rate (fig.~\ref{fig:scalars}, upper panels).  As the star gains mass, the radiative feedback becomes more important (lower
panels) and eventually ends the stellar accretion (upper panels).
Both feedback effects act first on small scales to diminish the
stellar growth rate, and later on large scales to expel mass from the
core.

\section{Outflow Evolution}
\label{sect:results-outflow}
As with the core, the bipolar outflows are governed by the injected
protostellar outflows at early times and later also shaped by the stellar
radiative forces.  Below we divide the discussion between the
outflows' dynamics, momentum and their opening angle
history.

\subsection{Outflow Dynamics}
The outflow dynamics at low latitudes is complex.  Basically, in the
runs without protostellar outflows, the polar regions' dynamics are
dominated by gravity and stellar radiation forces.  At low latitudes,
centrifugal forces also contribute, yielding lower effective
gravities.  The radiatively-driven outflow therefore starts first at
the lower latitudes, yielding a quadrupole or butterfly shape around
the protostar.  Shortly afterward, the radiation pressure also halts
the infall along the rotation axis.  In short order the flow reverses
all the way to the boundary.

In the runs with protostellar outflows, the injected momentum leads to
a disk wind at earlier epochs, before the onset of strong radiation
forces.  The centrifugal forces are unimportant because the outflow
is strong enough to drive the disk wind without help.  Once the
radiation pressure increases, the dynamics at low latitudes are
controlled by the kinematic feedback, the radiative feedback, and
gravity.
\vONE{
E.g.~for the transition from kinematically dominated to radiatively dominated outflows, see Fig.~\ref{fig:VisIt-WholeDomain7}, in which the white-colored region above and behind the accretion disk denotes the onset of flow reversal due to dominating radiative forces within the collapsing environment.
}

\subsection{Momentum Distribution}
Beginning soon after the protostellar outflows are first launched, and
continuing until stellar radiation forces become important, the outflow
momentum distribution is characterized by a collimated high-momentum
jet-like structure ending in a somewhat wider bow shock.  The degree
of collimation is determined by the angular weighting used when
injecting the outflows (Sect.~\ref{sect:methods}).

\begin{figure*}[p]
\begin{center}
\includegraphics[width=0.79\textwidth]{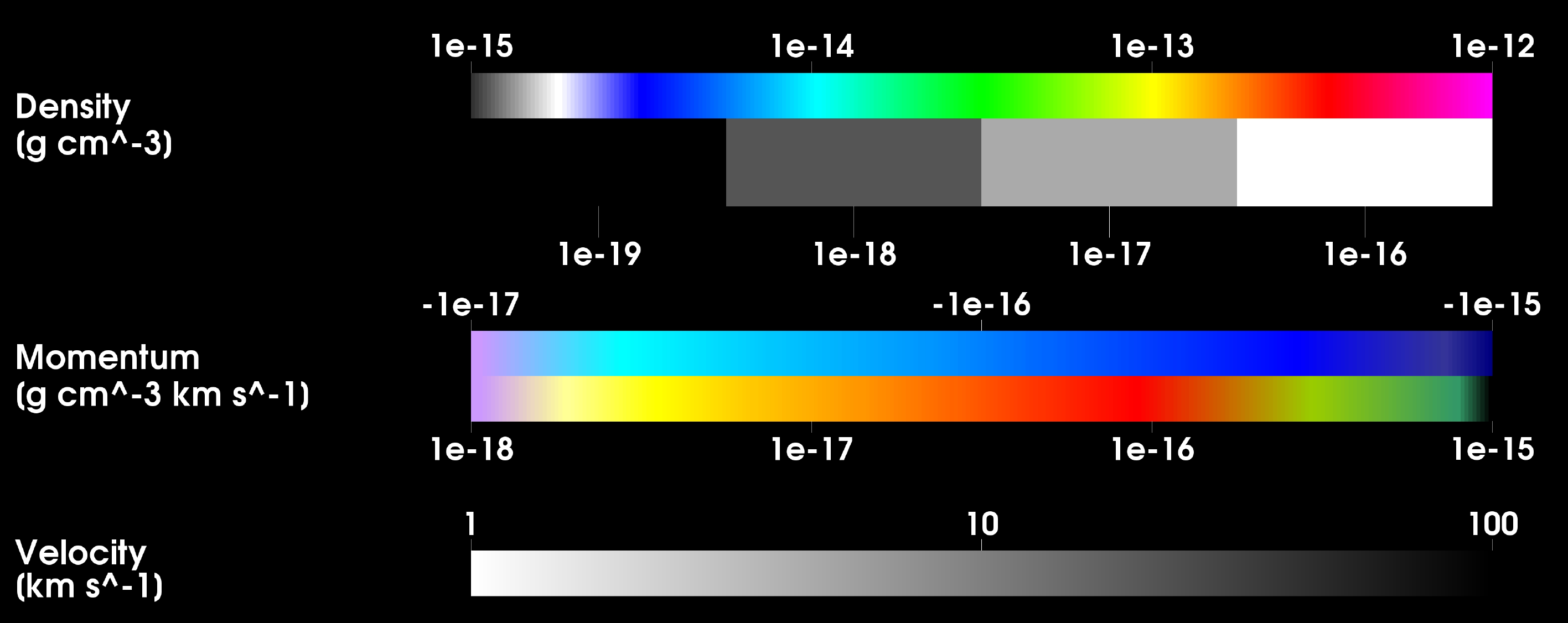}\\
\includegraphics[height=10cm]{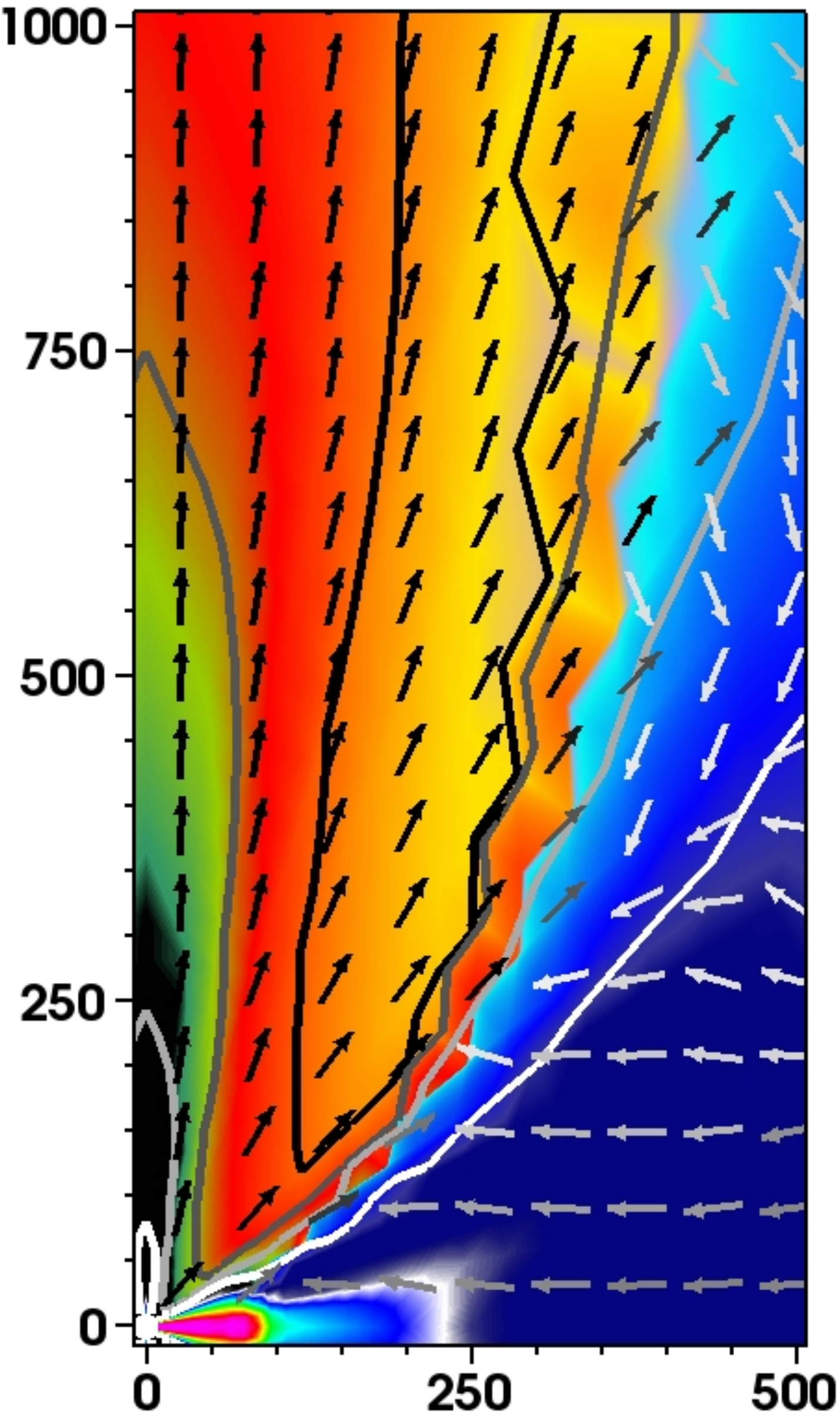}
\includegraphics[height=10cm]{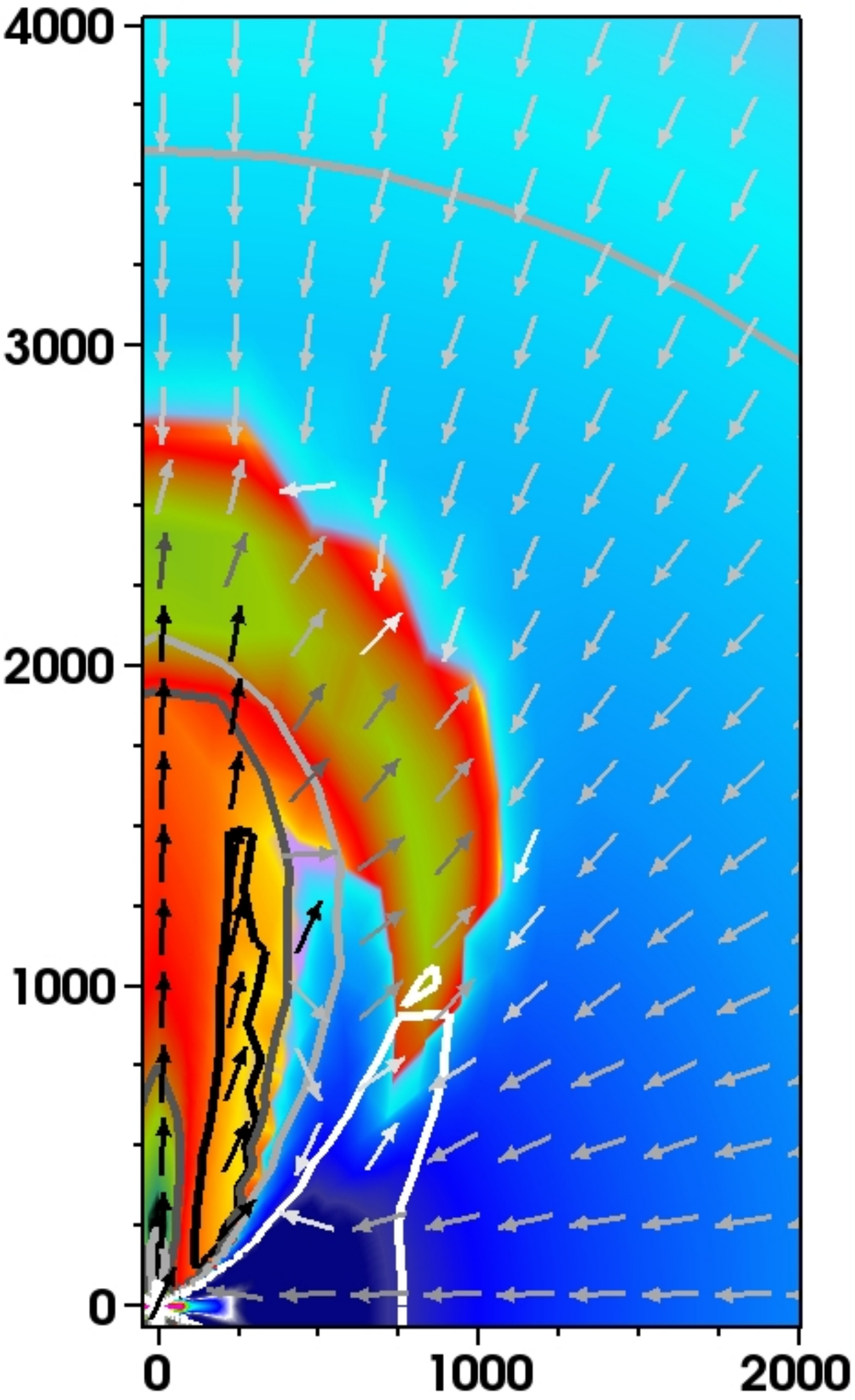}
\caption{ 
Gas density, momentum, and velocity in the vicinity of the protostar. 
The left panel covers a spatial extent of $500 \times 1000 \mbox{ AU}^2$. 
The right panel covers an extent four times larger along each axis, $2000 \times 4000 \mbox{ AU}^2$. 
Blue denotes radial infall, with higher momentum shown darker. 
Yellow to red to green to black denote radial outflow, with higher momentum again shown darker. 
Violet denotes regions of low momentum. 
The intermediate- and low-density gas ($\rho \le 10^{-16} \mbox{ g cm}^{-3}$) is marked with isodensity contours. 
The densest gas, found in the accretion disk, is overplotted in color ($\rho > 10^{-15} \mbox{ g cm}^{-3}$). 
The snapshot is from run rho2.0-PO at $t = 15$~kyr ($M_* \approx 24 \mbox{ M}_\odot$). 
\vONE{
According to the classification of Sect.~\ref{sect:classification}, the point in time belongs to the ``launching phase'' (stage I).
}
}
\label{fig:VisIt1}
\end{center}
\end{figure*}

\begin{figure*}[p]
\begin{center}
\includegraphics[height=10cm]{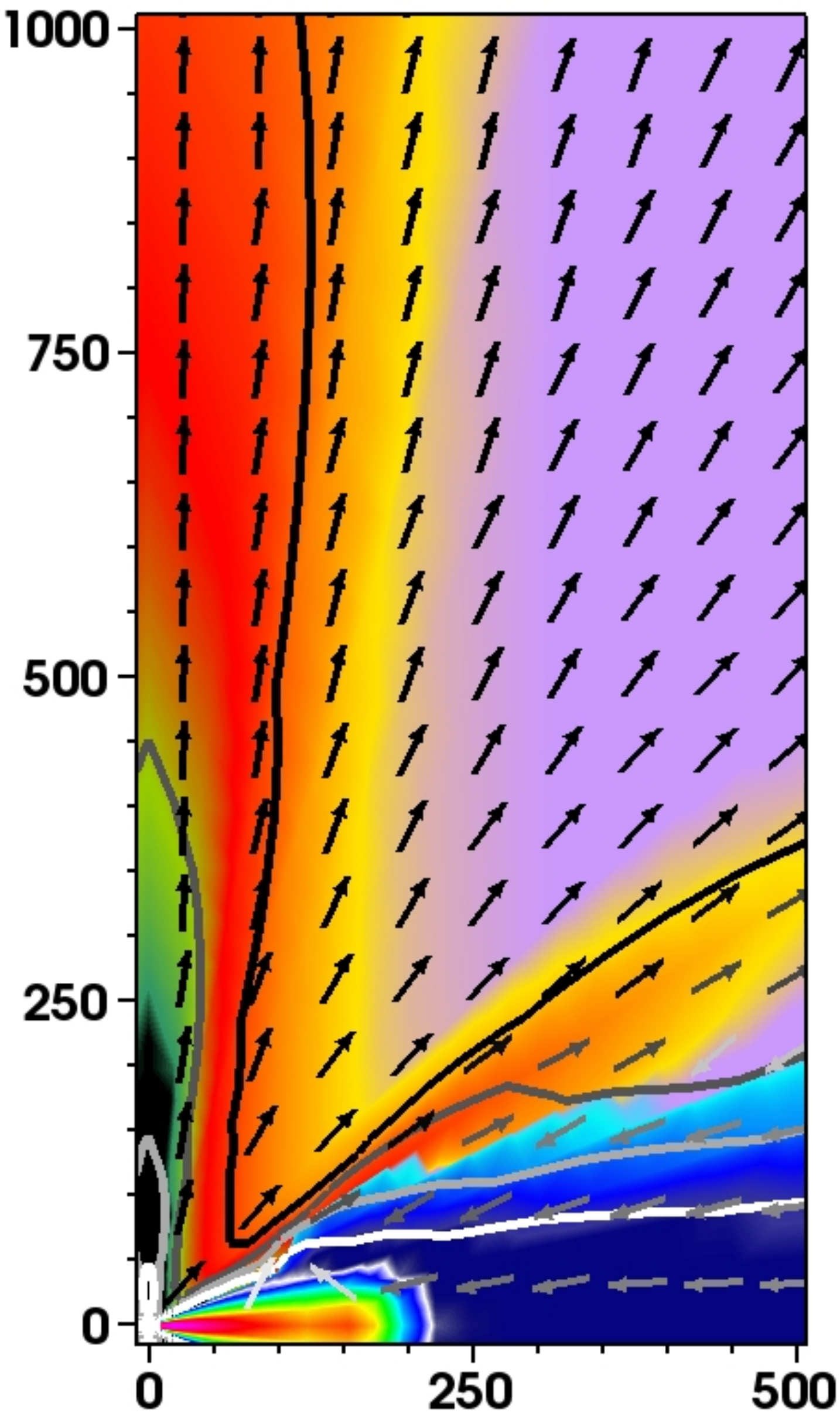}
\includegraphics[height=10cm]{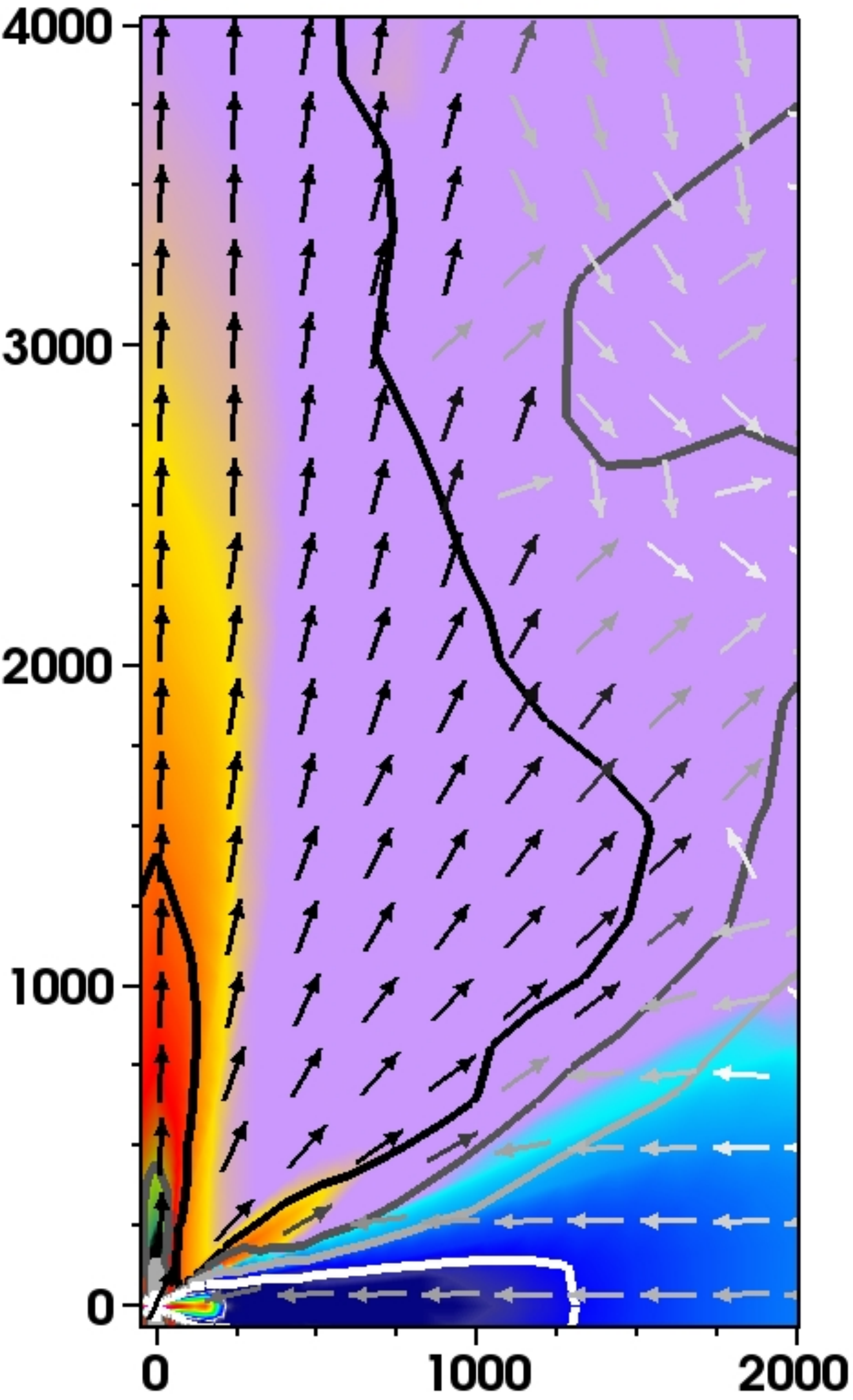}
\caption{
Same as Fig.~\ref{fig:VisIt1} but at $t=18$~kyr ($M_* \approx 28 \mbox{ M}_\odot$).
\vONE{
According to the classification of Sect.~\ref{sect:classification}, the point in time marks the end of the ``launching phase'' (stage I).
}
}
\label{fig:VisIt2}
\end{center}
\end{figure*}

\begin{figure*}[p]
\begin{center}
\includegraphics[height=10cm]{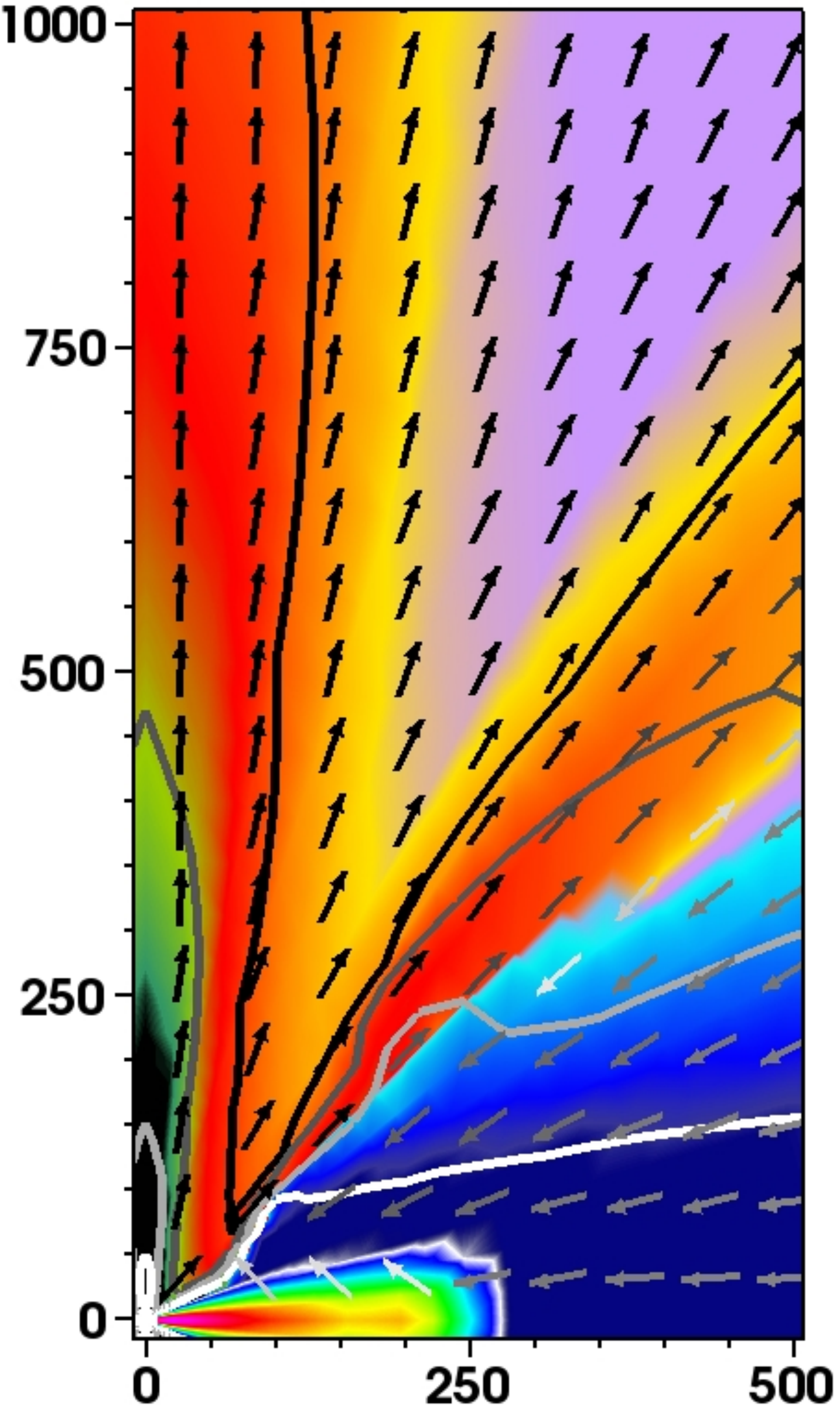}
\includegraphics[height=10cm]{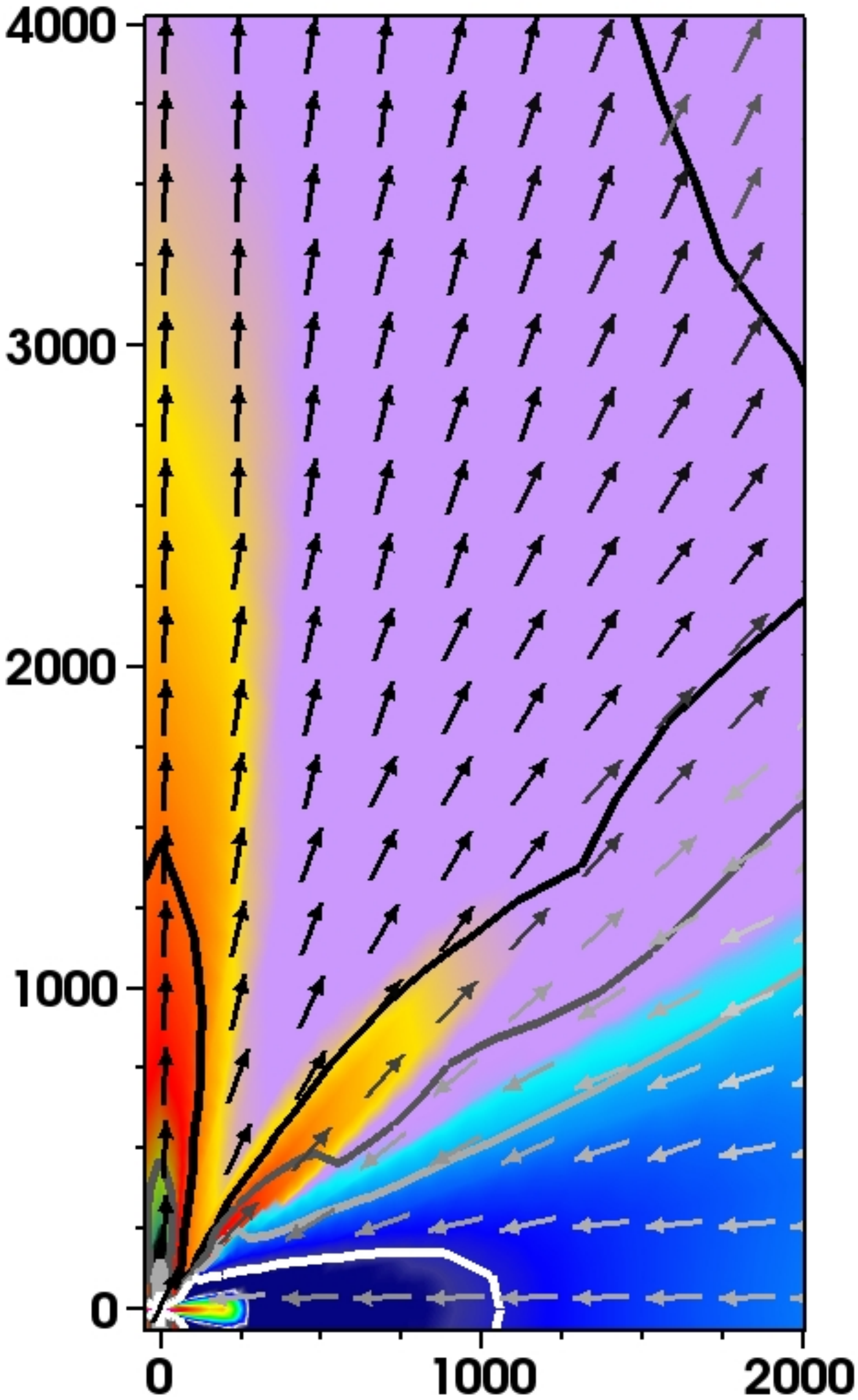}
\caption{
Same as Fig.~\ref{fig:VisIt1} but at $t=20$~kyr ($M_* \approx 30 \mbox{ M}_\odot$).
\vONE{
According to the classification of Sect.~\ref{sect:classification}, the point in time belongs to the ``quasi-stationary expansion phase'' (stage II).
}
}
\label{fig:VisIt3}
\end{center}
\end{figure*}

\begin{figure*}[p]
\begin{center}
\includegraphics[height=10cm]{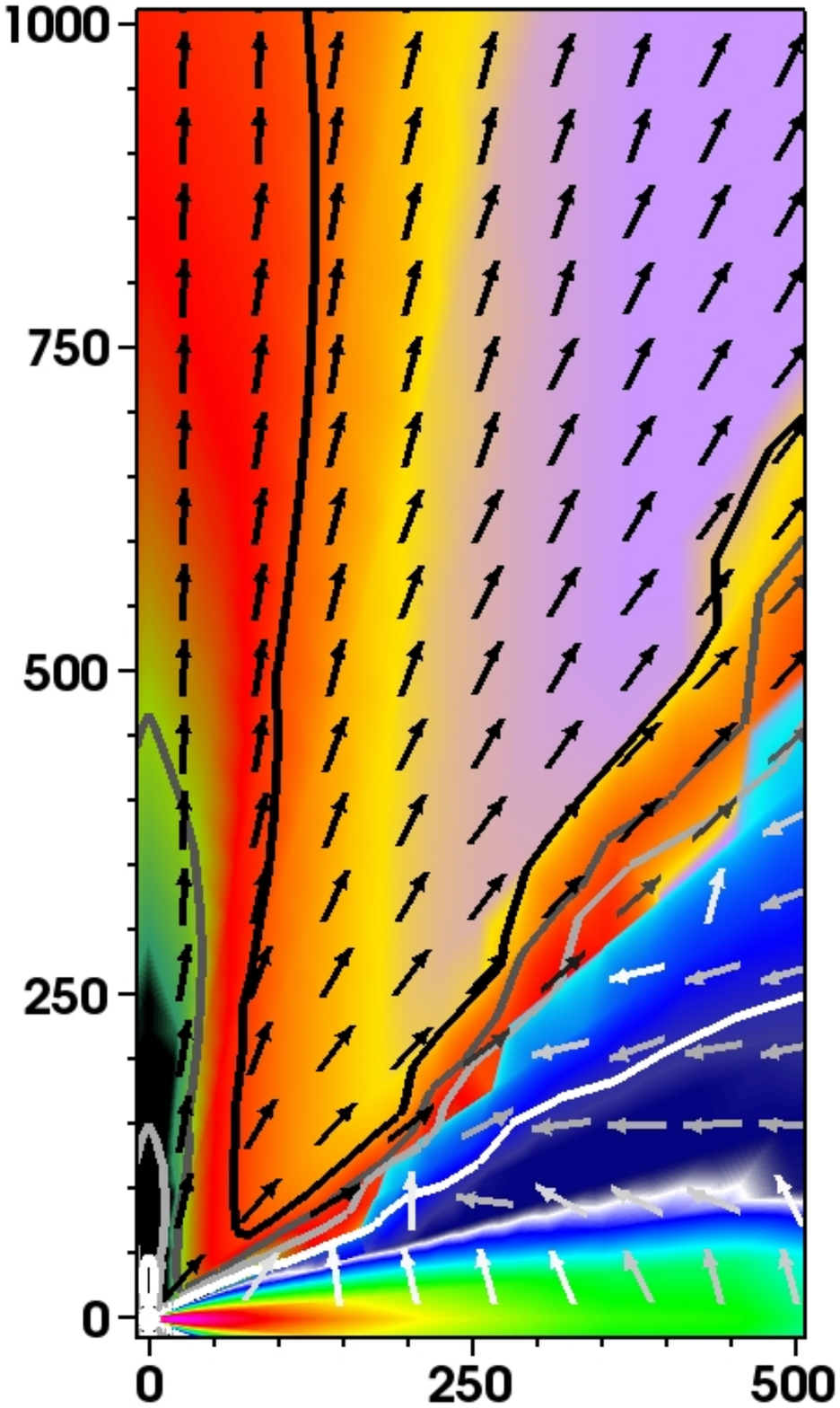}
\includegraphics[height=10cm]{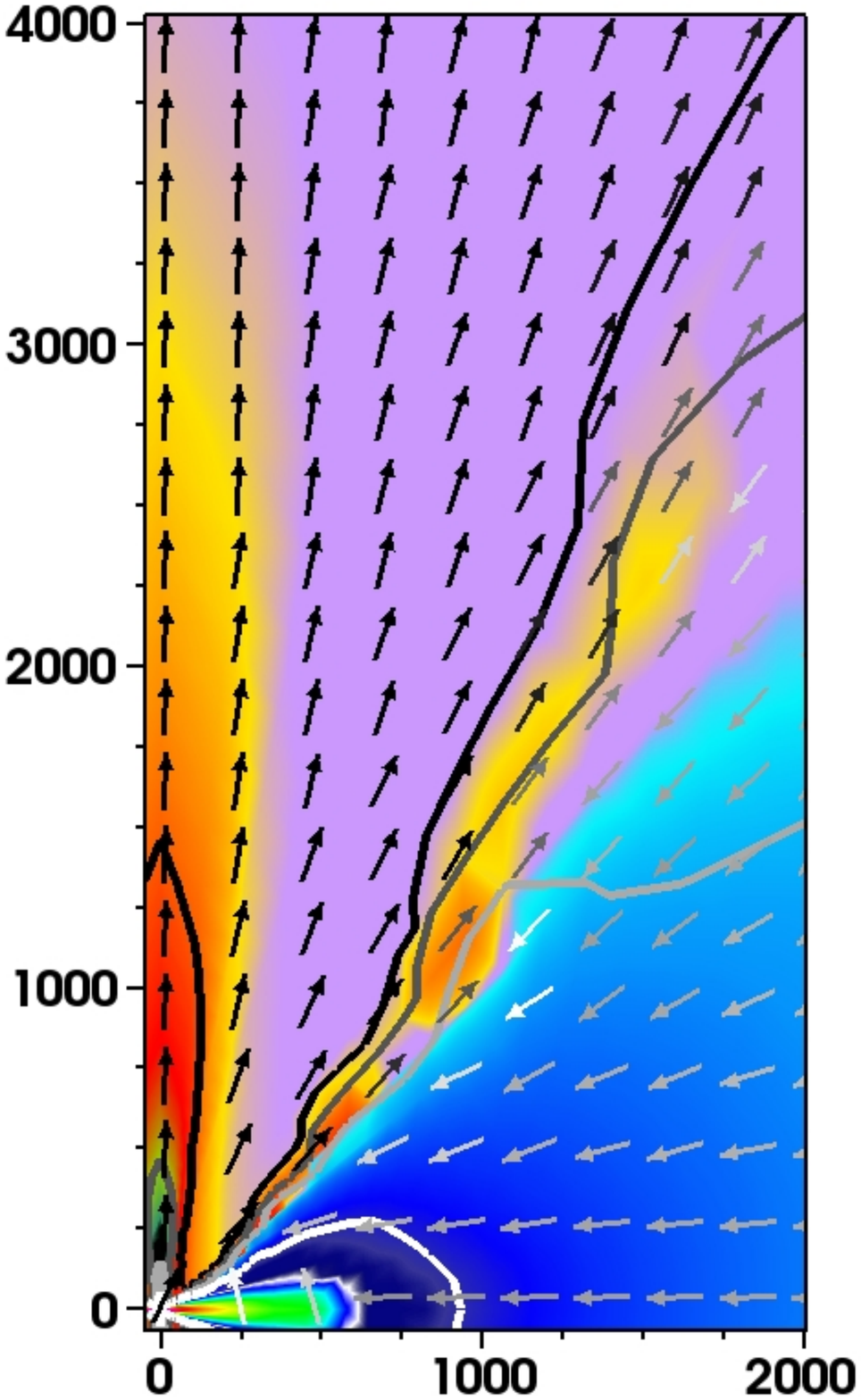}
\caption{
Same as Fig.~\ref{fig:VisIt1} but at $t=30$~kyr ($M_* \approx 35 \mbox{ M}_\odot$).
\vONE{
According to the classification of Sect.~\ref{sect:classification}, the point in time marks the end of the ``quasi-stationary expansion phase'' (stage II).
}
}
\label{fig:VisIt4}
\end{center}
\end{figure*}

\begin{figure*}[p]
\begin{center}
\includegraphics[height=10cm]{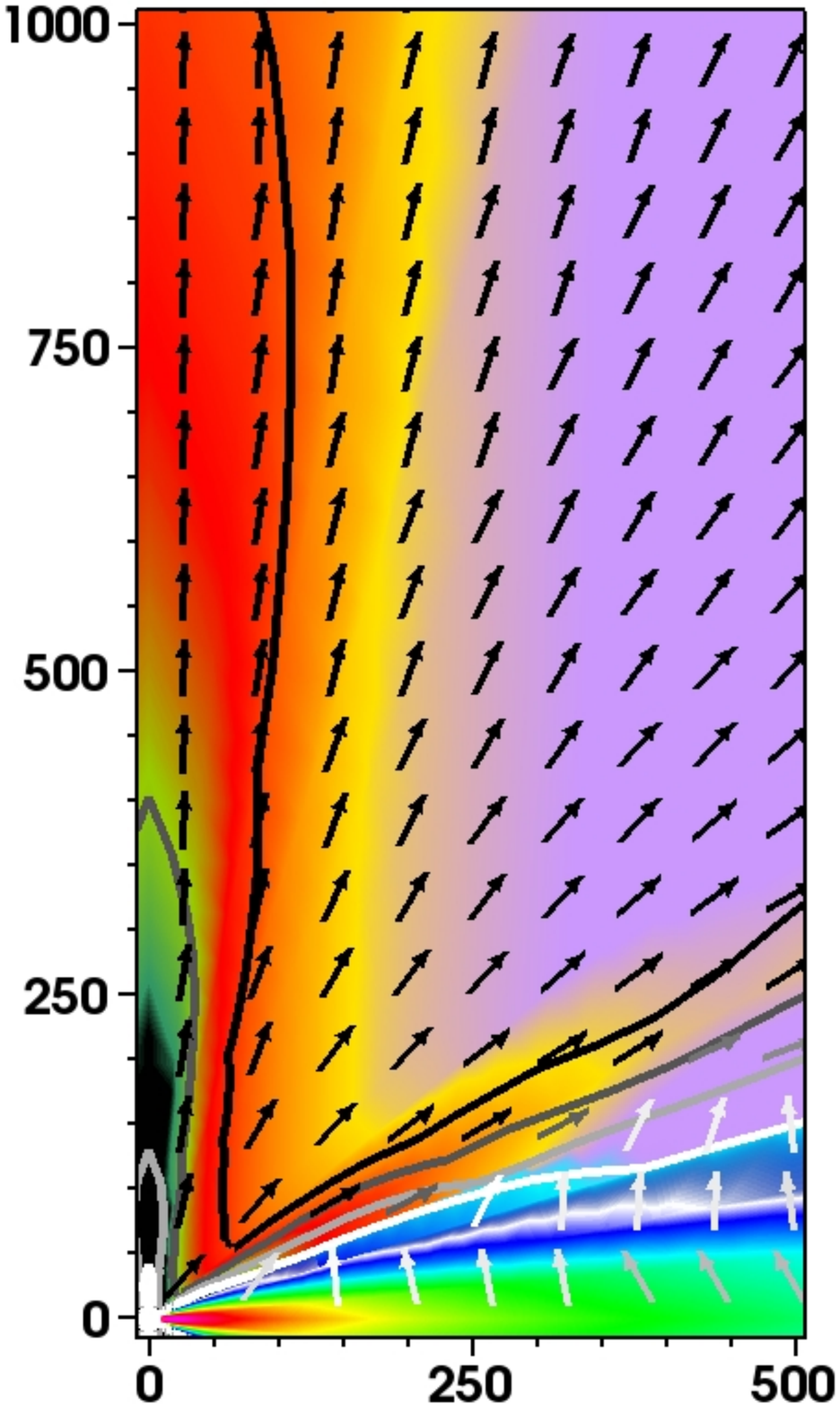}
\includegraphics[height=10cm]{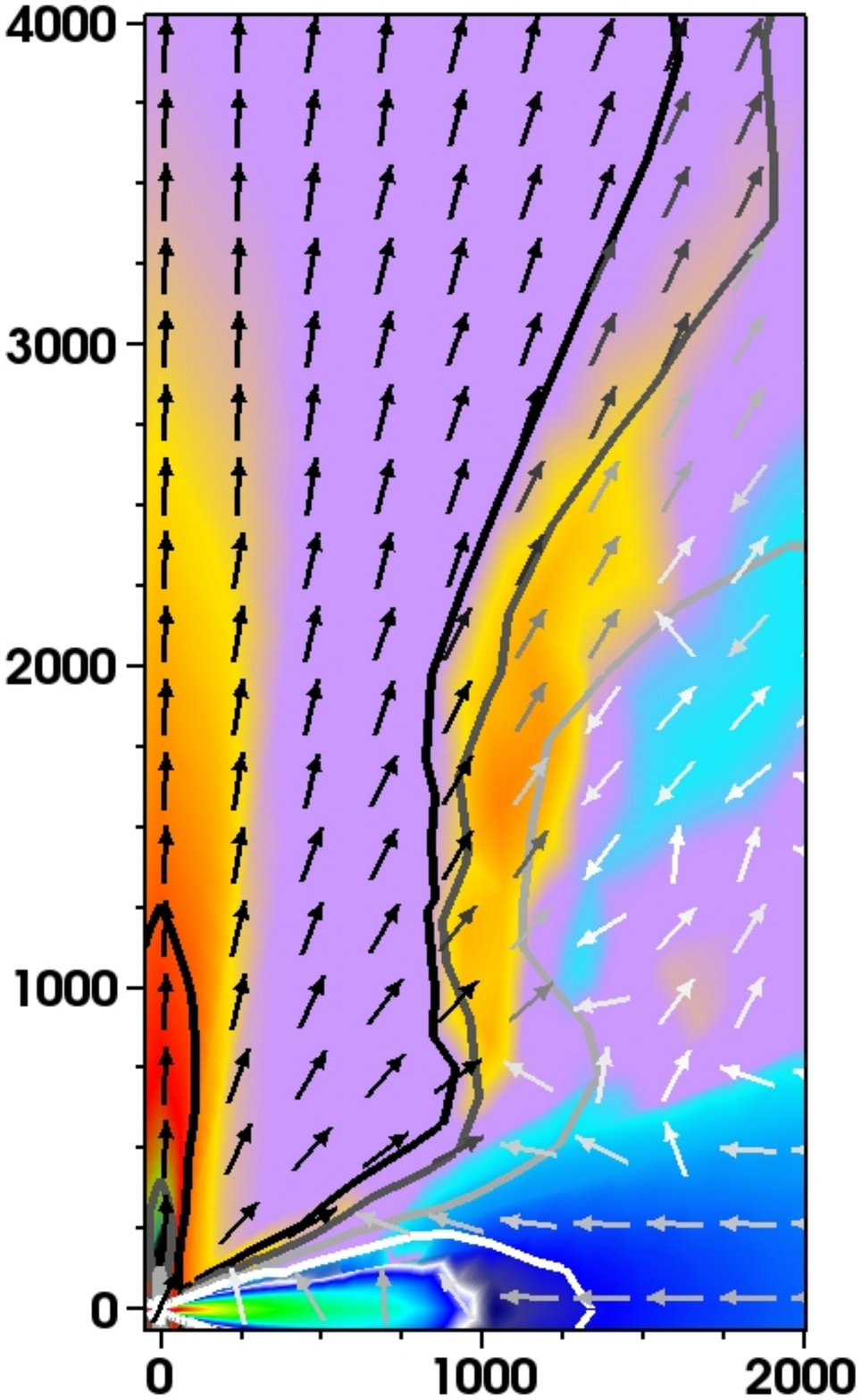}
\caption{
Same as Fig.~\ref{fig:VisIt1} but at $t=40$~kyr ($M_* \approx 40 \mbox{ M}_\odot$). 
\vONE{
According to the classification of Sect.~\ref{sect:classification}, the system is still at the beginning of the ``radiation-pressure-dominated phase'' (stage III).
}
}
\label{fig:VisIt5}
\end{center}
\end{figure*}

\begin{figure*}[p]
\begin{center}
\includegraphics[height=10cm]{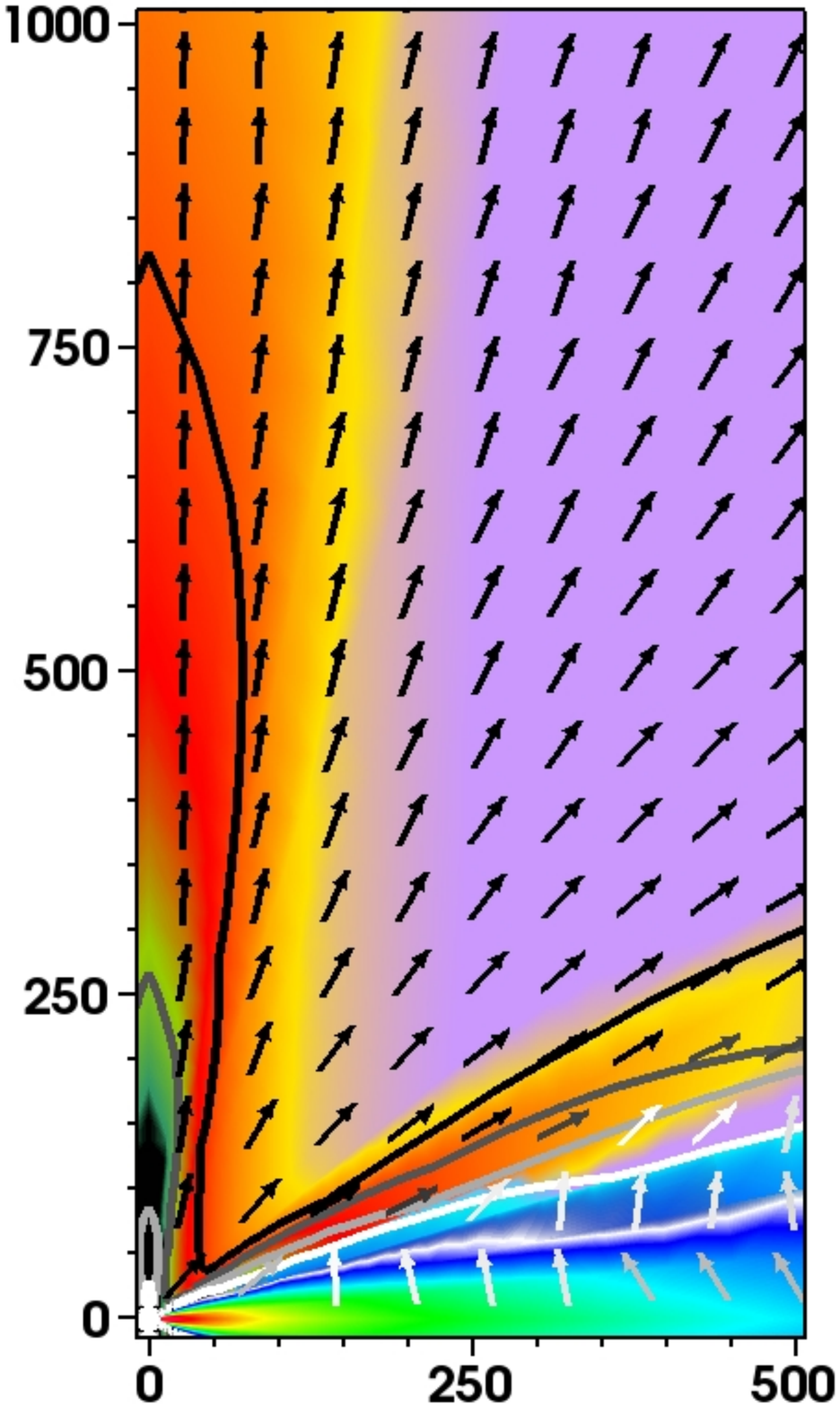}
\includegraphics[height=10cm]{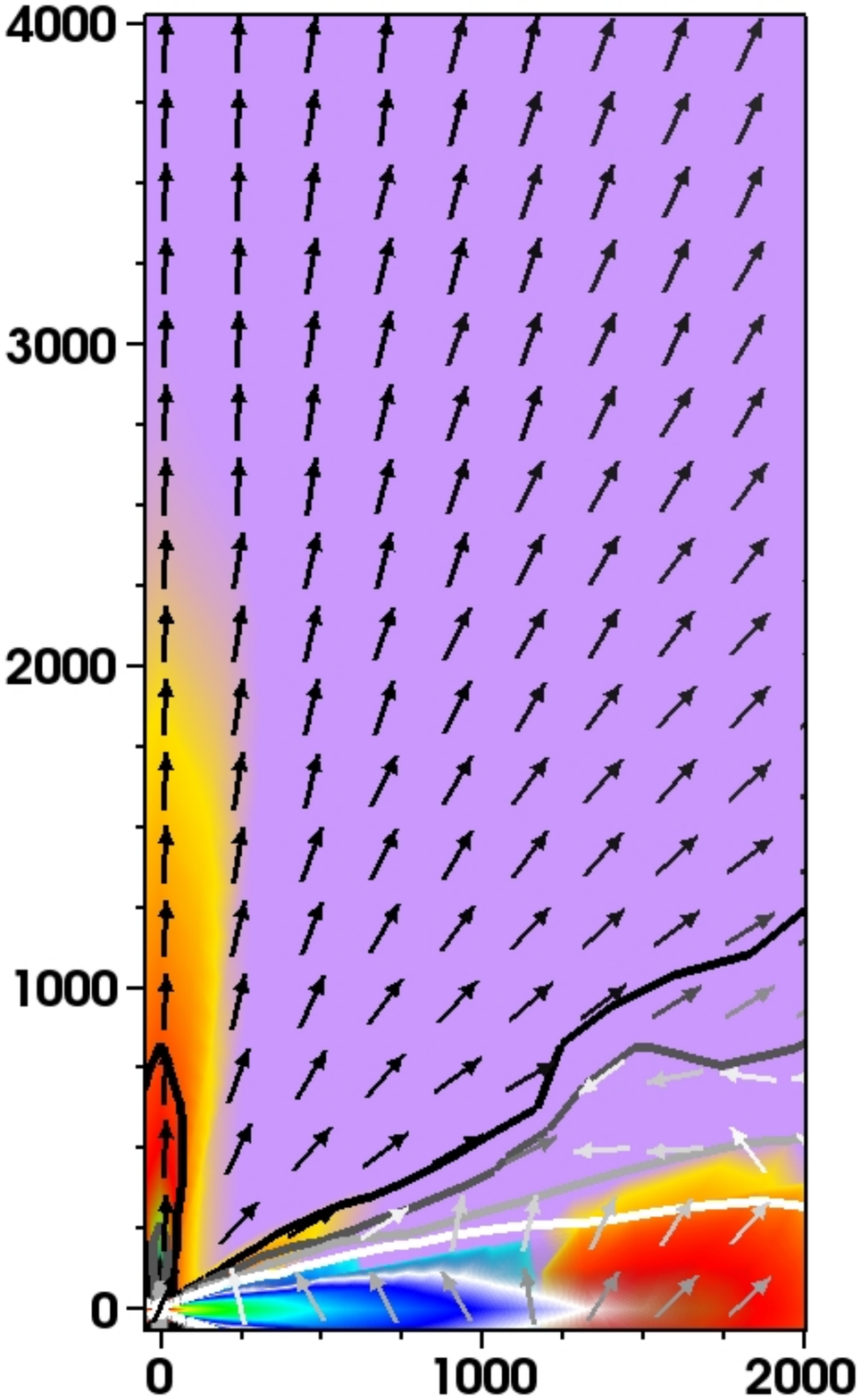}
\caption{
Same as Fig.~\ref{fig:VisIt1} but at  $t=60$~kyr ($M_* \approx 46 \mbox{ M}_\odot$).
\vONE{
According to the classification of Sect.~\ref{sect:classification}, the point in time belongs to the ``radiation-pressure-dominated phase'' (stage III).
}
}
\label{fig:VisIt6}
\end{center}
\end{figure*}

The outflow passes through several phases of evolution, shown by
snapshots of the momentum, velocity and density distributions in
Figs.~\ref{fig:VisIt1} to \ref{fig:VisIt6}. 
During all phases, the
outflow consists of two components (Figs.~\ref{fig:VisIt2} to
\ref{fig:VisIt6}). 
The first is a fast jet, confined to a small angle
around the symmetry axis.  Its extent is governed by the injection
rate and thus the stellar accretion rate.  The second component is a
slower wind, which transports intermediate-density material ($\rho \sim
10^{-19} - 10^{-17} \mbox{ g cm}^{-3}$) at lower latitudes in the
bipolar cavity and/or the top of the accretion disk atmosphere. 
\vONE{
The slower wind is initially driven by the kinematic feedback of the outflow and limited in size to a few 100 AU.
Over time the radiation forces become more and more important, and around the time they come to dominate, the wind briefly reaches out as far as 2000~AU (Fig.~\ref{fig:VisIt4} and Fig.~\ref{fig:VisIt5}).
}
In the later phase (Fig.~\ref{fig:VisIt6}), the protostellar radiation force eventually drags the intermediate-density gas to larger radii and clears the bipolar region over a large solid angle. 
Hence the wind component associated with the intermediate density gas vanishes on extents larger than roughly 500~AU once the stellar radiative forces dominate the outflow region.

\subsection{Opening Angle}
\label{sect:classification}
The bipolar outflow's opening angle passes through three stages of
evolution.  Over long timescales the opening angle widens as the
radiative feedback from the protostar increases.  Over shorter
timescales, however, the opening angle is affected by the distribution
of intermediate-density gas and dust above and below the disk.  The
material's opacity shields the environment from the central radiative
force.

When the protostellar outflow is first injected, a shock propagates
into the envelope, slowing and even reversing the infall on larger
scales.  The shock appears green in Fig.~\ref{fig:VisIt1} right panel.
Once the shock propagates past each radius, infall there resumes.  The
shock temporarily clears material, allowing the outflow opening angle
to increase up a maximum of about 60~degrees (measured from the
symmetry axis to one side).  This broadening of the outflow is
confined to within roughly $2000$~AU of the protostar and reaches its
maximum extent 18~kyr after the start of the collapse.

In a second stage, after the shock passes out of the domain, the
outflow within $4000$~AU of the protostar settles into a
quasi-stationary state.  The opening angle is governed by the injected
momentum and the infall from larger scales.  The quasi-stationary
opening angle is about 40~degrees starting at 25~kyr.  The outflow
reaches $14000$~AU along the symmetry axis at this time.

The quasi-stationary stage holds only up to about 30~kyr.  Then, in a
third stage, the stellar radiative force overcomes the gravitational
attraction of the protostar and the opening angle of the
radiation-pressure-dominated outflow broadens monotonically in time
(Figs.~\ref{fig:VisIt4} and \ref{fig:VisIt5}).  The maximum angle of
the radiation-pressure-driven outflow is controlled by the inner
accretion disk's variation of optical depth with angle.  The
radiatively-driven outflow has a larger opening angle than the early
protostellar outflow.

\newpage
\section{Disk Evolution and Stellar Accretion}
\label{sect:results-disk}
In this section we look at how the kinematic and radiative feedback
shape the star's growth.  The star grows by receiving mass from the
accretion disk.  Angular momentum conservation in the infalling gas
dictates that the disk is assembled from inside out.  Its growth is
governed by the mass fed from larger scales, which appears in
Figs.~\ref{fig:VisIt1} to \ref{fig:VisIt5} as a blue torus of inward
radial momentum.  
The disk's outer edge undergoes radial oscillations
and super-Keplerian azimuthal motion as the arriving material settles
into radial force balance.  
The balance involves contributions from
self-gravity, thermal pressure and radiation pressure.  
\vONE{
E.g.~the red-colored region (positive outward momentum) at the outer rim of the accretion disk (Fig.~\ref{fig:VisIt6}, right panel) denotes the interaction zone of the disk with the collapsing environment;
the process of in-falling material approaching its gravito-centrifugal balance (plus minor contributions of thermal and radiation pressure) yields small natural oscillations around the equilibrium point.
For further details of this disk-growing process, we refer the interested reader to \citet{Kuiper:2011p21204}, Sect.~4.2 and Fig.~4, where a 3-D calculation is compared with a series of axisymmetric disk models having varying shear viscosities..
}

The star's final mass is the time integral of the accretion rate
history and can be expressed as a mean accretion rate multiplied by
the duration of the epoch of disk accretion.  The disk's mass flow
rate and lifetime both depend on the supply of material falling from
the envelope onto the disk, which in turn is governed by the radiative
feedback. The surprise is that adding the
modest protostellar outflow makes the star's final mass larger.  The
outflow's kinematic feedback is more than balanced by a reduction in
the effectiveness of the thermal radiative feedback.  The bipolar cavity
cleared by the outflow amplifies the flashlight effect, letting both the stellar and thermal
radiation escape more easily near the poles and reducing the fluxes
into lower latitudes.  The anisotropy of the thermal radiation, in
particular, reduces the extent of the mass loss through the outer
boundary, allowing more of the envelope mass ultimately to fall onto the
disk.

\begin{figure*}[htbp]
\begin{center}
\subfigure[][15~kyr]{\includegraphics[width=0.44\textwidth]{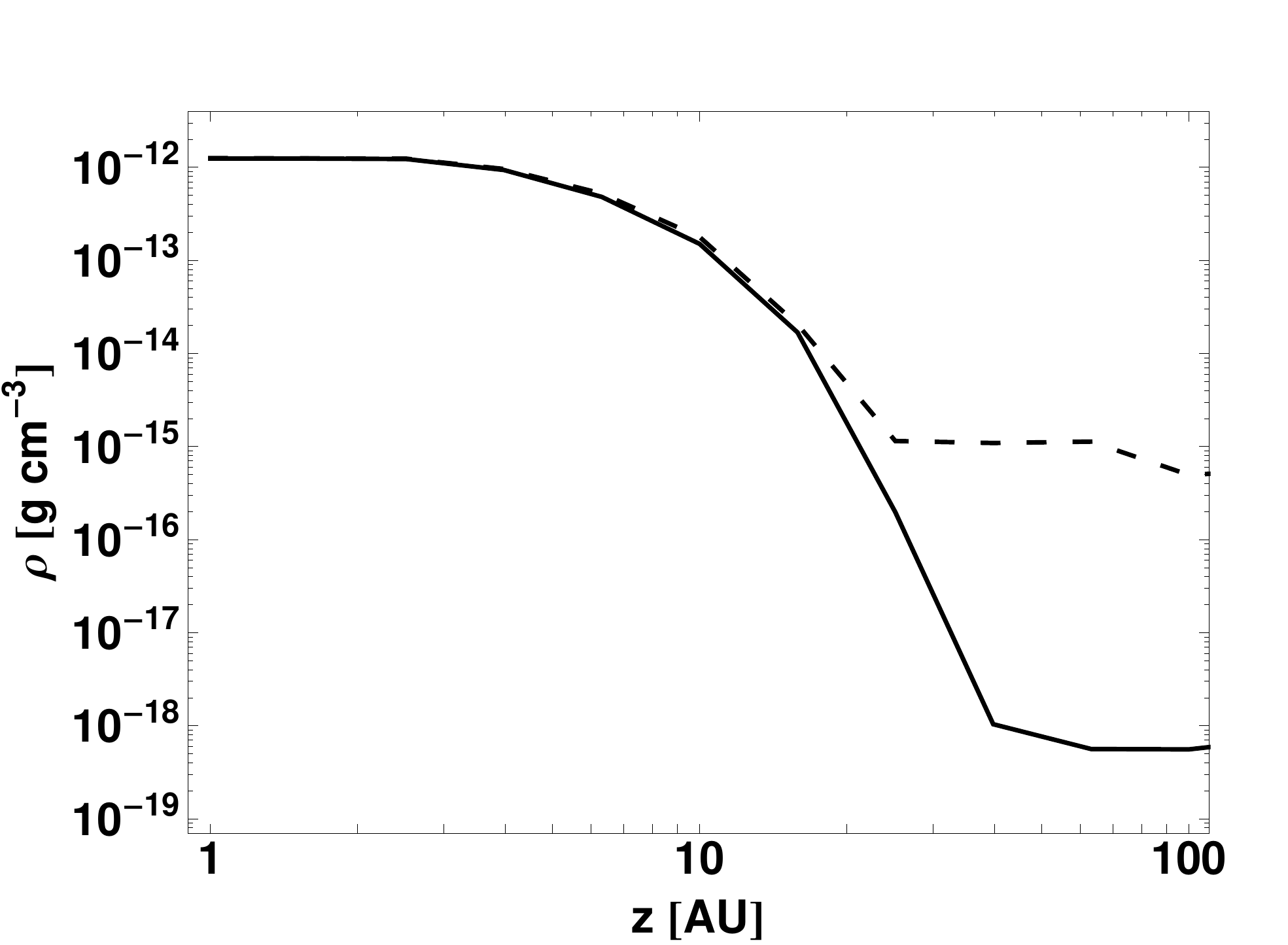}}
\subfigure[][20~kyr]{\includegraphics[width=0.44\textwidth]{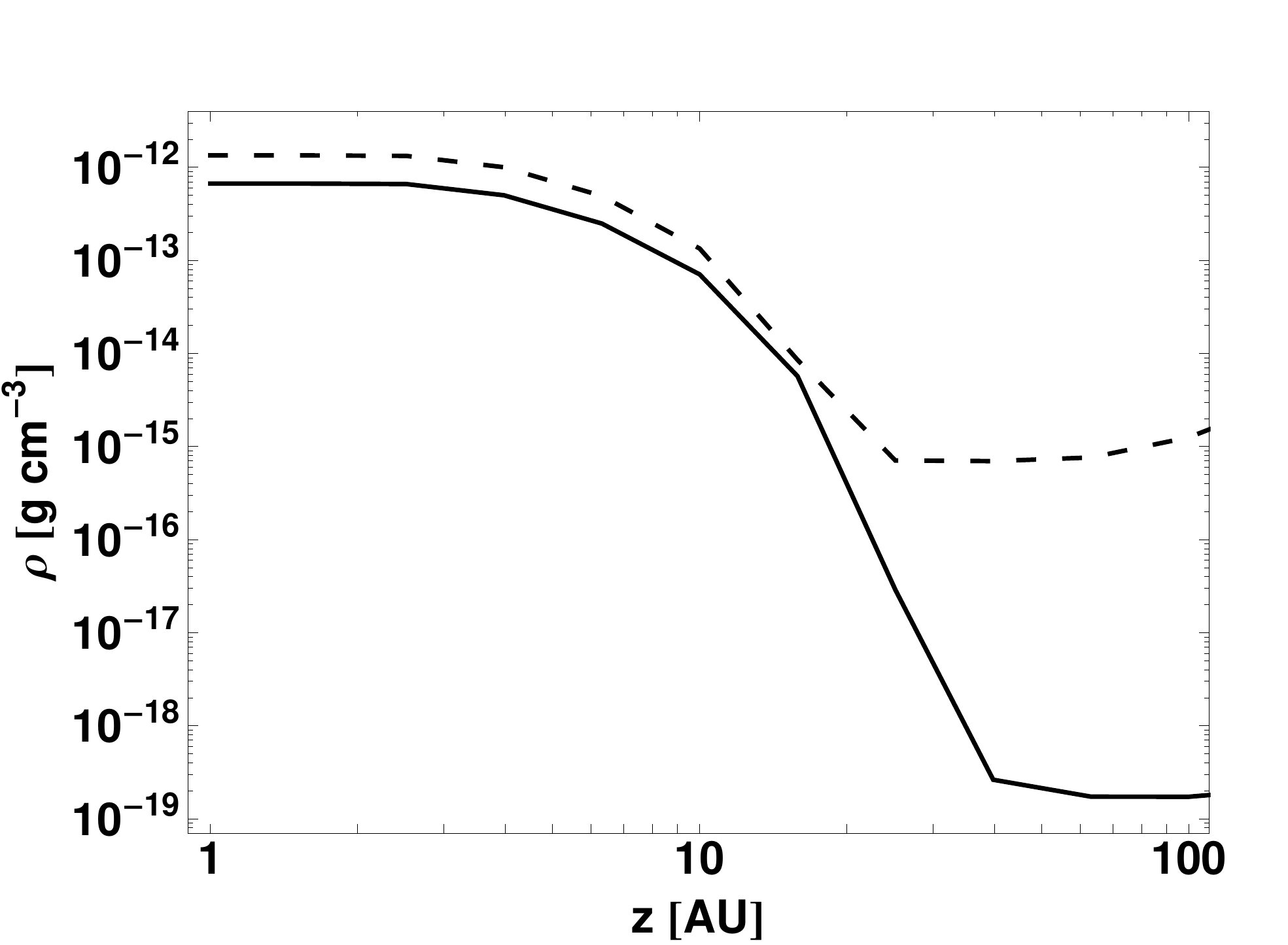}}
\subfigure[][25~kyr]{\includegraphics[width=0.44\textwidth]{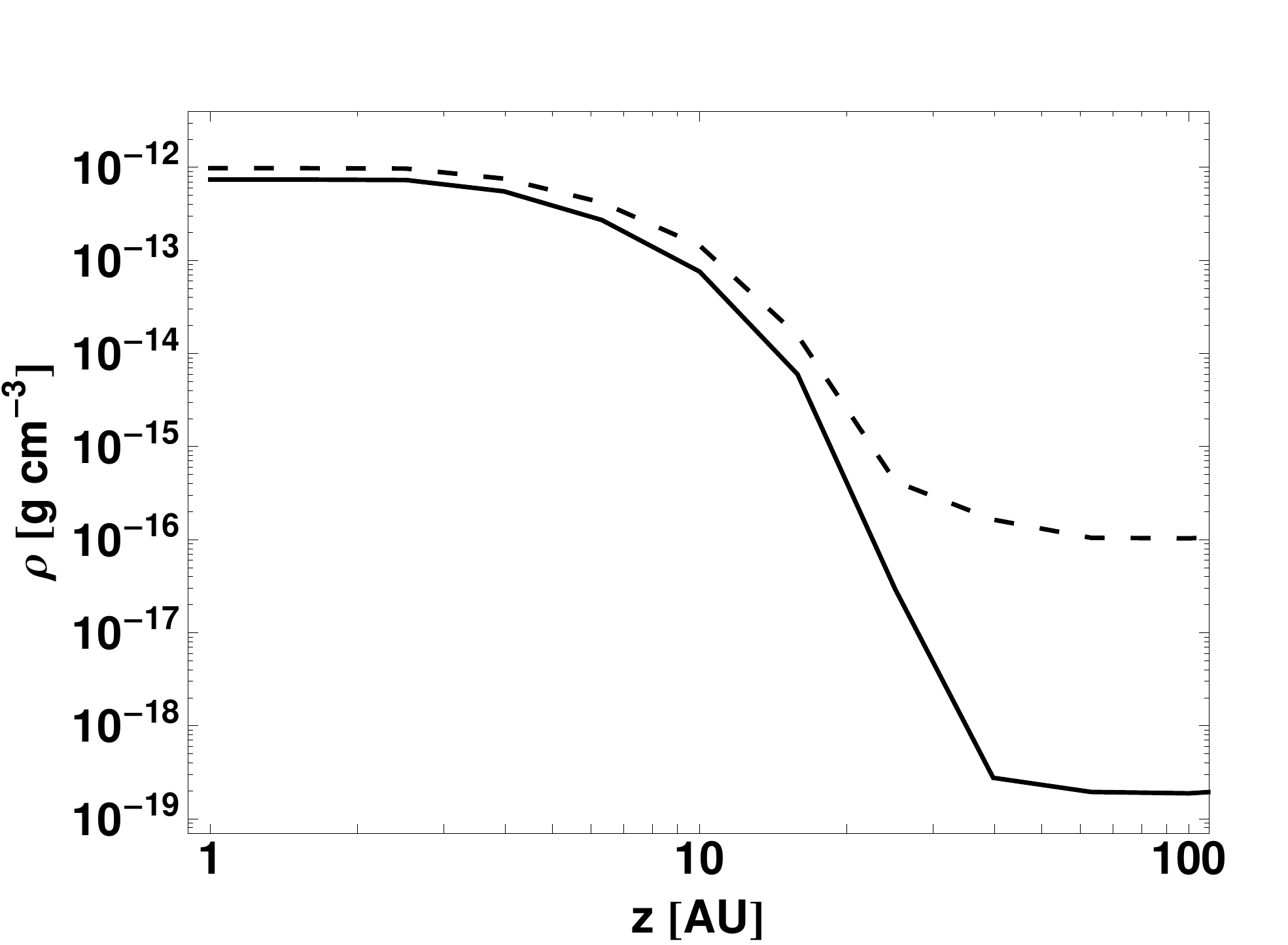}}
\subfigure[][30~kyr]{\includegraphics[width=0.44\textwidth]{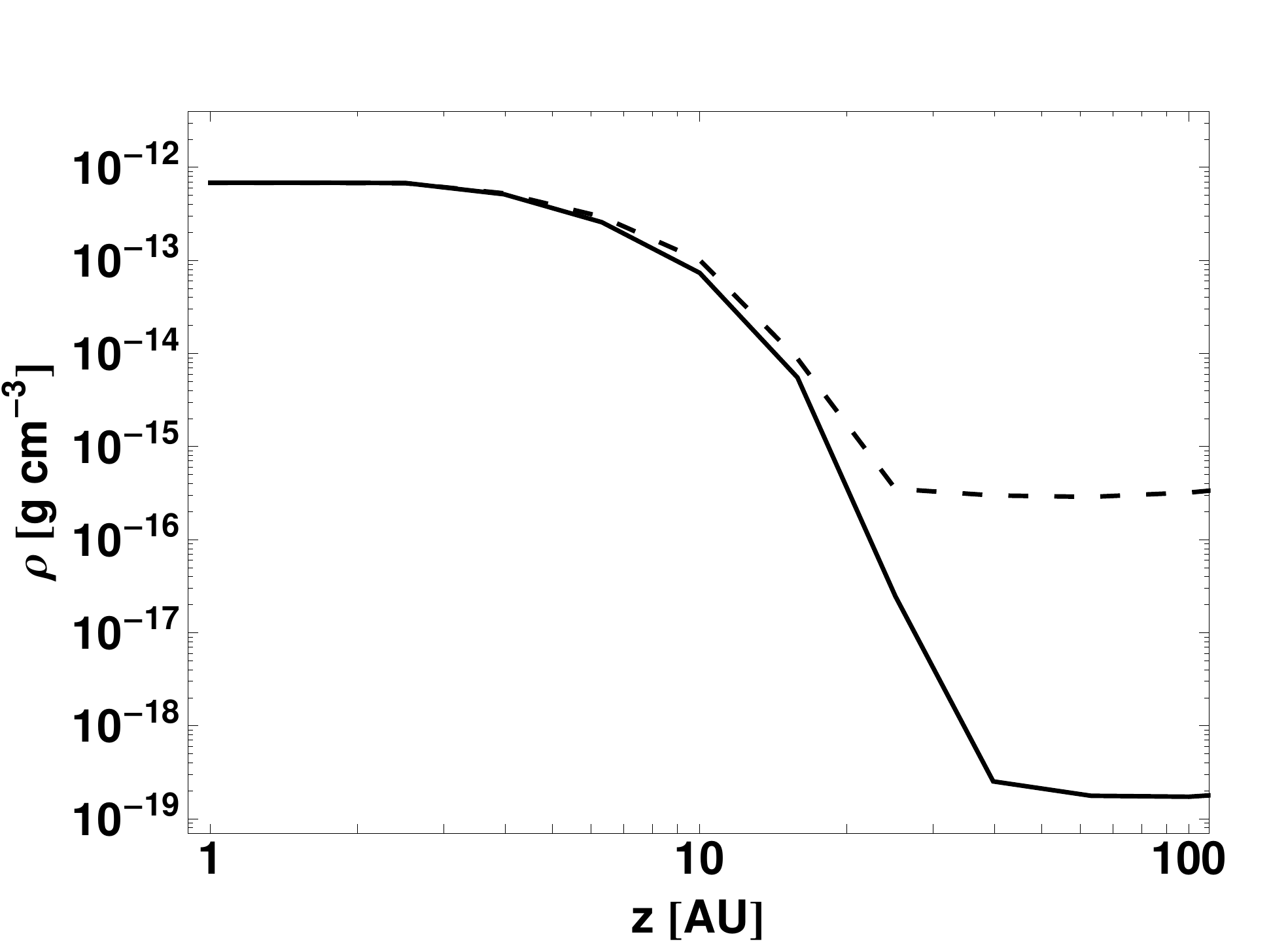}}
\subfigure[][40~kyr]{\includegraphics[width=0.44\textwidth]{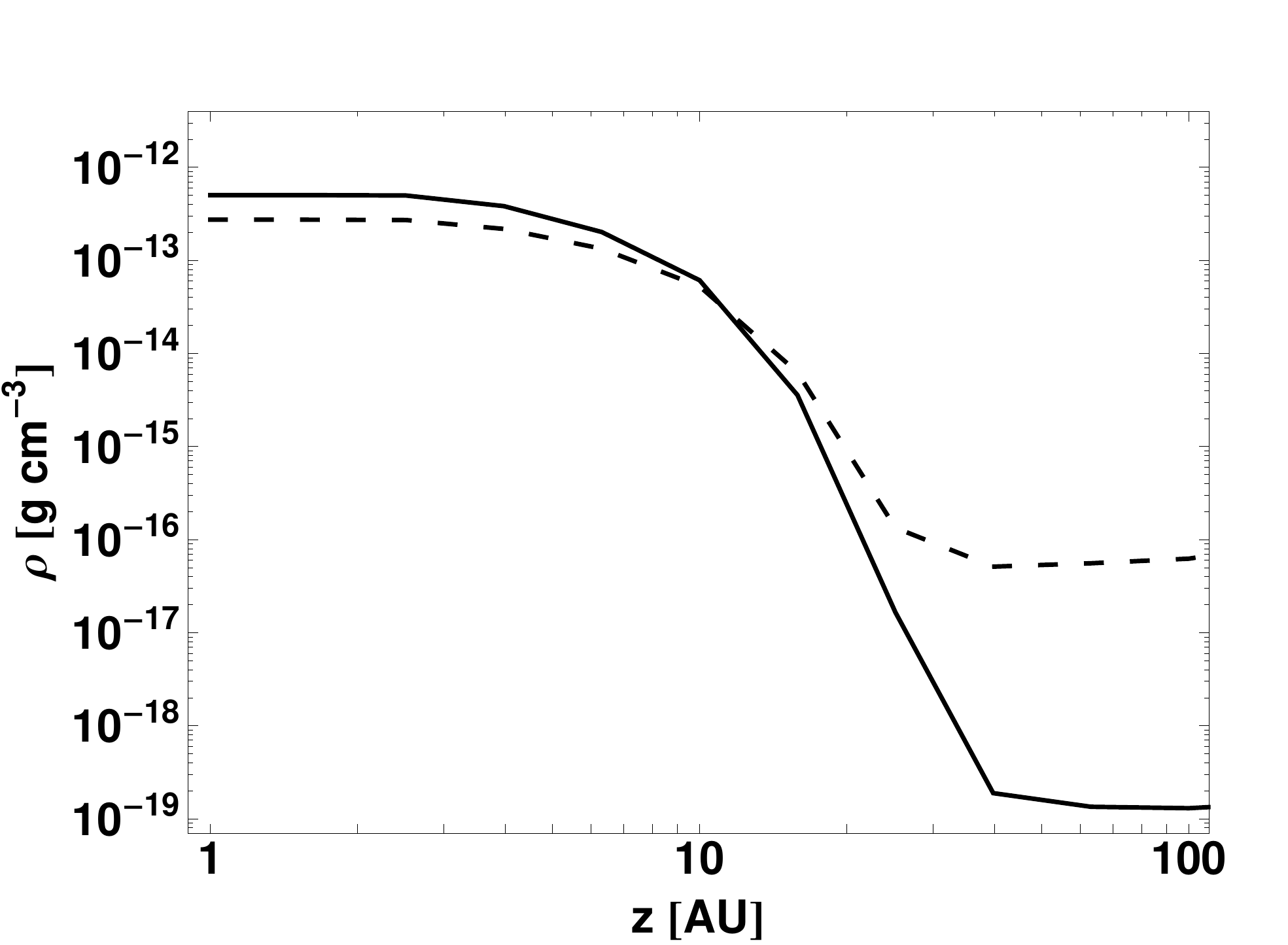}}
\subfigure[][60~kyr]{\includegraphics[width=0.44\textwidth]{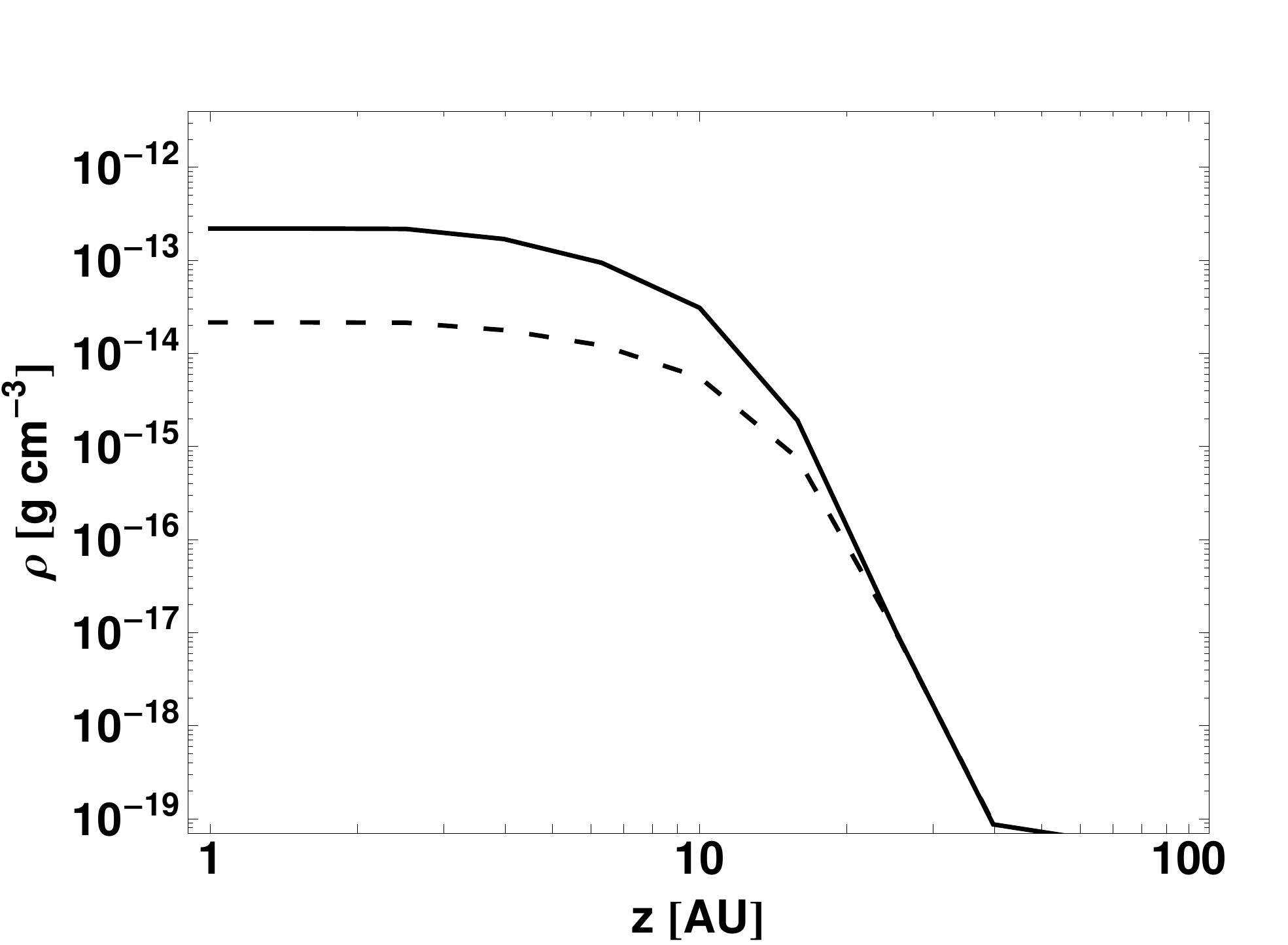}}
\caption{ Gas density profile at radius $R=50$~AU vs.\ the height $z$
  above the disk's mid-plane.  The data are from the collapse case
  $\beta=-2.0$ at selected evolutionary times as labeled.  Solid
  (dashed) lines depict the cases with (without) the protostellar
  outflow.  }
\label{fig:Density_vs_z}
\end{center}
\end{figure*}

\begin{figure*}[htbp]
\begin{center}
\subfigure[][15~kyr]{\includegraphics[width=0.42\textwidth]{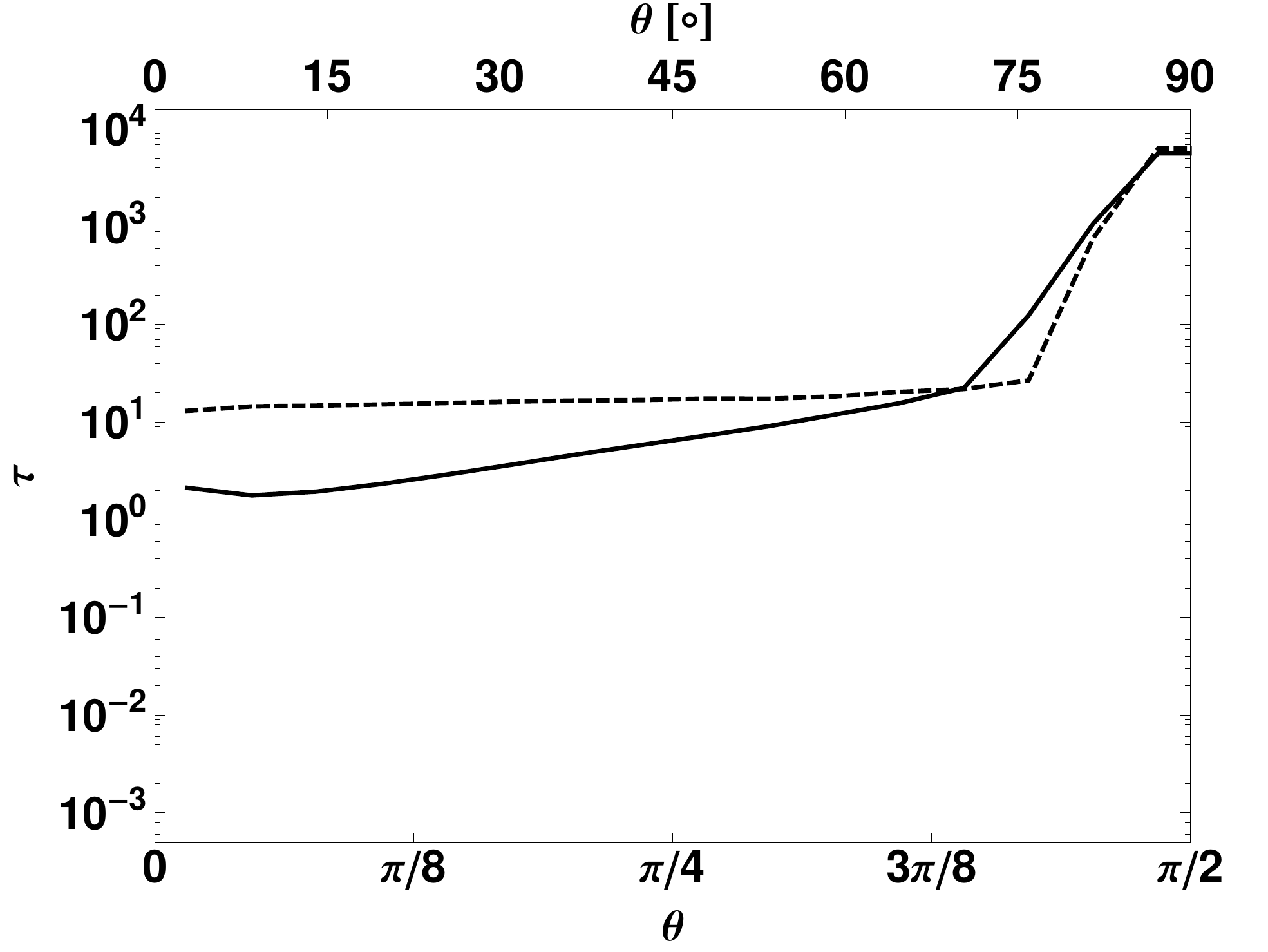}
\label{fig:Tau_vs_theta-a}}
\subfigure[][20~kyr]{\includegraphics[width=0.42\textwidth]{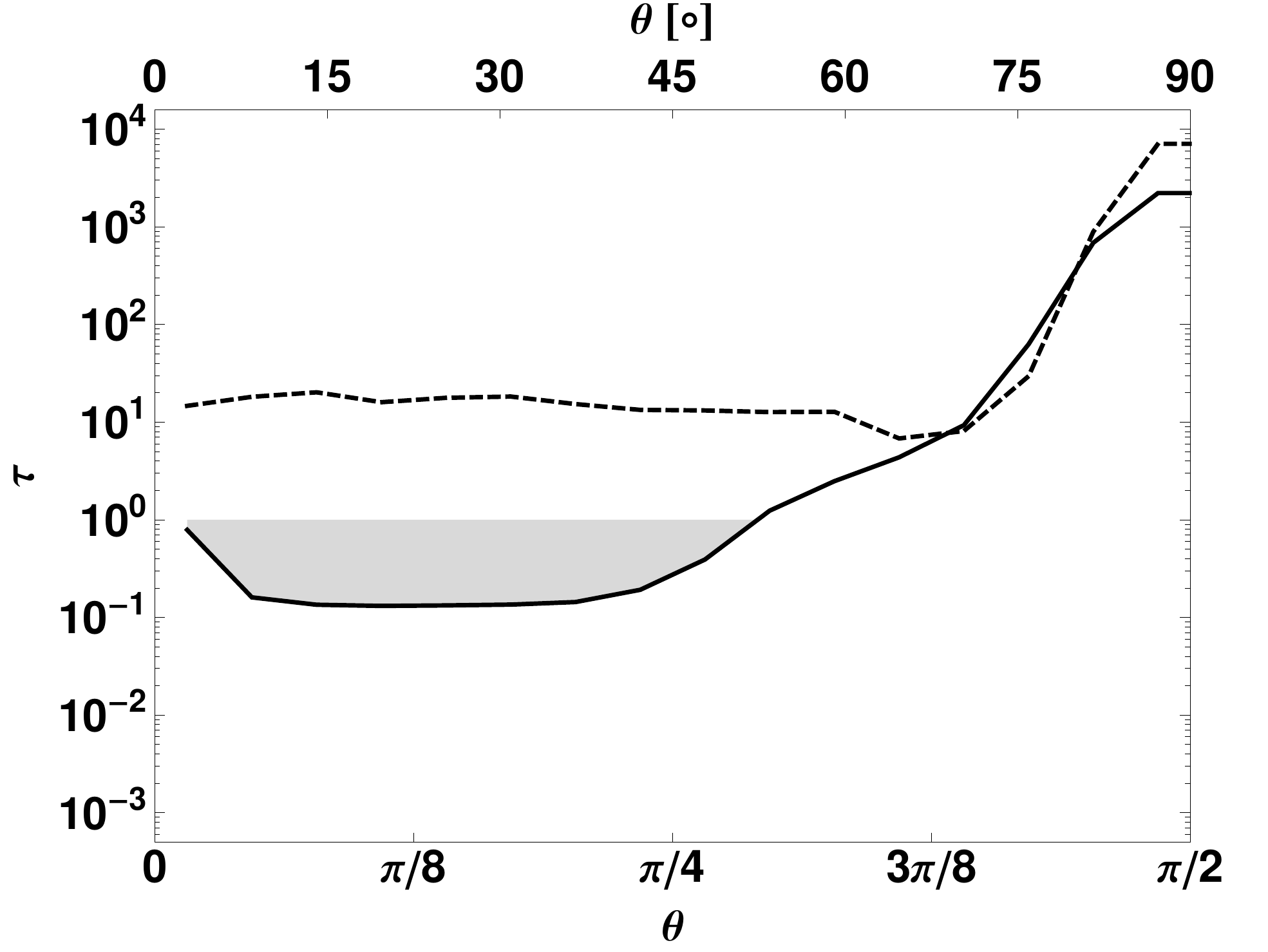}
\label{fig:Tau_vs_theta-b}}\\
\vspace{5mm}
\subfigure[][25~kyr]{\includegraphics[width=0.42\textwidth]{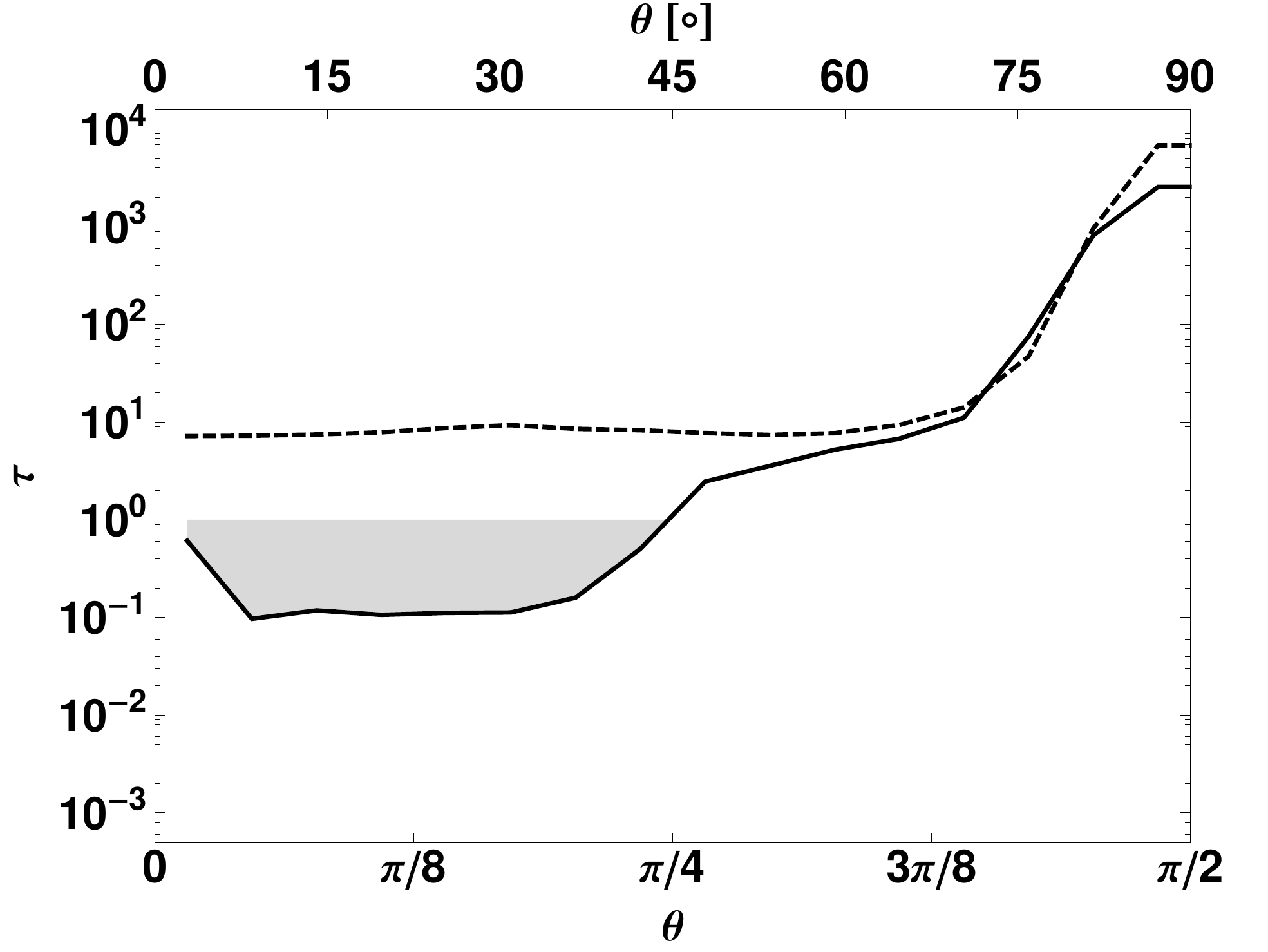}
\label{fig:Tau_vs_theta-c}}
\subfigure[][30~kyr]{\includegraphics[width=0.42\textwidth]{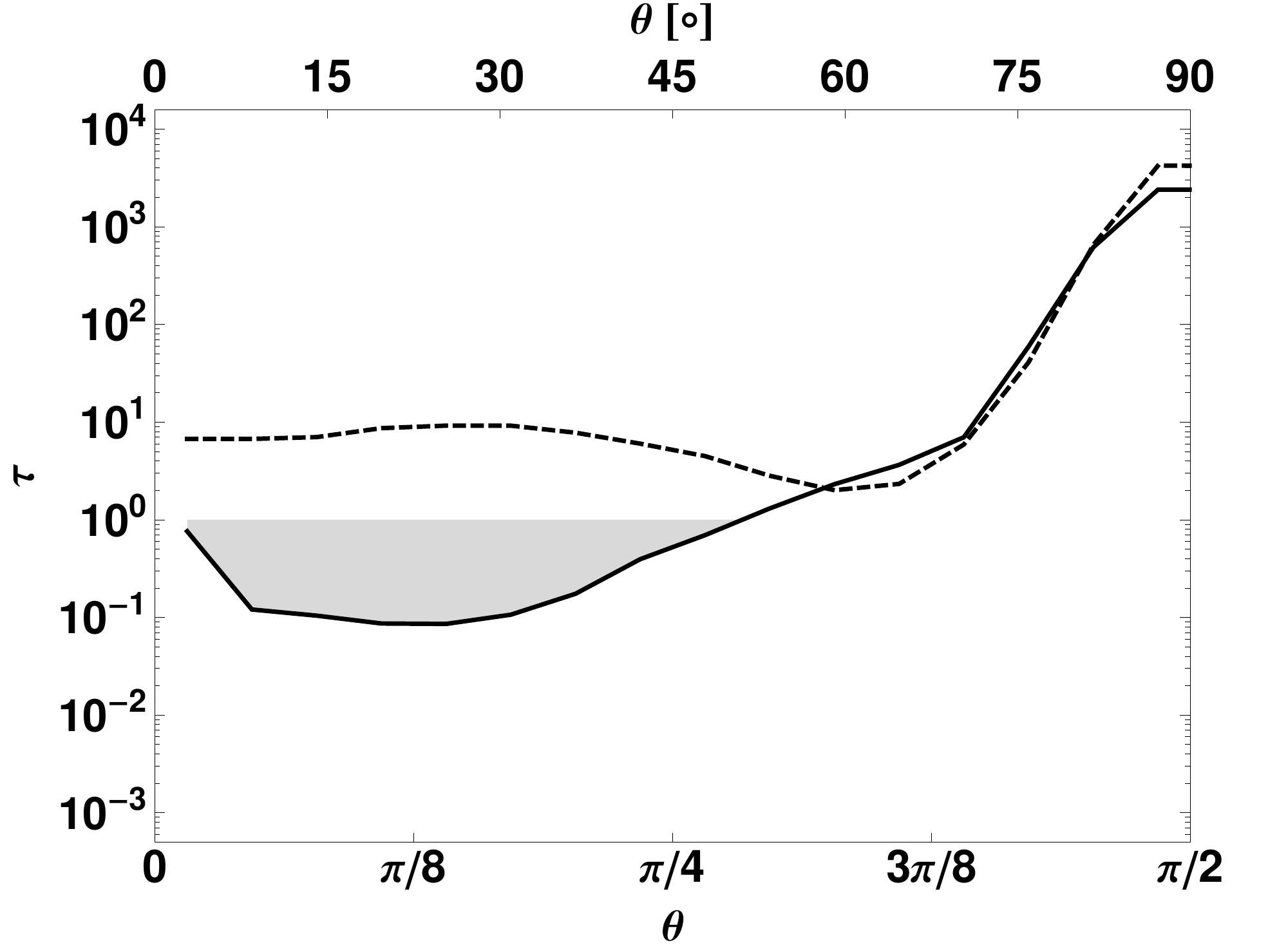}
\label{fig:Tau_vs_theta-d}}\\
\vspace{5mm}
\subfigure[][40~kyr]{\includegraphics[width=0.42\textwidth]{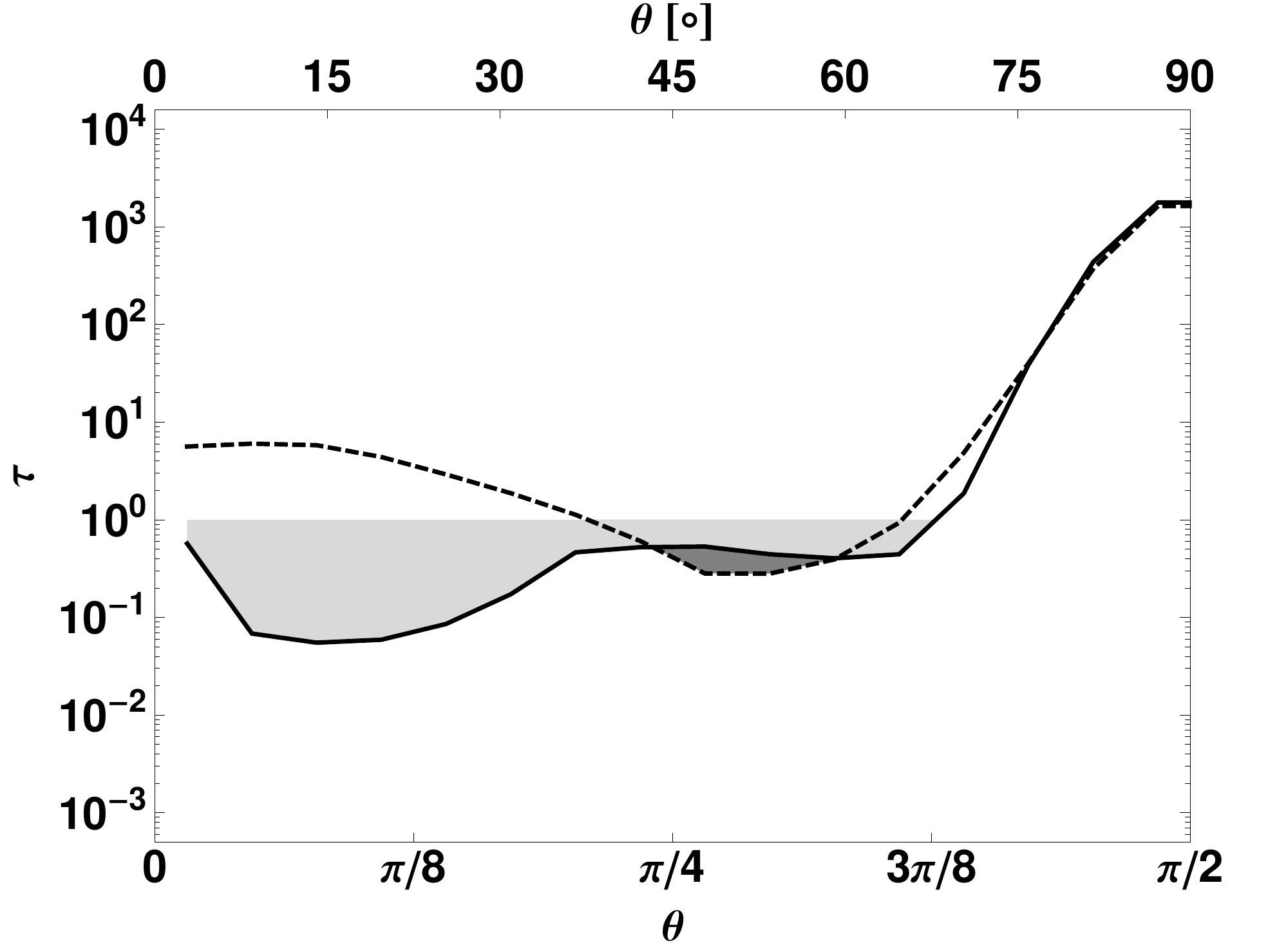}
\label{fig:Tau_vs_theta-e}}
\subfigure[][60~kyr]{\includegraphics[width=0.42\textwidth]{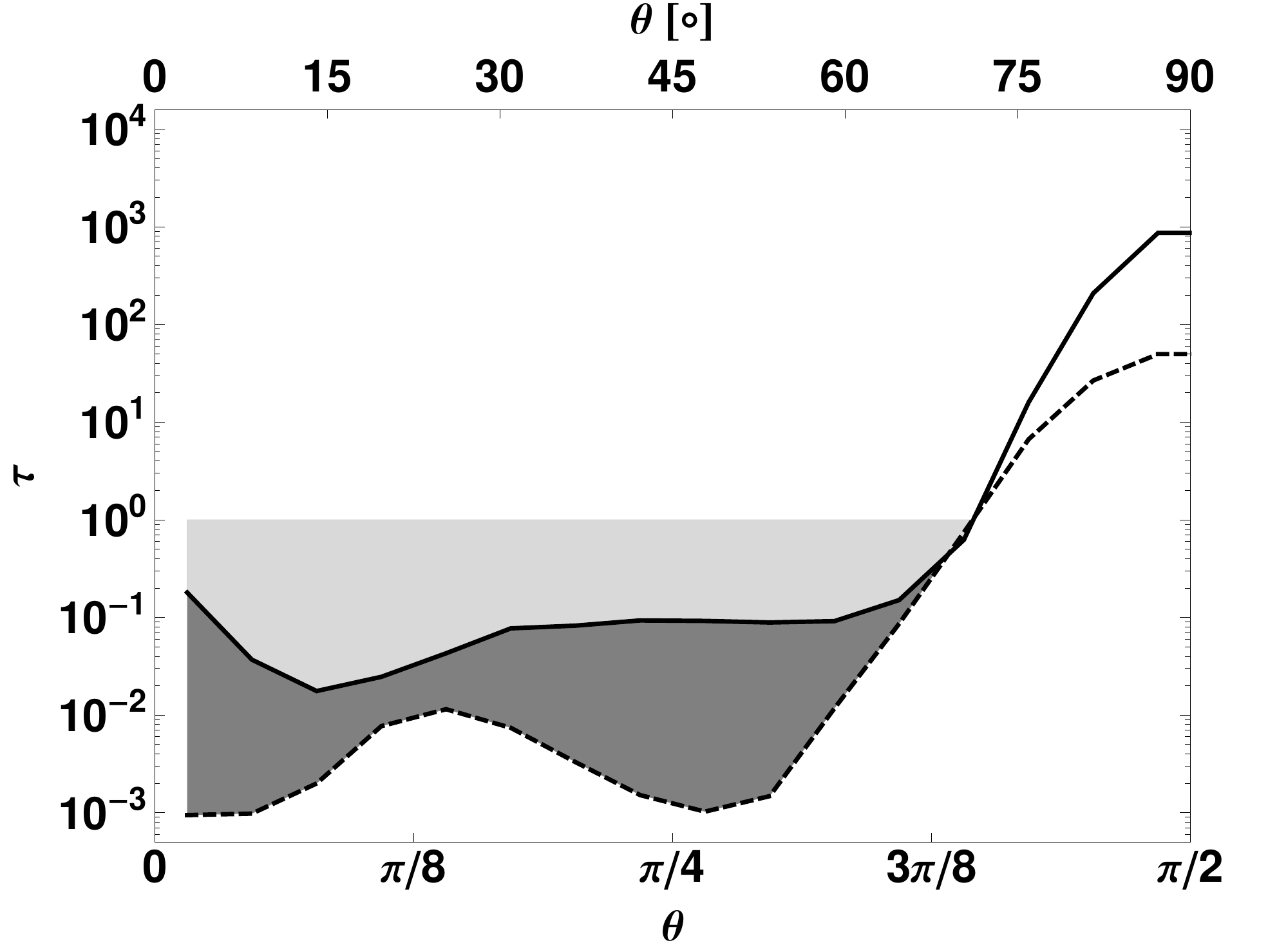}
\label{fig:Tau_vs_theta-f}}
\caption{
Optical depth 
$\tau=\int_{R_*}^{R_\mathrm{max}} \kappa_\mathrm{R}(T(r))~\rho(r)~dr$ 
from core's center up to its outer radius as a function of the polar angle $\theta$ 
(pole at $\theta=0\degr$; midplane at $\theta=90\degr$).
The data are from the collapse case $\beta=-2$ at selected evolutionary times as labeled.
Solid (dashed) lines mark the cases with (without) the protostellar outflow.
Grey shading indicates the optically thin regions where $\tau<1$.
}
\label{fig:Tau_vs_theta}
\end{center}
\end{figure*}

A measure of the flashlight effect is the density contrast between the
disk and its surroundings.  The contrast is several orders of
magnitude greater with the protostellar outflow
(Fig.~\ref{fig:Density_vs_z}).  The flashlight effect's strength can
also be seen in the contrast from pole to equator in the optical depth
for thermal radiation, $\tau=\int_{R_*}^{R_\mathrm{max}}
\kappa_\mathrm{R}\left[T(r)\right]~\rho(r)~dr$
(Fig.~\ref{fig:Tau_vs_theta}).  The early evolution in simulations
without protostellar outflow is characterized by an optical depth
independent of the polar angle $\theta$ --- that is, the protostar's
surroundings are isotropic.  By contrast, simulations including the
protostellar outflow reach the same point in time with the poles
cleared so that the optical depth for the thermal radiation $\tau<1$.
The angular extent of the optically thin region grows in time (gray
shading in Figs.~\ref{fig:Tau_vs_theta-a} to
\ref{fig:Tau_vs_theta-f}).  Along the axis, the optical depth is due
to the protostellar outflow itself and hence remains at higher values.

Even in the calculations without the protostellar outflows, the
radiative forces eventually strengthen, slowing the accretion and at
certain angles reversing it.  Inside the optically thick accretion
disk, the flashlight effect diminishes the thermal radiative forces.  At
angles immediately above the disk, the dynamics of the infalling
envelope are governed by gravity, centrifugal and radiative forces,
while at angles near the poles, centrifugal forces are negligible.
Therefore the inflow is most easily halted along radial lines passing
just outside the body of the disk.  In our calculations without the
protostellar outflows, the radiative and centrifugal forces drive
winds at angles roughly $30\degr$ above the midplane, leading to
decreased optical depths (Figs.~\ref{fig:Tau_vs_theta-b},
\ref{fig:Tau_vs_theta-d}, and \ref{fig:Tau_vs_theta-e}).  But the
winds are not continuously present in time.  The winds' erratic
interaction with the envelope means during some phases the disk is
mass-starved, while in others the disk receives infalling envelope
material (compare the extent of the blue-colored disk feeding region
in Figs.~\ref{fig:VisIt1} to \ref{fig:VisIt3}).  At late epochs, when the star is luminous and
radiative forces become important, simulations lacking the outflow are
characterized by mass loss from the protostellar core due to radiative
feedback on the previously isotropic environment.  The optical depth
decreases in a large range of viewing angles around the pole (see
Fig.~\ref{fig:Tau_vs_theta-f}).  At the same time, the disk's optical
depth decreases due to limited accretion from the envelope.

\begin{figure*}[htbp]
\begin{center}
\subfigure[][15~kyr]{\includegraphics[width=0.44\textwidth]{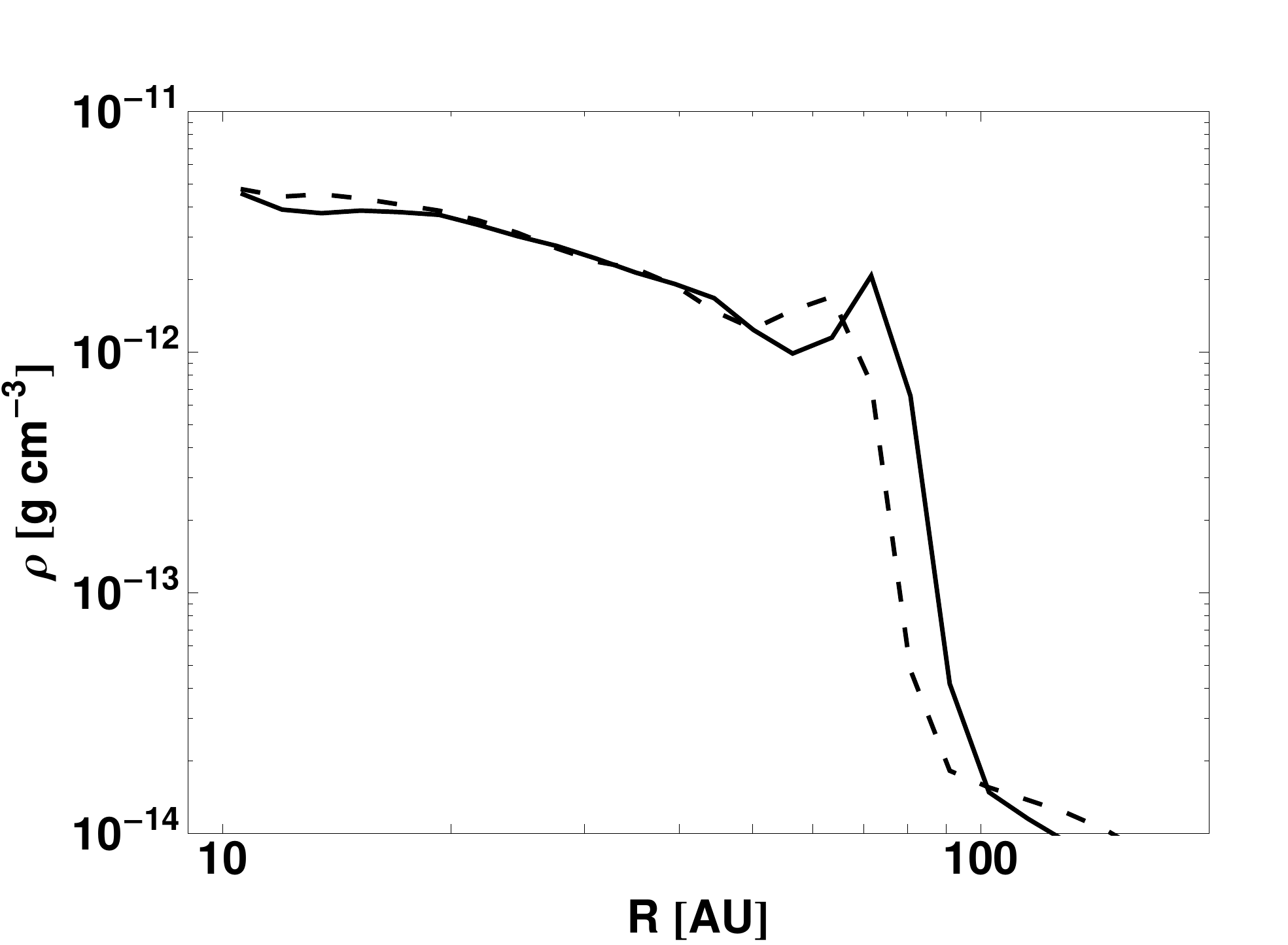}}
\subfigure[][20~kyr]{\includegraphics[width=0.44\textwidth]{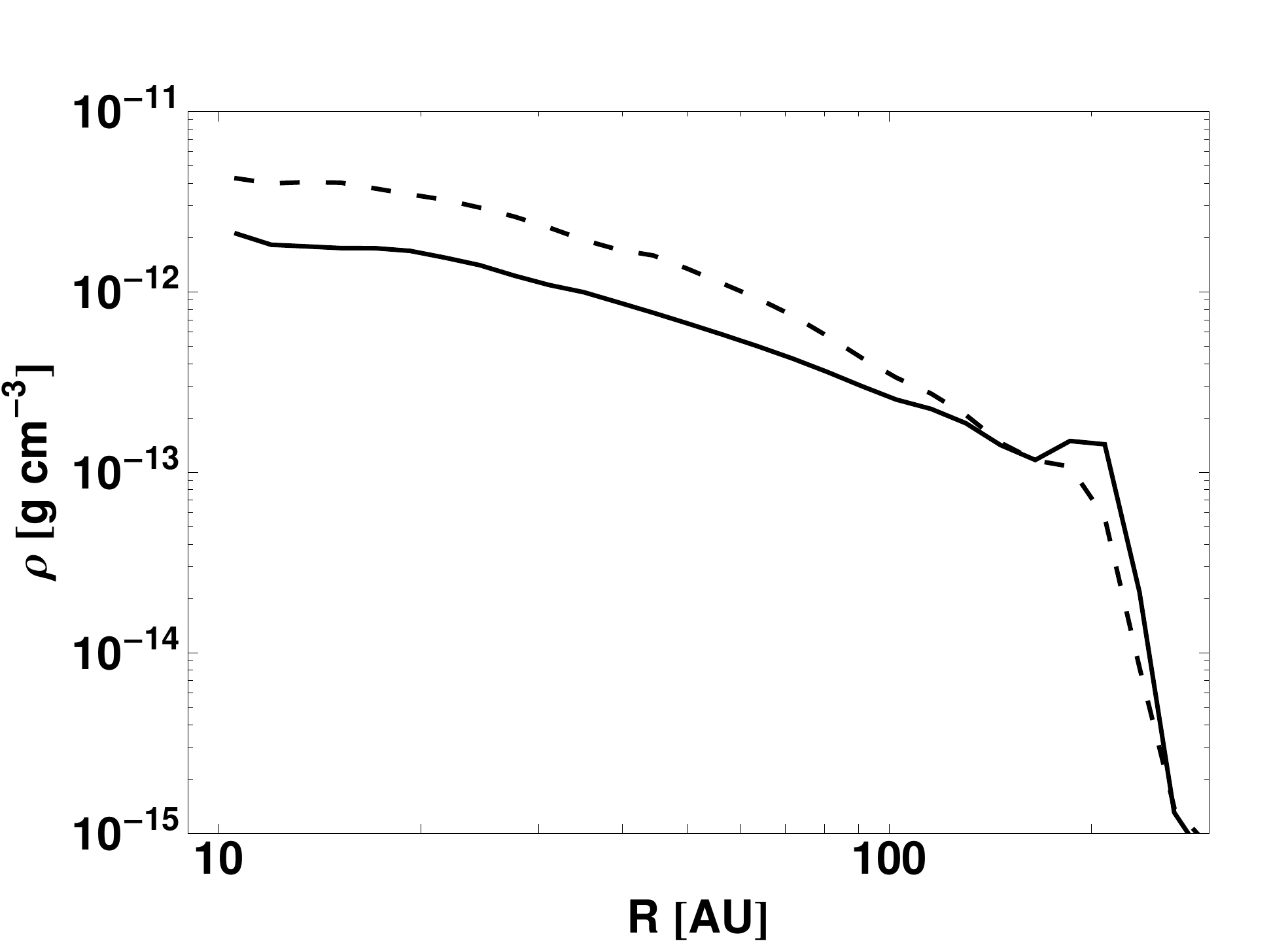}}
\subfigure[][25~kyr]{\includegraphics[width=0.44\textwidth]{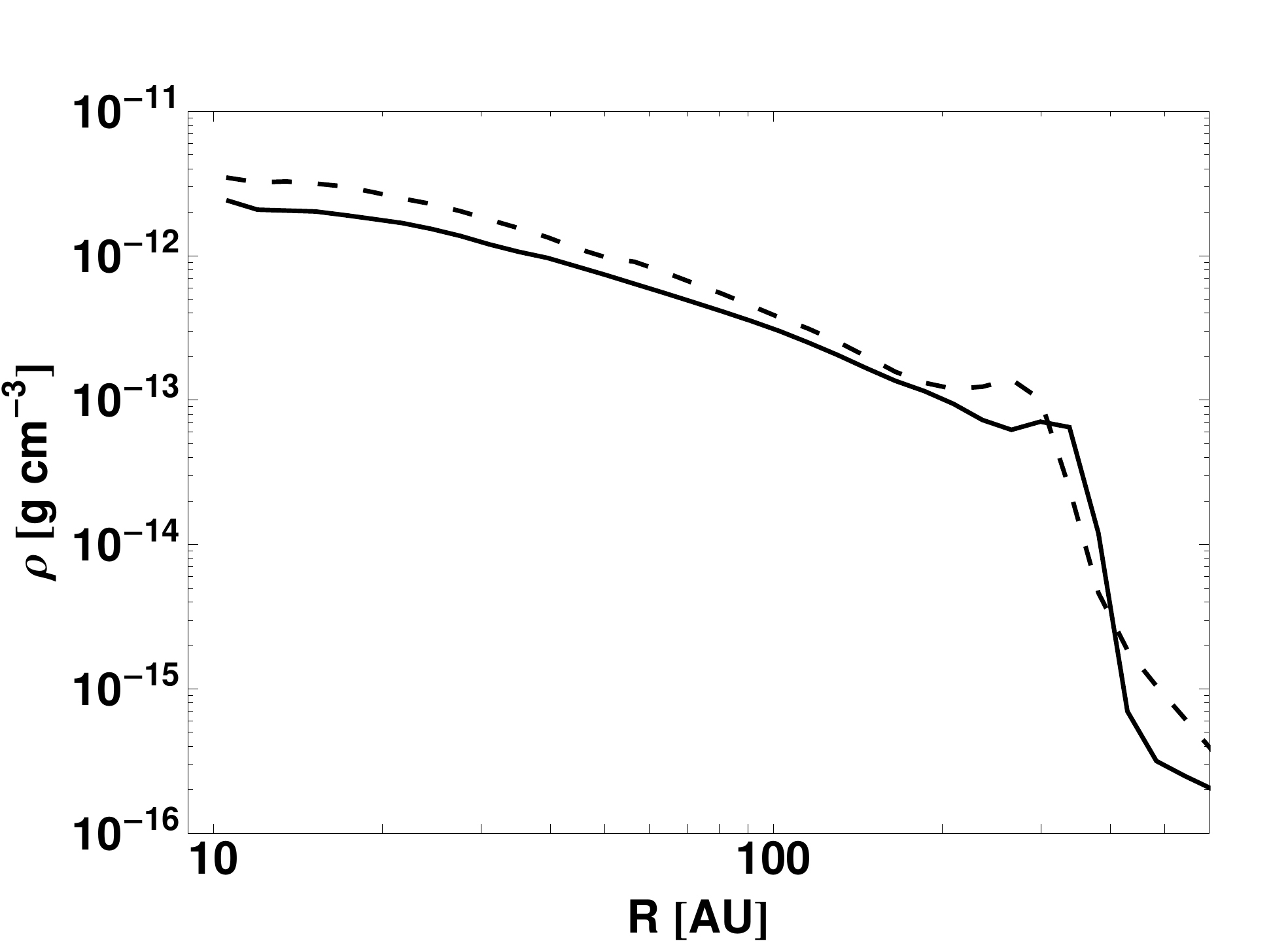}}
\subfigure[][30~kyr]{\includegraphics[width=0.44\textwidth]{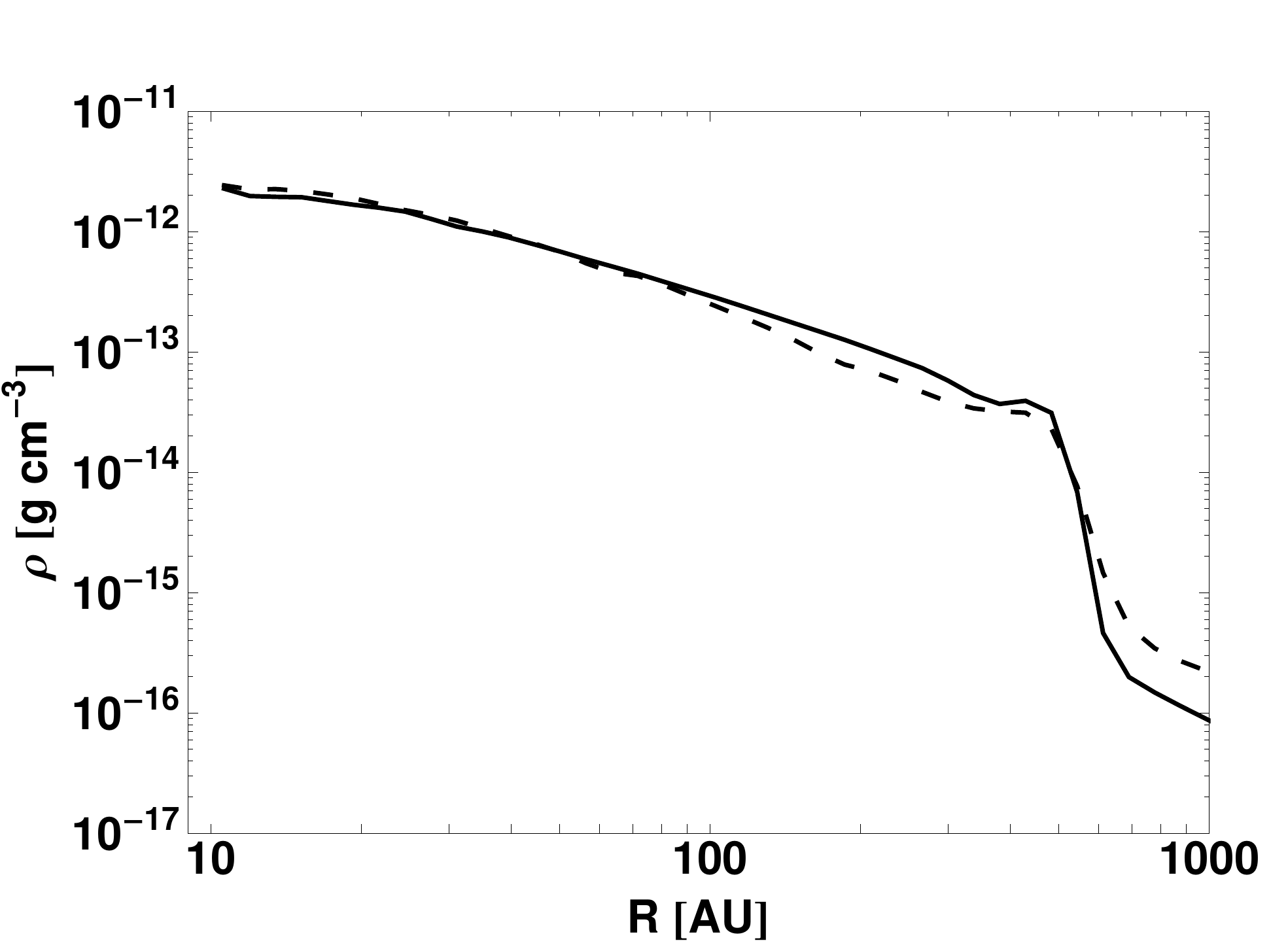}}
\subfigure[][40~kyr]{\includegraphics[width=0.44\textwidth]{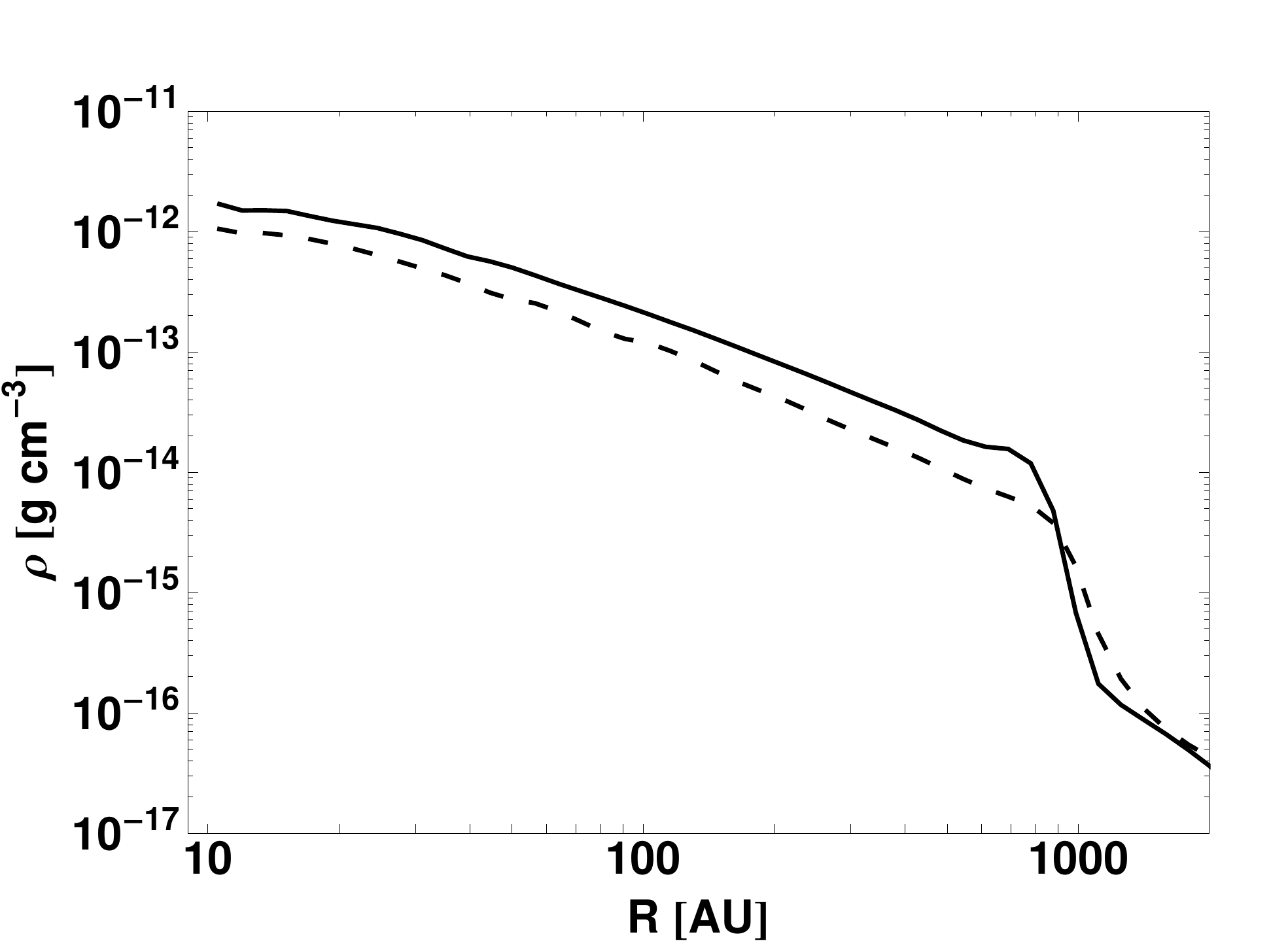}}
\subfigure[][60~kyr]{\includegraphics[width=0.44\textwidth]{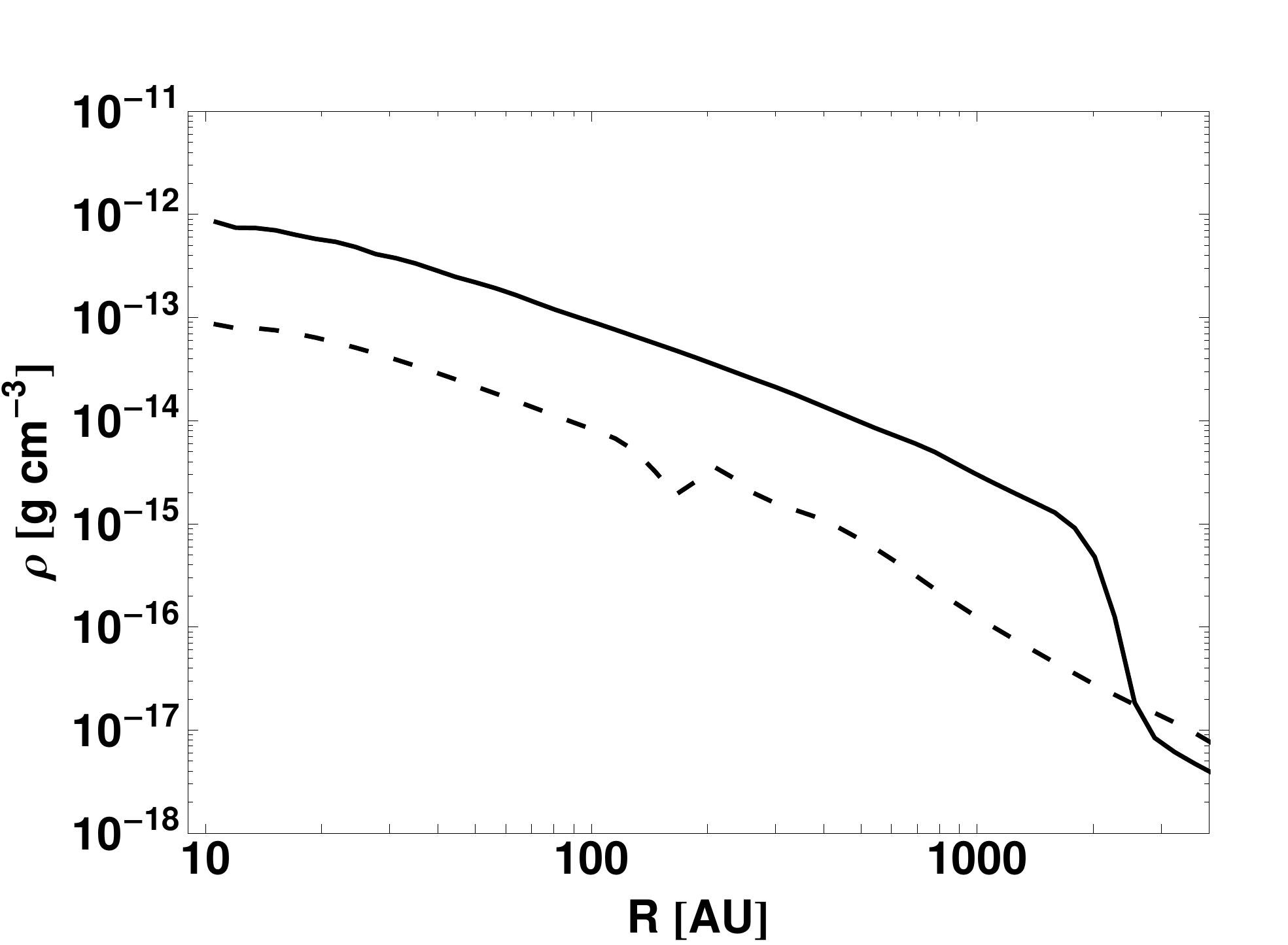}}
\caption{ Midplane gas density vs.\ radius $R$.  The data are from the
  collapse case $\beta=-2$ at selected evolutionary times as labeled.
  Solid (dashed) lines mark the cases with (without) the protostellar
  outflow.  Note that the spatial scale expands from one panel to the
  next, following the disk's growth.  }
\label{fig:Density_vs_R}
\end{center}
\end{figure*}

When we include the protostellar outflow, both the stellar and the thermal radiative flux is
concentrated near the poles, and the bulk of the envelope elsewhere
receives less radiation.  The weaker radiation forces on the envelope
mean a larger reservoir of mass is available to accrete onto the disk
and ultimately onto the star.  Although the disk mass flow rate is
lower, the flow continues for longer.  The slower depletion of the
disk is evident in midplane density cuts
(Fig.~\ref{fig:Density_vs_R}).  The epoch of disk accretion lasts up
to four times longer in the simulations with protostellar outflows
(Table~\ref{tab:run-table}).  However the relatively low accretion
rates in the later stages yield only a few Solar masses of extra
material.  The final masses for the cases with protostellar outflow
are $47 \mbox{ M}_\odot$ and $55 \mbox{ M}_\odot$ for initial density
distributions $\rho \propto r^{-1.5}$ and $\rho \propto r^{-2.0}$,
respectively.  The final masses in the reference runs without the
protostellar outflow are $42 \mbox{ M}_\odot$ and $52 \mbox{
  M}_\odot$, respectively.

\section{Observational Implications}
\label{sect:observations}
The phase of rapid mass accretion onto a protostar is particularly
important for understanding the origins of massive stars.  The
challenge is to capture observations during this brief phase.  As the
envelope collapses under its own gravity, the gas density above and
below the disk decreases, allowing the radiation-driven outflow to
broaden.  The increasing opening angle makes the central protostar
directly visible for observers over a larger portion of the sky.  The
cases we discuss here indicate that we could observe high-mass
protostars either directly through their outflow cavities, if the
viewing angle is nearly pole-on, or indirectly via scattered light \citep[see also][]{Bruderer:2009p28194}.
Several candidate massive protostars appear to be bloated
\citep{Linz:2009p3056, Bik:2012p3198, Palau:2013p16699} suggesting
they are accreting gas at high rates \citep{Hosokawa:2010p690, Kuiper:2013p19987}.  For
more quantitative predictions of the observability and expected
signatures of massive protostars during rapid mass accretion,
hydrodynamical models like those presented here ought to be
complemented by detailed radiative transport calculations.

Radiative feedback reduces the star formation efficiency of the
pre-stellar core collapse to roughly 50\%.  In our simulations, about
half the initial mass is expelled through the outer boundary at 0.1~pc
and therefore does not contribute to the growth of the central star.
The lost mass is mostly envelope material entrained in the
protostellar outflow (at early times) and accelerated by
radiation (at later times).  There are minor contributions from disk
winds and the redirection of the disk accretion flow into the
protostellar outflow.  This outcome is consistent with the tight
correlation observed between outflow mass and core mass over many
orders of magnitude by \citet{Beuther:2002p3040}.  High-mass molecular
outflows could resemble their counterparts around lower-mass stars in
resulting from entrainment in a collimated protostellar outflow, as
suggested by the match of the IRAS~05358+3543 outflow system to a
shock entrainment model \citet{Beuther:2002p3046}, and by the
correlation of outflow mass with source bolometric luminosity across
outflows from both low-mass and high-mass stars \citet{Wu:2004p20604}.

Finally, we note that since the outflowing gas entrains part of the envelope, we expect that the observed outflow rates exceed
the sum of the masses being lost in the protostellar outflow and
radiation-driven winds.

\section{Limitations \& Outlook}
\label{sect:limitations}
An important limitation of this study is the fact that the
protostellar outflows are injected following a sub-grid prescription,
rather than calculated from first principles.  Unfortunately,
determining the outflow launching forces accurately would require
significantly better spatial resolution of the central regions.
Achieving higher resolution with an explicit dynamical code requires
short timesteps, so that the long-term evolution we studied here is so
far accessible only using such an outflow prescription.

With a clear focus on disentangling the kinematic protostellar outflow
feedback from the radiation pressure feedback, we considered an
outflow-to-accretion-rate conversion factor of~0.01 only in this
exploratory study.  Because protostellar outflows and radiation
interact, and do not simply add to each other's effects, a range of
outflow strengths should be considered in more detailed follow-up
studies.

Also, shock waves driven into the surroundings when the outflow is launched
would be even more important if the outflow were episodic.  In our
calculations the outflow momentum is determined from the stellar
accretion rate, which in turn is governed mostly by the disk's
evolution (sect.~\ref{sect:results-disk} above).  Hence the outflow
evolves smoothly in time and weakens at later epochs.  In contrast to
this gradual evolution, similar collapse scenarios investigated in 3-D
show long-term variations in the accretion rate due to gravitational
torques from nonaxisymmetric density disturbances in the disk
\citep{Kuiper:2011p21204}.  Rapid accretion bursts also occur if the
protostar and envelope are coupled together by the inward accretion
flow and outward radiation forces \citep{Kuiper:2013p19987}.  Such an
accretion history could mean episodic protostellar outflows driving
repeated shocks into the surroundings.  In this scenario, the outflow
may never reach a quasi-stationary stage and its opening angle could
be episodic also.  This picture is best investigated including
magnetic fields, so that the outflow launching is coupled to the
magneto-hydrodynamical evolution of the infall, disk formation, and
accretion flow.

The protostellar evolution of the forming star in the sink cell relies on the assumption that the accretion flow over the 10~AU sink cell radius denotes the actual stellar accretion rate.
The potential formation of a binary or multiple stellar system within this region has not been taken into account.
Hence, the resulting final stellar masses obtained in the simulations denote an upper mass limit of the most massive star of such a multiple system.

The stars' growth continues much longer in our models including the
protostellar outflows.  The cause is the enhanced flashlight effect
resulting from the clearing of the bipolar regions.  The anisotropy in
the thermal radiation could be stronger still if we included the
accretion disk's innermost optically thick dust-free region.
\citet{Kuiper:2013p17358} treated this region using temperature- and
density-dependent mean gas opacities, while our simpler but
longer-term calculations involve a constant (and generally lower) gas
opacity.  Accounting for both effects would likely result in an even
longer disk accretion phase and lower accretion rates.

Finally, the disk lifetime is influenced by effects other than
radiative acceleration, not accounted for in our simulations,
including ionization, disk fragmentation under self-gravity, and the
winds and UV radiation from nearby sources.

\section{Summary}
\label{sect:summary}
We numerically model the whole accretion phase of massive star
formation, including the effects of protostellar outflows, stellar and thermal
radiative feedback.  We examine both the outflows' direct kinematic
feedback, and the indirect consequences for the later epoch of
radiative feedback.  To enable such long-term calculations, we
restrict the simulations to (1)~axially- and midplane-symmetric
configurations and (2)~launching the protostellar outflows using a
sub-grid model.

The protostellar outflow and radiative forces together drive the
surrounding material through two stages of evolution.  At early times,
kinematic feedback yields a bipolar outflow cavity whose shape comes
from the collimation in the sub-grid launching model.  At later times,
radiative acceleration becomes significant and the outflow broadens.
In addition, the protostellar outflow and radiative forces drive a
high-momentum-flux wind along the disk's upper (and lower) surface.
These outflows ultimately entrain part of the envelope, reversing its
infall and expelling it from the system.  The outflow rates measured
at the core's outer edge are thus greater than those near the star and
accretion disk.

The protostellar outflow's kinematic feedback reduces the accretion
rate onto the disk and ultimately onto the star, compared with
calculations lacking the outflow.  In later stages, however, the
protostellar outflow cavities strengthen the flashlight effect,
extending it from the starlight near the disk to the thermal radiation
throughout the envelope.  Although the thermal radiation forces are
still significant, as they must be for such a luminous protostar, the
flashlight effect means the thermal radiation ejects less of the
envelope.  For the cases considered here, the retained envelope
material allows the star a longer accretion phase, more than balancing
the decrease in accretion rate from the kinematic feedback.  For the
future it is desirable to examine whether this result holds for other
initial conditions and ratios of outflow to accretion rate.  Moreover,
the final mass of the forming star depends on the lifetime of the
circumstellar accretion disk, which could be influenced by several
processes we have not included, such as ionization, magnetic fields,
disk fragmentation, and UV radiation and winds from nearby sources.
Nonetheless, the results of this study clearly show that long-term
effects must be accounted for when investigating the final feedback
efficiency of protostellar outflows.

\acknowledgments
We thank Takashi Hosokawa for many critical and fruitful discussions. 
R.~K.~acknowledges financial support by the German Academy of Science Leopoldina within the Leopoldina Fellowship Programme, grant No.~LPDS 2011-5.
R.~K.~further acknowledges funding from the Max Planck Research Group Star formation throughout the Milky Way Galaxy at the Max Planck Institute for Astronomy.
Major portions of this work were conducted at the Jet Propulsion Laboratory, California Institute of Technology, operating under a contract with the National Aeronautics and Space Administration (NASA).

\vONE{
\appendix
\section{Convergence}
For the simulations presented, we use a 2D static grid in spherical coordinates with maximum resolution $\Delta r \times \Delta \theta= 1.27 \times 1.04 \mbox{ AU}^2$.
To check for numerical convergence, 
we compare one of the simulations with three additional simulation runs using a different resolution, namely with a factor of two lower, with a factor of two higher, and with a factor of four higher resolution in each dimension, respectively.
The initial condition for these simulations are identical to the one with $M_\mathrm{core} = 100 \mbox{ M}_\odot$ and an exponent of the density slope of $\beta=-2$.
The factor two higher resolution run was computed up to 34~kyr of evolution.
The factor four higher resolution run was computed up to 14.2~kyr of evolution, which marks the onset of the protostellar outflow launching.
Due to the high speed of the outflow, the Courant-Friedrich-Levy condition on the numerical hydrodynamical timestep limits the realizability of longer-term and/or higher resolution studies.

The resulting mass growth of the central star, the maximum velocity of the protostellar outflow, and the accumulated injected mass by the outflow without the entrainment is presented in Fig.~\ref{fig:Convergence}.
\begin{figure*}[p]
\begin{center}
\includegraphics[width=8cm]{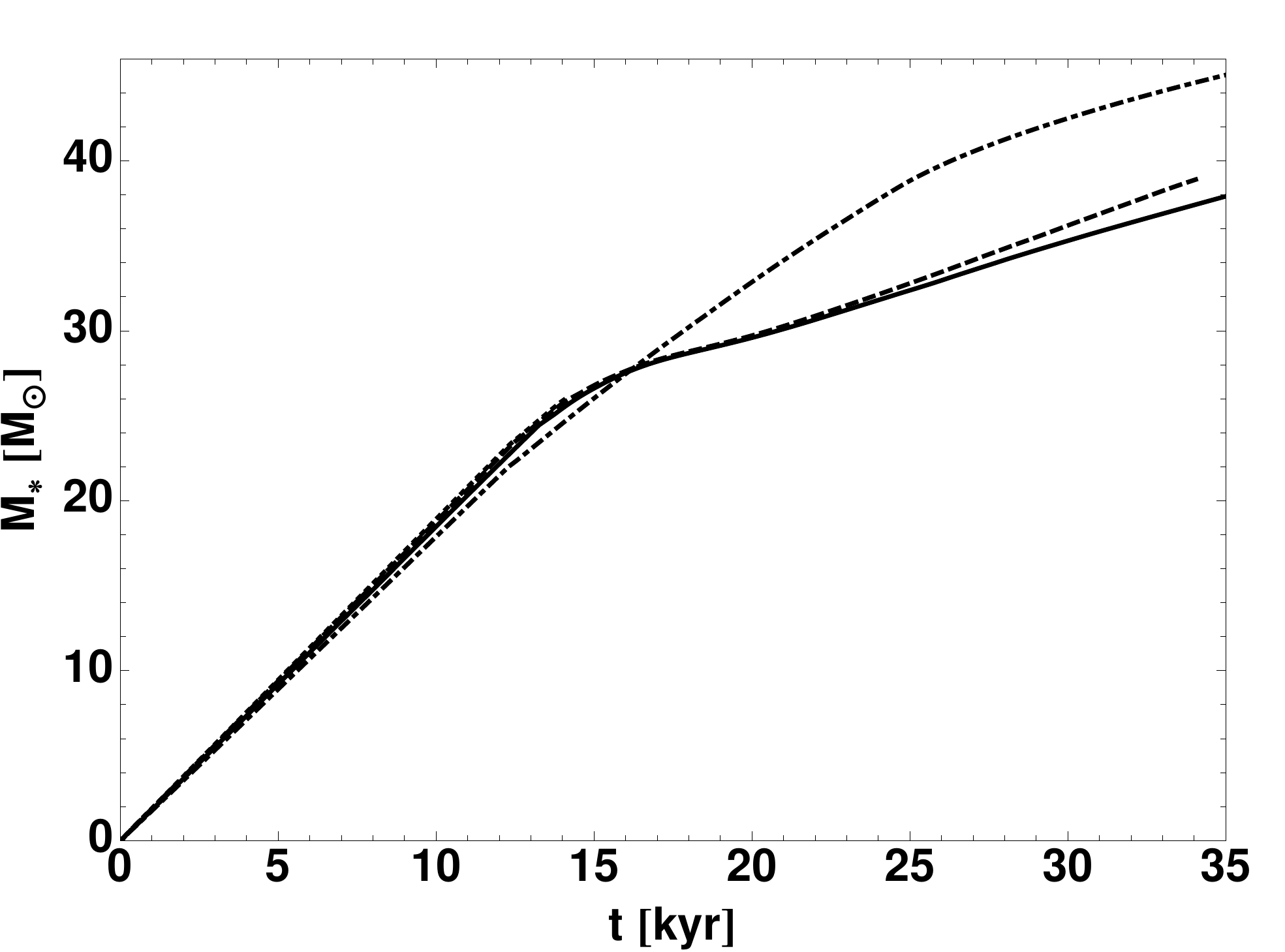}
\includegraphics[width=8cm]{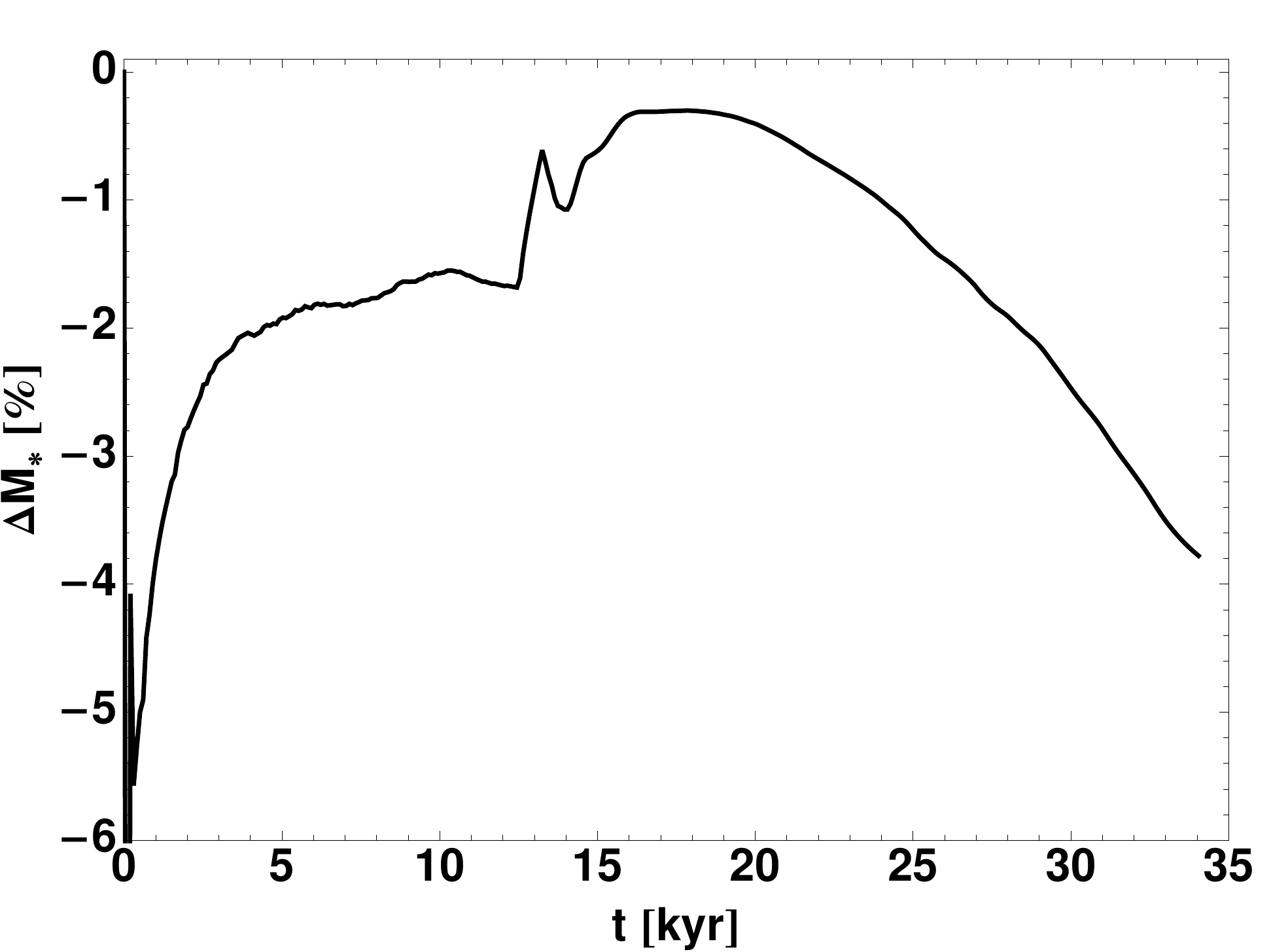}\\
\includegraphics[width=8cm]{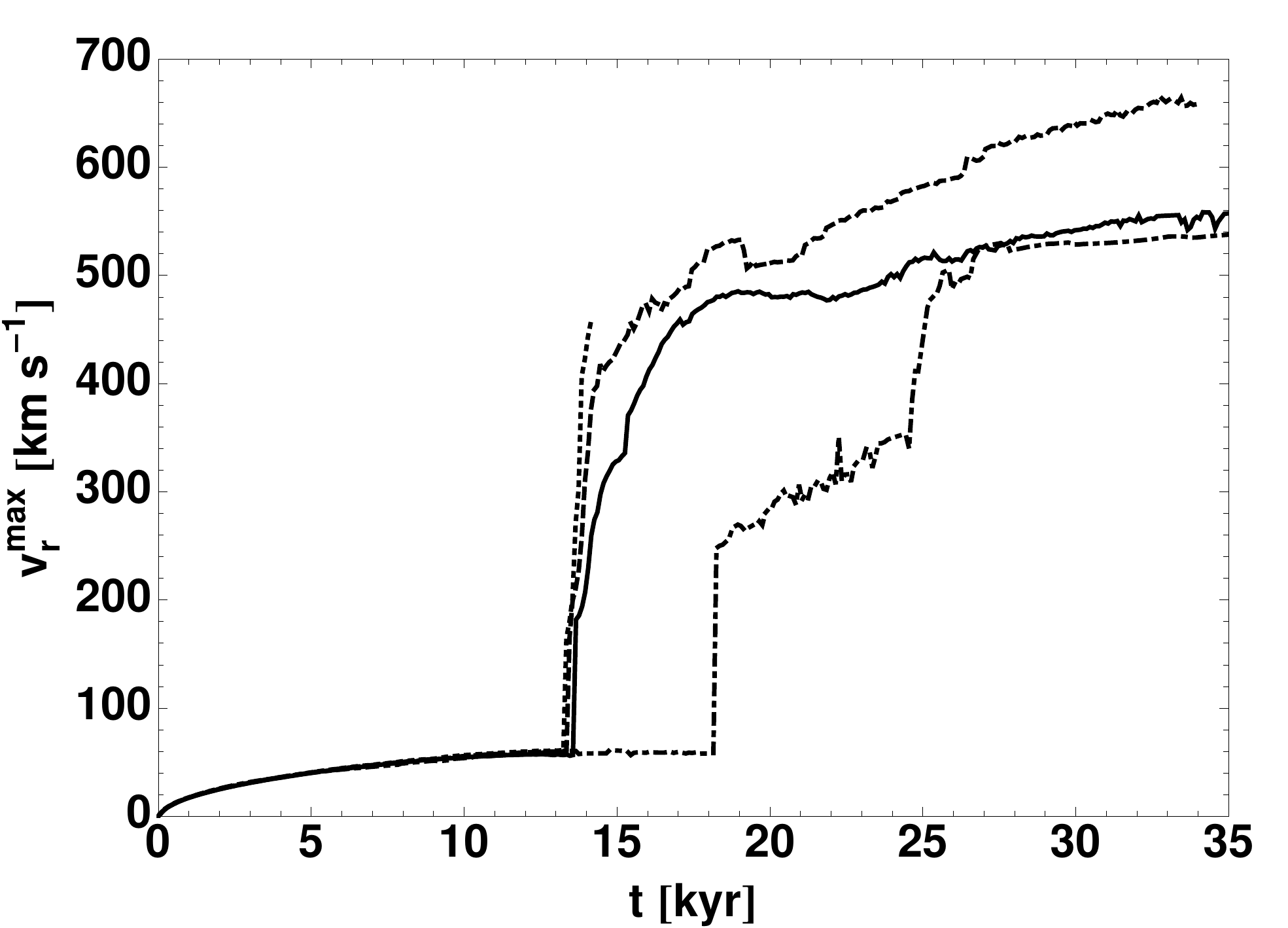}
\includegraphics[width=8cm]{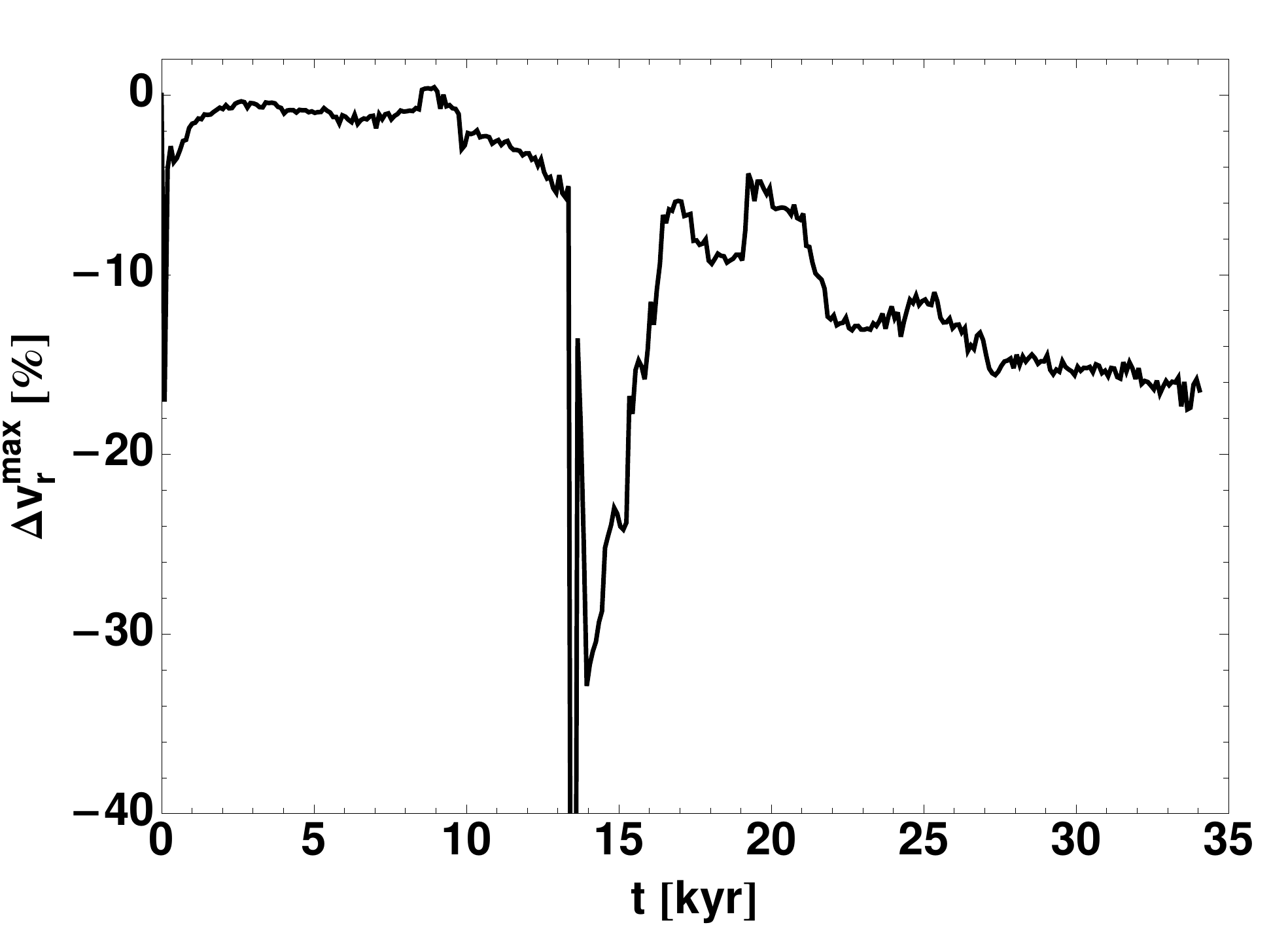}\\
\includegraphics[width=8cm]{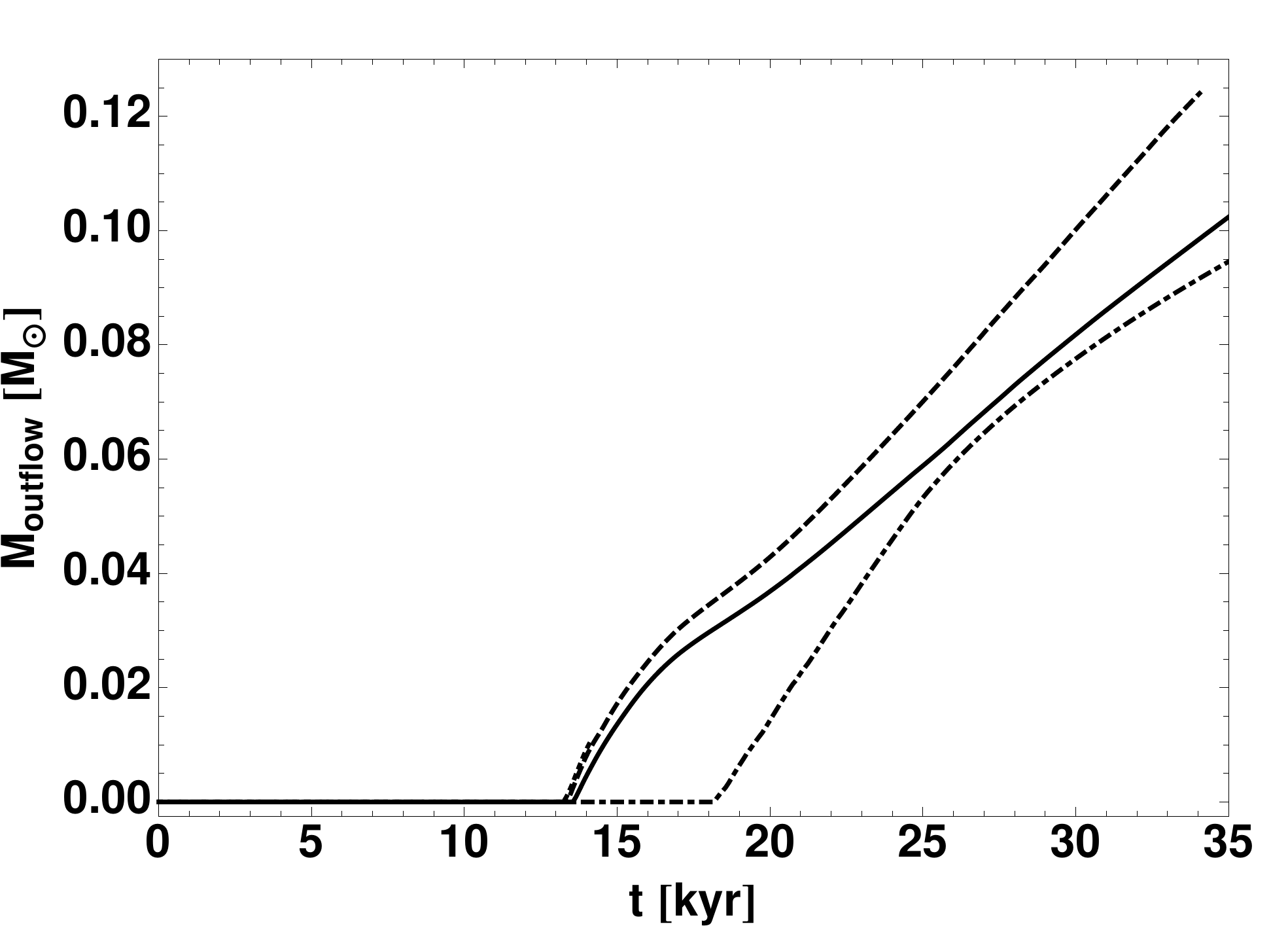}
\includegraphics[width=8cm]{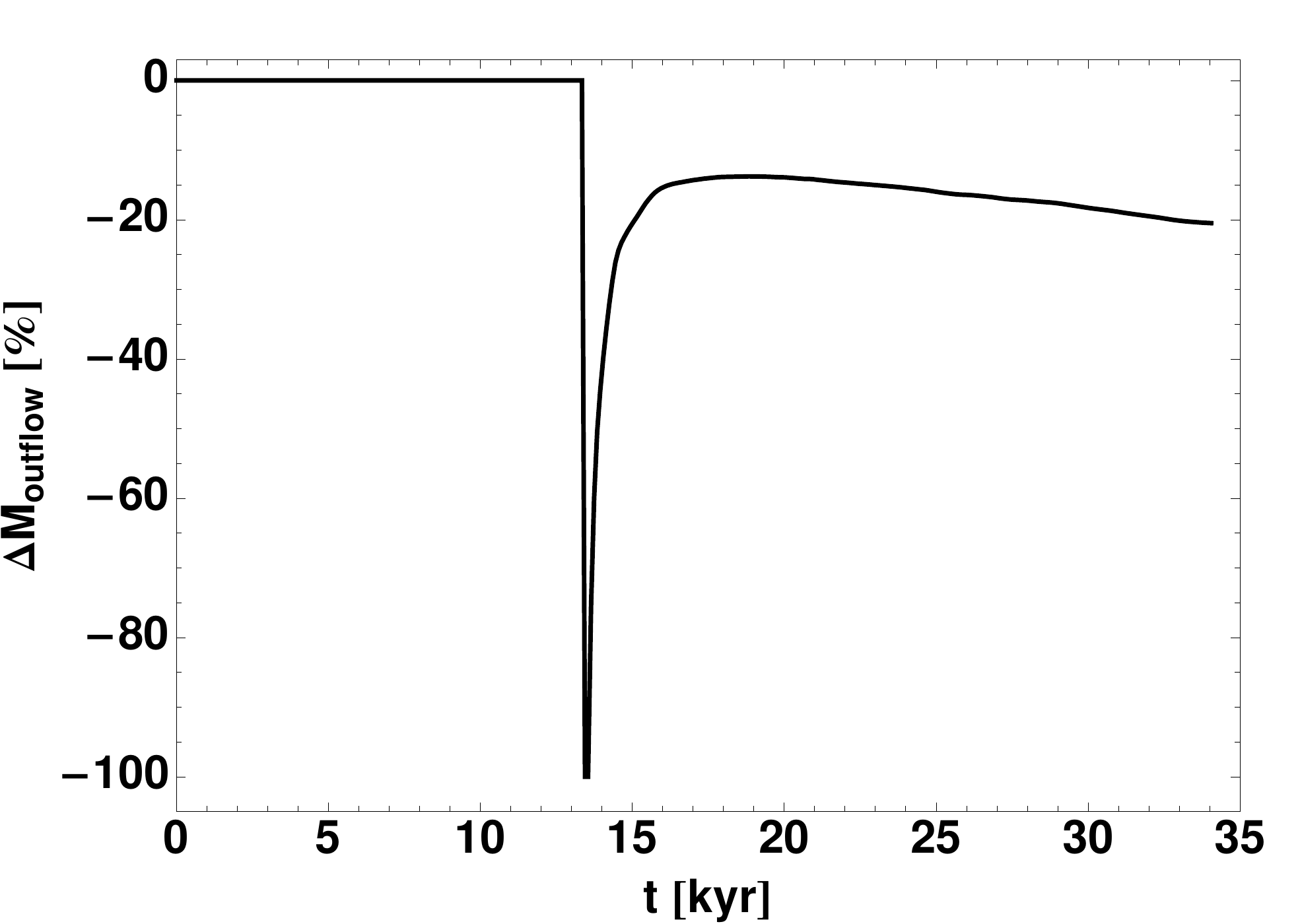}
\caption{
Convergence study of the simulations including the launching of early protostellar outflows.
Left panels show from top to bottom
the stellar mass growth, 
the maximum velocity of the protostellar outflow, and 
the accumulated injected mass of the outflow. 
Different line styles denote 
the default (solid),
the factor two lower (dot-dashed),
the factor two higher (dashed), and
the factor four higher resolution run (dotted).
The highest resolution simulations were computed up to 14.2~kyr of evolution only and, hence, are only marginally visible.
Right panels show the difference of the quantities between the default and the factor two higher resolution simulation.
}
\label{fig:Convergence}
\end{center}
\end{figure*}
The stellar accretion rate is controlled by the large scale collapse at early times and the accretion disk physics at later times.
The default resolution of our simulation runs is already high enough to follow the pressure scale height of the forming accretion disk; 
as a result, the differences in the stellar mass growth $M_*$ between the simulations with default or higher resolution stay below a few percent.

With respect to the outflow properties, the a priori defined angular distribution of the outflow velocity (Eq.~\ref{eq:angularweighting}) yields a faster outflow at the poles with increasing polar resolution 
as well as an earlier launching time (due to the higher velocity/momentum at the pole).
Both effects are clearly visible in Fig.~\ref{fig:Convergence}, middle and lower panel, at about 13 -18~kyr of evolution (the onset of outflow launching).
After this difference in the launching time, the maximum velocity of the outflow $v_\mathrm{r}^\mathrm{max}$ remains within a 20\% difference between the default resolution and the factor of two higher resolution run.
As a side effect of the delayed outflow launching in the low resolution run, also the stellar mass growth of the low resolution differs visibly from the other runs in default and higher resolution due to the fact that accretion along the polar regions sustains for a longer period in time.
The accumulated injected mass by the outflow $M_\mathrm{outflow}$ agrees within 20\% as well.

The dependence of the injected outflow velocity at the poles (see Eq.~\ref{eq:angularweighting} within model section) on the numerical polar resolution in use makes it difficult to derive further conclusions.
In principle, this dependence could have been avoided in the convergence study by computing the injected outflow velocity as the integral value of Eq.~\ref{eq:angularweighting} over the polar extent of a grid cell rather than using the grid cell center as a reference point.

In conclusion, the quantitative results such as the outflow velocity and mass load differ on the 20\% level and clearly depend on the sub-grid model of the outflow injection.
A scan of the broad parameter space goes beyond the scope of this pioneering study.
We also have to admit that a quantitative determination of the outflow quantities cannot be the aim of this study and would e.g.~better be tackled within MHD simulations of the outflow launching process.

Our focus is the protostellar outflow feedback's long-term consequences, including its influence on the radiation-pressure-dominated epoch.  
In a nutshell, the main conclusions are 
(a) the outflow amplifies the flashlight effect and the large-scale anisotropy of the core's radiation field; 
as a result, 
(b) the core loses less mass, 
(c) a large reservoir of gas in the core's outer region remains available for sustained feeding of the disk, and 
(d) radiation feedback is lessened, compensating for the outflow's kinematic feedback so that growth in the super-Eddington regime is easier than expected.
These findings are of general and qualitative interest, and, hence, adequately represented in the runs with the default grid resolution.
}

\bibliographystyle{apj}
\bibliography{Papers}

\end{document}